\documentclass[12pt,a4paper]{JHEP}

\usepackage{amsmath,amsfonts}
\usepackage{epic,eepic,epsfig}

\newcommand{\eq}[1]{eq.(\ref{#1})}
\newcommand{\sssection}[1]{\vspace{3mm}\noindent{\bf{#1}}}
\newcommand{\bphi}{\boldsymbol{\phi}}
\newcommand{\weight}{\boldsymbol{\lambda}}
\newcommand{\aroot}{\boldsymbol{\alpha}}
\newcommand{\real}[1]{\text{Re}(#1)}
\newcommand{\imaginary}[1]{\text{Im}(#1)}
\newcommand{\sign}{\text{sign}}
\newcommand{\arcsinh}{\text{Arcsinh}}
\newcommand{\integer}{\mathbb{Z}}

\newcommand{\drawcenteredtext}[3]{\put(#1,#2){\makebox(0,0){#3}}}%
\newcommand{\drawlefttext}[3]{\put(#1,#2){\makebox(0,0)[l]{#3}}}%
\newcommand{\drawrighttext}[3]{\put(#1,#2){\makebox(0,0)[r]{#3}}}%

\newcommand{\drawpath}[4]{\path(#1,#2)(#3,#4)}%
\newcommand{\drawdotline}[4]{\dottedline[.]{1}(#1,#2)(#3,#4)}%

\newcommand{\drawleftbrace}[3]%
{\drawcenteredtext{#1}{#2}{$\left\{ \rule[0mm]{0mm}{#3mm} \right.$}}%

\newcommand{\drawrightbrace}[3]%
{\drawcenteredtext{#1}{#2}{$\left\} \rule[0mm]{0mm}{#3mm} \right.$}}%

\newcommand{\drawoverbrace}[3]%
{\drawcenteredtext{#1}{#2}{$\overbrace{\rule[0mm]{#3mm}{0mm}}$}}%

\newcommand{\drawunderbrace}[3]%
{\drawcenteredtext{#1}{#2}{$\underbrace{\rule[0mm]{#3mm}{0mm}}$}}%

\newcommand{\drawarc}[5]%
{\put(#1,#2){\arc{#3}{#4}{#5}}}%

\title{Restricting affine Toda theory to the half-line}

\author{Gustav W Delius \\
Department of Mathematics, King's College London,\\
Strand, London WC2R 2LS, UK.\\
E-mail: \email{delius@mth.kcl.ac.uk}\\
\href{http://www.mth.kcl.ac.uk/~delius/}
{WWW: \tt{http://www.mth.kcl.ac.uk/$\sim$delius/}}
}

\abstract{
We restrict affine Toda field theory to the half-line by imposing
certain boundary conditions at $x=0$. The resulting theory possesses
the same spectrum of solitons and breathers as affine Toda theory
on the whole line. The classical solutions describing the reflection
of these particles off the boundary are obtained from those on the whole
line by a kind of method of mirror images. Depending on the boundary
condition chosen, the mirror must be placed either at, in front, or behind
the boundary. We observe that incoming solitons are converted
into outgoing antisolitons during reflection.
Neumann boundary conditions allow additional solutions
which are interpreted as boundary excitations (boundary breathers). 
For $a_n^{(1)}$ and $c_n^{(1)}$ Toda theories, on which we concentrate 
mostly, the boundary conditions which we study are among the integrable
boundary conditions classified by Corrigan et.al.
As applications of our work we study the vacuum solutions
of real coupling Toda theory on the half-line and we perform
semiclassical calculations which support recent conjectures for
the $a_2^{(1)}$ soliton reflection matrices by Gandenberger.
}

\begin{document}

\section{Introduction and Overview}

Affine Toda theories are integrable relativistic field theories in one space
and one time dimension. They have
been widely studied, both classically and at the quantum level, because of
their remarkable properties and interesting algebraic structure. For a
review see \cite{Cor94}. The simplest affine Toda theory is the sine-Gordon
model.

\EPSFIGURE{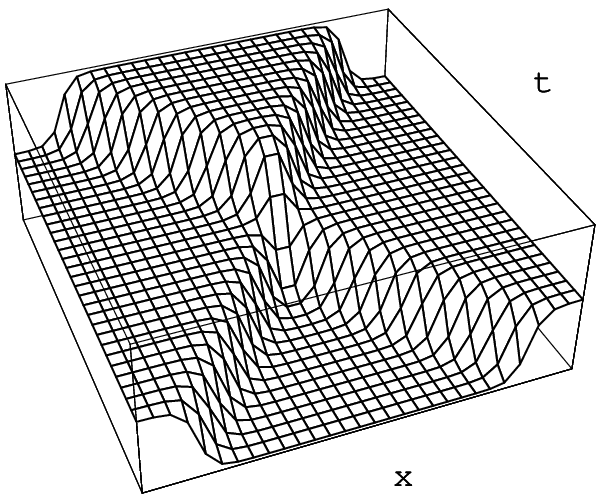,width=6.5cm}
{The interaction of a sine-Gordon soliton with an antisoliton.
\label{fig1}}

The affine Toda equations of motion possess
soliton solutions \cite{Hol92,Oli93b}. 
These are kink configurations which interpolate
between the different vacua of the Toda potential. The characteristic
property of solitons is that they propagate without dispersion and
that even after collision with other solitons they regain their original
shape. As an example figure \ref{fig1} shows a solution of the sine-Gordon
model which describes an antisoliton and a soliton moving in opposite
directions. One sees how they move through each other, the only effect of
their interaction being a time delay.

These solitons can be quantized and then their scattering is described
by factorizable S-matrices obeying the axioms of crossing symmetry,
unitarity, Yang-Baxter equation and the bootstrap principle
\cite{Zam}. These S-matrices can be obtained by exploiting the
affine quantum group symmetry of affine Toda theory \cite{Ber}.

Interestingly, certain of the classical 
multi-soliton solutions of affine Toda theory seem to make
physical sense also if one views them only on the left half-plane.
To illustrate this we have taken the solution of figure \ref{fig1}
and have restricted it to $x<0$ to arrive at figure \ref{fig2}.
Now it describes a reflection process off a boundary at $x=0$ . 
The right-moving antisoliton
hits the wall at $x=0$ and is reflected back as a soliton with
opposite velocity.

\EPSFIGURE{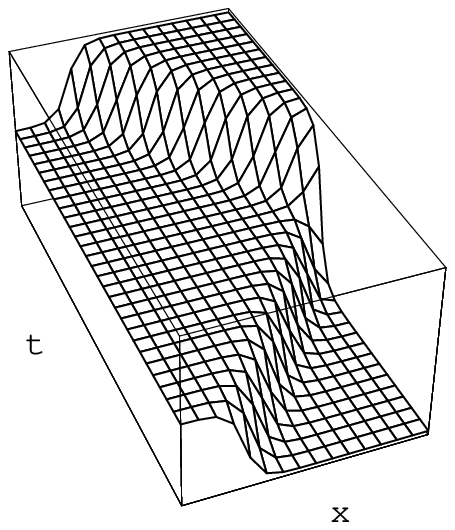,width=4.4cm}
{The solution of figure \ref{fig1} interpreted as describing
the elastic reflection of an antisoliton off a boundary at $x=0$.
\label{fig2}}

In this paper we will investigate this possibility of restricting 
affine Toda theory to
the left half-line by putting a reflecting boundary at $x=0$. In other
words we will impose boundary conditions at $x=0$ which among
all multi-soliton solutions of Toda theory on the whole line select
only those which, when viewed on the left half-line, describe the
purely elastic reflection of solitons off the boundary. 

The existence of these classical solutions describing solitons
on the half line opens up the challenge of deriving the corresponding
quantum theory. In addition to the S-matrix describing the scattering
of the solitons among themselves one will also have to construct
a reflection matrix describing the reflection of the solitons off the
boundary. The complete set of axioms which such reflection matrices 
have to satisfy was given by Ghoshal and Zamolodchikov \cite{Gho}
extending the early work of Cherednik \cite{Che}. Gandenberger \cite{Gan98}
has found solutions to these axioms for the case of $a_2^{(1)}$ Toda theory.

In this paper we will prepare the ground for extending these investigations
into the quantum theory of solitons on the half-line
by performing classical and semiclassical calculations.
By using the method of images we are
able to complement the investigations which have already 
been performed on this subject. Corrigan et.al. \cite{Bow95} have classified 
boundary conditions which preserve integrability. We see that
many of them descend from just the three boundary conditions
which we study. 
The Durham group and in particular Bowcock \cite{Bow96} have
studied the classical solutions of real coupling Toda theory
on the half-line. There are conjectures for particle reflection
matrices for real coupling Toda theory on the half-line \cite{Fri94b,
Sas93,Kim96}.
In order to check these conjectures and to connect them with
specific boundary conditions it is necessary to determine the
classical vacuum solution \cite{Cor94a}. 
Our analysis simplifies these calculations.
Gandenberger \cite{Gan98} has determined quantum soliton reflection
matrices by solving the boundary Yang-Baxter equation. We
verify his results semiclassically by comparing with the time
delays computed from our solutions. 
The idea of obtaining soliton solutions on the half-line from
those on the whole line is of course not new. It has been
applied to the sine-Gordon model in \cite{Sal94}, to
Toda lattice models in \cite{Fuj95} and to real coupling
Toda theory \cite{Bow96}.

There is one affine Toda theory $T(\hat{g})$ for
every affine simple complex Lie algebra $\hat{g}$. It describes an $n$%
-component bosonic field $\bphi$, where $n$ is the rank of $\hat{g}$.
Let $\aroot_i,\,i=1,\dots ,n$ be the simple roots of the
finite-dimensional Lie algebra $g$ underlying $\hat{g}$ and let $\aroot_0
=-\sum_{i=1}^n\eta _i\aroot_i$ be the extra simple root
that needs to be added to obtain the extended Dynkin diagram of $\hat{g}$.
Define $\eta _0=1$. 
Then the corresponding affine Toda theory $T(\hat{g}%
) $ is described by the equations of motion 
\begin{equation}
\ddot{\bphi}-\bphi^{\prime \prime }+m^2\sum_{i=0}^n\eta _i%
\aroot_ie^{\aroot_i\cdot \bphi}=0.
\label{phieom}
\end{equation}
We have normalized the field so that the coupling constant does not
appear in the equations of motion.
We will use units where $m=c=1$.
The sine-Gordon model is obtained by choosing
$\hat{g}=a_1^{(1)}=\widehat{sl(2)}$.

This first section will explain our general ideas without going into
calculational detail. Many of the arguments presented here apply to any
affine Toda theory. In later sections we will do the calculations
for the case of $a_n^{(1)}$ Toda theory. Section \ref{sect:review}
reviews the solutions to $a_n^{(1)}$ Toda equations of motion in the Hirota
formalism. In section \ref{sect:3} we will impose three different
boundary conditions and identify the solutions of the equations
of motion which satisfy these. We will use the results obtained to
learn about real coupling Toda theory on the half-line in section 
\ref{sect:real} and to perform semiclassical checks of conjectured
quantum reflection matrices in section \ref{sect:semiclassical}.
Section \ref{sect:discussion} contains discussions.

\subsection{Neumann boundary condition}

\FIGURE{
\setlength{\unitlength}{0.5mm}
\begin{picture}(76,98)
\thinlines
\drawpath{36.0}{94.0}{36.0}{4.0}
\drawdotline{36.0}{38.0}{66.0}{94.0}
\drawdotline{6.0}{94.0}{36.0}{38.0}
\drawdotline{36.0}{60.0}{66.0}{4.0}
\drawdotline{6.0}{4.0}{36.0}{60.0}
\thicklines
\path(6.0,94.0)(6.0,94.0)(6.71,92.62)(7.42,91.29)(8.11,89.98)(8.81,88.69)(9.51,87.44)(10.18,86.2)(10.86,85.01)(11.53,83.83)
\path(11.53,83.83)(12.19,82.66)(12.85,81.54)(13.51,80.43)(14.15,79.33)(14.8,78.27)(15.43,77.23)(16.06,76.2)(16.68,75.2)(17.31,74.22)
\path(17.31,74.22)(17.93,73.26)(18.54,72.3)(19.14,71.38)(19.75,70.48)(20.35,69.58)(20.94,68.69)(21.53,67.83)(22.12,67.0)(22.7,66.16)
\path(22.7,66.16)(23.28,65.33)(23.86,64.54)(24.43,63.74)(25.0,62.95)(25.56,62.18)(26.13,61.43)(26.7,60.68)(27.26,59.95)(27.8,59.22)
\path(27.8,59.22)(28.37,58.5)(28.92,57.79)(29.46,57.08)(30.02,56.38)(30.56,55.68)(31.12,55.0)(31.65,54.31)(32.2,53.65)(32.75,52.97)
\path(32.75,52.97)(33.29,52.31)(33.83,51.63)(34.36,50.97)(34.9,50.31)(35.45,49.65)(35.99,49.0)(36.54,48.33)(37.08,47.66)(37.61,47.0)
\path(37.61,47.0)(38.15,46.34)(38.7,45.68)(39.24,45.0)(39.77,44.34)(40.33,43.66)(40.86,42.99)(41.41,42.29)(41.97,41.61)(42.52,40.9)
\path(42.52,40.9)(43.06,40.2)(43.61,39.49)(44.18,38.77)(44.72,38.04)(45.29,37.29)(45.84,36.56)(46.41,35.79)(46.99,35.02)(47.56,34.25)
\path(47.56,34.25)(48.13,33.45)(48.7,32.65)(49.29,31.82)(49.86,31.0)(50.45,30.14)(51.04,29.29)(51.63,28.4)(52.24,27.51)(52.84,26.6)
\path(52.84,26.6)(53.45,25.68)(54.06,24.72)(54.68,23.77)(55.29,22.78)(55.91,21.78)(56.54,20.76)(57.18,19.7)(57.83,18.64)(58.47,17.55)
\path(58.47,17.55)(59.13,16.45)(59.79,15.31)(60.45,14.15)(61.11,12.97)(61.79,11.77)(62.47,10.55)(63.16,9.28)(63.86,8.01)(64.56,6.69)
\path(64.56,6.69)(65.27,5.36)(65.99,4.0)(66.0,4.0)
\path(6.0,4.0)(6.0,4.0)(6.71,5.36)(7.42,6.69)(8.11,8.01)(8.81,9.3)(9.51,10.56)(10.18,11.8)(10.86,13.01)(11.53,14.19)
\path(11.53,14.19)(12.19,15.35)(12.85,16.5)(13.51,17.62)(14.15,18.71)(14.8,19.79)(15.43,20.86)(16.06,21.88)(16.68,22.9)(17.31,23.9)
\path(17.31,23.9)(17.93,24.88)(18.54,25.85)(19.14,26.79)(19.75,27.71)(20.35,28.62)(20.94,29.53)(21.53,30.4)(22.12,31.28)(22.7,32.13)
\path(22.7,32.13)(23.28,32.97)(23.86,33.79)(24.43,34.59)(25.0,35.4)(25.56,36.18)(26.13,36.97)(26.7,37.74)(27.26,38.5)(27.8,39.25)
\path(27.8,39.25)(28.37,39.99)(28.92,40.72)(29.46,41.45)(30.02,42.15)(30.56,42.86)(31.12,43.58)(31.65,44.27)(32.2,44.97)(32.75,45.65)
\path(32.75,45.65)(33.29,46.34)(33.83,47.02)(34.36,47.7)(34.9,48.38)(35.45,49.06)(35.99,49.74)(36.54,50.41)(37.08,51.09)(37.61,51.77)
\path(37.61,51.77)(38.15,52.45)(38.7,53.11)(39.24,53.79)(39.77,54.49)(40.33,55.16)(40.86,55.86)(41.41,56.54)(41.97,57.25)(42.52,57.95)
\path(42.52,57.95)(43.06,58.66)(43.61,59.38)(44.18,60.11)(44.72,60.84)(45.29,61.58)(45.84,62.31)(46.41,63.08)(46.99,63.84)(47.56,64.62)
\path(47.56,64.62)(48.13,65.41)(48.7,66.2)(49.29,67.01)(49.86,67.83)(50.45,68.66)(51.04,69.51)(51.63,70.38)(52.24,71.26)(52.84,72.15)
\path(52.84,72.15)(53.45,73.05)(54.06,73.98)(54.68,74.93)(55.29,75.88)(55.91,76.86)(56.54,77.86)(57.18,78.87)(57.83,79.9)(58.47,80.94)
\path(58.47,80.94)(59.13,82.01)(59.79,83.12)(60.45,84.23)(61.11,85.37)(61.79,86.52)(62.47,87.7)(63.16,88.91)(63.86,90.15)(64.56,91.41)
\path(64.56,91.41)(65.27,92.69)(65.99,93.98)(66.0,94.0)
\thinlines
\drawpath{4.0}{22.0}{70.0}{22.0}
\drawpath{36.0}{92.0}{40.0}{94.0}
\drawpath{36.0}{90.0}{40.0}{92.0}
\drawpath{36.0}{88.0}{40.0}{90.0}
\drawpath{36.0}{86.0}{40.0}{88.0}
\drawpath{36.0}{84.0}{40.0}{86.0}
\drawpath{36.0}{82.0}{40.0}{84.0}
\drawpath{36.0}{80.0}{40.0}{82.0}
\drawpath{36.0}{78.0}{40.0}{80.0}
\drawpath{36.0}{76.0}{40.0}{78.0}
\drawpath{36.0}{74.0}{40.0}{76.0}
\drawpath{36.0}{72.0}{40.0}{74.0}
\drawpath{36.0}{70.0}{40.0}{72.0}
\drawpath{36.0}{68.0}{40.0}{70.0}
\drawpath{36.0}{66.0}{40.0}{68.0}
\drawpath{36.0}{64.0}{40.0}{66.0}
\drawpath{36.0}{62.0}{40.0}{64.0}
\drawpath{36.0}{60.0}{40.0}{62.0}
\drawpath{36.0}{58.0}{40.0}{60.0}
\drawpath{36.0}{56.0}{40.0}{58.0}
\drawpath{36.0}{54.0}{40.0}{56.0}
\drawpath{36.0}{52.0}{40.0}{54.0}
\drawpath{36.0}{50.0}{40.0}{52.0}
\drawpath{36.0}{48.0}{40.0}{50.0}
\drawpath{36.0}{46.0}{40.0}{48.0}
\drawpath{36.0}{44.0}{40.0}{46.0}
\drawpath{36.0}{42.0}{40.0}{44.0}
\drawpath{36.0}{40.0}{40.0}{42.0}
\drawpath{36.0}{38.0}{40.0}{40.0}
\drawpath{36.0}{36.0}{40.0}{38.0}
\drawpath{36.0}{34.0}{40.0}{36.0}
\drawpath{36.0}{32.0}{40.0}{34.0}
\drawpath{36.0}{30.0}{40.0}{32.0}
\drawpath{36.0}{28.0}{40.0}{30.0}
\drawpath{36.0}{26.0}{40.0}{28.0}
\drawpath{36.0}{22.0}{40.0}{24.0}
\drawpath{36.0}{24.0}{40.0}{26.0}
\drawpath{36.0}{20.0}{40.0}{22.0}
\drawpath{36.0}{18.0}{40.0}{20.0}
\drawpath{36.0}{16.0}{40.0}{18.0}
\drawpath{36.0}{14.0}{40.0}{16.0}
\drawpath{36.0}{12.0}{40.0}{14.0}
\drawpath{36.0}{10.0}{40.0}{12.0}
\drawpath{36.0}{8.0}{40.0}{10.0}
\drawpath{36.0}{6.0}{40.0}{8.0}
\drawpath{36.0}{4.0}{40.0}{6.0}
\path(6.0,94.0)(6.0,94.0)(6.46,93.08)(6.94,92.16)(7.38,91.23)(7.84,90.31)(8.27,89.4)(8.69,88.48)(9.11,87.55)(9.52,86.65)
\path(9.52,86.65)(9.93,85.73)(10.31,84.8)(10.68,83.9)(11.06,82.98)(11.42,82.06)(11.77,81.15)(12.1,80.23)(12.44,79.33)(12.77,78.41)
\path(12.77,78.41)(13.07,77.5)(13.38,76.58)(13.68,75.66)(13.96,74.76)(14.22,73.84)(14.5,72.94)(14.75,72.02)(15.0,71.12)(15.22,70.2)
\path(15.22,70.2)(15.46,69.3)(15.67,68.38)(15.88,67.48)(16.06,66.58)(16.26,65.66)(16.44,64.76)(16.61,63.84)(16.77,62.95)(16.92,62.04)
\path(16.92,62.04)(17.04,61.13)(17.18,60.22)(17.29,59.31)(17.4,58.41)(17.51,57.52)(17.61,56.61)(17.69,55.7)(17.76,54.79)(17.81,53.9)
\path(17.81,53.9)(17.87,53.0)(17.92,52.09)(17.95,51.2)(17.97,50.29)(17.98,49.4)(18.0,48.5)(17.98,47.59)(17.97,46.7)(17.95,45.79)
\path(17.95,45.79)(17.92,44.9)(17.87,44.0)(17.81,43.09)(17.76,42.2)(17.69,41.31)(17.61,40.4)(17.52,39.52)(17.4,38.61)(17.29,37.72)
\path(17.29,37.72)(17.18,36.83)(17.04,35.93)(16.92,35.04)(16.77,34.15)(16.61,33.25)(16.44,32.36)(16.26,31.46)(16.07,30.57)(15.88,29.68)
\path(15.88,29.68)(15.67,28.79)(15.46,27.89)(15.22,27.01)(15.0,26.12)(14.75,25.22)(14.5,24.34)(14.22,23.45)(13.96,22.55)(13.68,21.68)
\path(13.68,21.68)(13.38,20.79)(13.07,19.89)(12.77,19.01)(12.44,18.12)(12.11,17.23)(11.77,16.35)(11.42,15.47)(11.06,14.57)(10.68,13.69)
\path(10.68,13.69)(10.31,12.81)(9.93,11.93)(9.52,11.05)(9.11,10.15)(8.69,9.27)(8.27,8.39)(7.84,7.51)(7.38,6.63)(6.94,5.76)
\path(6.94,5.76)(6.46,4.88)(6.0,4.0)(6.0,4.0)
\drawcenteredtext{68.0}{26.0}{$x$}
\drawcenteredtext{32.0}{92.0}{$t$}
\end{picture}
\setlength{\unitlength}{1mm}
\caption{Space-time diagram for the soliton-antisoliton process
satisfying the Neumann boundary condition.\label{figst1}}
}

One boundary condition which selects solutions of the form 
discussed above among
all the multi-soliton solutions is the Neumann condition
\begin{equation}\label{neumann}
\partial_x\bphi=0~~~\mbox{ at }~~x=0.
\end{equation}
It is clear that any solution which is invariant under parity reversal
$x\rightarrow -x$
will satisfy this boundary condition. Thus any solution composed
of any number of soliton-antisoliton pairs with exactly equal but opposite
velocities will satisfy the boundary condition
provided the center of mass of each pair is at $x=0$.  It is also
easy to convince oneself that there are no other solutions which
satisfy \eq{neumann}. The solution
plotted in figure \ref{fig1} is an example of such a
configuration with one soliton-antisoliton pair.

Figure \ref{figst1} is a space-time diagram showing the worldlines
of the centers of mass of the solitons in such a solution satisfying
the Neumann boundary condition. Of course close to the interaction point
the solitons loose their individual existence, thus the worldlines
drawn in the figure are meant for illustrative purposes only. The fat solid
lines are the trajectories of the two solitons on the whole line. The dotted
lines are their asymptotic trajectories. The fine line will be explained
below

The fact that the incoming and outgoing
asymptotic trajectories do not meet the time axis at the same point
indicates that the solitons experience a time delay
during their interaction. 
We notice from the diagram
that the time delay is negative, it is really a time advance.
The fact that the solitons on the whole line are experiencing 
a time advance while they pass through each other shows that
they are attractive to each other.

As a consequence of this time advance in soliton-antisoliton
scattering also the reflected soliton on the half-line
experiences a time advance during its interaction with the boundary. 
There are two possible interpretations for this time advance.
One can imagine
that the boundary is attractive and that the reflected soliton is 
reaching the boundary along the fat trajectory in figure
\ref{figst1}. The soliton is accelerated by the attractive boundary
while it is approaching and that is the reason for the time advance.
Alternatively one can imagine that the boundary is repulsive and
that the reflected soliton is moving along the
fine  trajectory in figure \ref{figst1}. Because it is repelled
by the boundary the soliton already turns around
some time before it actually reaches $x=0$ and this is the cause
of the time advance. 
For reasons which we will explain we believe the boundary to be
attractive and thus prefer the first interpretation.

In $a_n^{(1)}$ Toda theory there are $n$ types of solitons with
masses
\begin{equation}
M_a=\frac{2(n+1)}{(i\beta)^2}m_a\text{~~~with~~~}
m_a=2\sin\frac{a\pi}{n+1},~~a=1,\cdots,n.
\end{equation}
The time delay which solitons of type $a$ experience during
reflection is given by (see \eq{rt})
\begin{equation}
\Delta t=\frac{\sqrt{1-v^2}}{m_a v }
\log\left(1-\frac{m_a^2}{4}(1-v^2)\right).
\end{equation}
$\Delta t$ is always negative and its magnitude increases with
the mass of the soliton while decreasing with its velocity.
Thus light and fast solitons are affected less by the boundary
than heavy and slow ones.

\EPSFIGURE{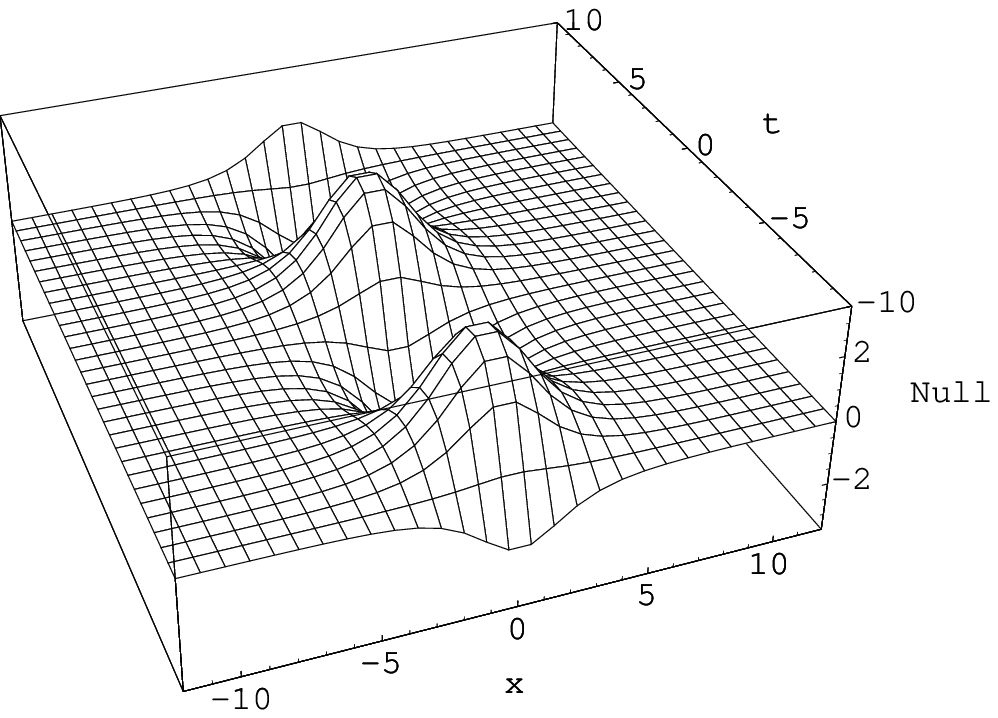,width=7cm}
{A breather solution
\label{fig3}}

One can allow the relative rapidities between solitons to be complex
(as long as one ensures that the corresponding solution does not
develop singularities \cite{Har94}). A configuration of two solitons whose relative
rapidity is imaginary and which therefore are oscillating around each
other in a bound state is called a breather, see figure \ref{fig3}. 
(If the breather has non-vanishing
topological charge some authors call it "breathing soliton"
or "excited soliton").

Again one can describe breathers on the half-line by restricting a
solution describing breather-antibreather pairs on the whole line. Thus
we see that the spectrum of affine Toda theory on the half-line
with a Neumann boundary condition is as 
rich as that on the whole line, consisting of any combination of solitons
and breathers. \footnote{
A slight caveat is provided by the fact that solitons on the whole
line satisfy a kind of exclusion principle which forbids two solitons
of the same type to have the same velocity. On the half-line
this implies that it is not possible
to have both a stationary soliton of some type $a$ \textit{and}
a stationary soliton of the conjugate type $n+1-a$.}

But Toda theory on the half-line contains even some new states not
present in the spectrum on the whole line. These are boundary states.
Consider a stationary breather as in figure
\ref{fig3} on the whole line
centered at $x=0$. On the half-line
this will be interpreted as a soliton bound to the boundary. We will
call these configurations boundary breathers.
The fact that the boundary can bind a soliton is the reason why
we interpret it as an attractive boundary.

\subsection{Other boundary conditions}

In this paper we also study more general boundary conditions of the
form
\begin{equation}
\partial _x\bphi|_{x=0}=\varepsilon\sum_{i=0}^n\ 
\eta _i\,\aroot_i\ e^{\aroot_i\cdot 
\bphi/2}|_{x=0}  \label{bc}
\end{equation}
$\varepsilon=0$ corresponds to
the Neumann condition already discussed.

It is easy to see that a configuration describing a
soliton and an antisoliton with opposite velocities on the whole line
will satisfy the boundary
conditions \eq{bc} in the far past and in the far future
by the following consideration. 

At $t=-\infty$ the solitons
are located at $x=-\infty$ and $x=\infty$ respectively and thus
at $x=0$ the field takes the right asymptotic value of the right-moving
soliton. The asymptotic values of a finite-energy solution of affine
Toda theory must always be equal to a vacuum value of the Toda potential
and are therefore of the form $2\pi i\weight$ 
where $\weight$ is some coweight of the
underlying Lie algebra $g$.
The difference between the left asymptotic value and
the right asymptotic value (divided by $2\pi i$)
is called the \textit{topological charge} 
of the solution. Only the difference between the left and
right asymptotic values is of physical relevance because
the constant shift $\bphi\rightarrow\bphi+
2\pi i\weight$ is a symmetry of affine Toda theory for any coweight
$\weight$ and can
be used to shift the asymptotic value at both ends simultaneously without
changing the physics. 
It follows that
we can always shift the soliton-antisoliton pair solution so that 
in the far past it takes the constant value $\bphi=0$ near
$x=0$. It then satisfies the boundary condition \eq{bc} because
$\sum_{i=0}^n\eta _i\aroot_i=0$.

At $t=\infty$ the solitons will have passed through each other and
will again be infinitely separated. The value of the field at $x=0$
will be equal to $2\pi i (\weight_2-\weight_1)$ where $\weight_1$ and
$\weight_2$ are the topological charges of the right-moving and the
left-moving solitons respectively. This satisfies the boundary
condition if $\weight_1=\pm\weight_2$, i.e., if the two solitons are
identical or if they are antiparticles of each other.

\FIGURE{
\setlength{\unitlength}{0.6mm}
\begin{picture}(56,98)
\thinlines
\drawpath{28.0}{94.0}{28.0}{4.0}
\drawpath{28.0}{92.0}{32.0}{94.0}
\path(8.0,94.0)(8.0,94.0)(8.7,92.62)(9.41,91.29)(10.12,89.98)(10.81,88.69)(11.51,87.43)(12.19,86.19)(12.87,84.98)(13.54,83.79)
\path(13.54,83.79)(14.19,82.62)(14.86,81.48)(15.51,80.37)(16.16,79.26)(16.8,78.19)(17.44,77.12)(18.06,76.09)(18.69,75.08)(19.3,74.08)
\path(19.3,74.08)(19.93,73.09)(20.54,72.13)(21.15,71.19)(21.75,70.26)(22.34,69.36)(22.94,68.45)(23.52,67.58)(24.12,66.7)(24.69,65.86)
\path(24.69,65.86)(25.27,65.01)(25.86,64.19)(26.43,63.38)(27.0,62.59)(27.56,61.79)(28.13,61.02)(28.69,60.25)(29.26,59.49)(29.8,58.74)
\path(29.8,58.74)(30.37,58.0)(30.91,57.27)(31.47,56.54)(32.01,55.83)(32.56,55.11)(33.12,54.4)(33.66,53.7)(34.2,53.02)(34.75,52.33)
\path(34.75,52.33)(35.29,51.63)(35.83,50.95)(36.37,50.27)(36.91,49.59)(37.44,48.91)(37.98,48.25)(38.54,47.56)(39.08,46.9)(39.62,46.22)
\path(39.62,46.22)(40.16,45.54)(40.69,44.86)(41.23,44.18)(41.77,43.5)(42.33,42.81)(42.86,42.13)(43.41,41.43)(43.97,40.74)(44.52,40.02)
\path(44.52,40.02)(45.07,39.31)(45.61,38.61)(46.17,37.88)(46.72,37.15)(47.28,36.4)(47.85,35.66)(48.41,34.9)(48.99,34.15)(49.55,33.36)
\drawpath{28.0}{90.0}{32.0}{92.0}
\drawpath{28.0}{88.0}{32.0}{90.0}
\drawpath{28.0}{86.0}{32.0}{88.0}
\drawdotline{8.0}{4.0}{38.0}{60.0}
\drawdotline{8.0}{94.0}{38.0}{38.0}
\drawpath{28.0}{84.0}{32.0}{86.0}
\path(8.0,4.0)(8.0,4.0)(8.7,5.3)(9.41,6.57)(10.12,7.84)(10.81,9.06)(11.51,10.27)(12.19,11.46)(12.87,12.61)(13.54,13.76)
\path(13.54,13.76)(14.19,14.86)(14.86,15.96)(15.51,17.04)(16.16,18.09)(16.8,19.12)(17.44,20.12)(18.06,21.12)(18.69,22.1)(19.3,23.05)
\path(19.3,23.05)(19.93,24.0)(20.54,24.93)(21.15,25.84)(21.75,26.72)(22.34,27.6)(22.94,28.46)(23.52,29.31)(24.12,30.14)(24.69,30.96)
\path(24.69,30.96)(25.27,31.78)(25.86,32.58)(26.43,33.36)(27.0,34.13)(27.56,34.9)(28.13,35.66)(28.69,36.4)(29.26,37.15)(29.8,37.88)
\path(29.8,37.88)(30.37,38.61)(30.91,39.31)(31.47,40.02)(32.01,40.74)(32.56,41.43)(33.12,42.13)(33.66,42.81)(34.2,43.5)(34.75,44.18)
\path(34.75,44.18)(35.29,44.86)(35.83,45.54)(36.37,46.22)(36.91,46.9)(37.44,47.56)(37.98,48.24)(38.54,48.91)(39.08,49.59)(39.62,50.27)
\path(39.62,50.27)(40.16,50.95)(40.69,51.63)(41.23,52.33)(41.77,53.02)(42.33,53.7)(42.86,54.4)(43.41,55.11)(43.97,55.83)(44.52,56.54)
\path(44.52,56.54)(45.07,57.27)(45.61,58.0)(46.17,58.74)(46.72,59.49)(47.28,60.25)(47.85,61.02)(48.41,61.79)(48.99,62.58)(49.55,63.38)
\drawpath{28.0}{82.0}{32.0}{84.0}
\drawpath{28.0}{80.0}{32.0}{82.0}
\drawpath{28.0}{78.0}{32.0}{80.0}
\drawpath{28.0}{76.0}{32.0}{78.0}
\drawpath{28.0}{74.0}{32.0}{76.0}
\drawpath{28.0}{72.0}{32.0}{74.0}
\drawpath{28.0}{70.0}{32.0}{72.0}
\drawpath{28.0}{68.0}{32.0}{70.0}
\drawpath{28.0}{66.0}{32.0}{68.0}
\drawpath{28.0}{64.0}{32.0}{66.0}
\drawpath{28.0}{62.0}{32.0}{64.0}
\drawpath{28.0}{60.0}{32.0}{62.0}
\drawpath{28.0}{58.0}{32.0}{60.0}
\drawpath{28.0}{56.0}{32.0}{58.0}
\drawpath{28.0}{54.0}{32.0}{56.0}
\drawpath{28.0}{52.0}{32.0}{54.0}
\drawpath{28.0}{50.0}{32.0}{52.0}
\drawpath{28.0}{48.0}{32.0}{50.0}
\drawpath{28.0}{46.0}{32.0}{48.0}
\drawpath{28.0}{46.0}{32.0}{48.0}
\drawpath{28.0}{44.0}{32.0}{46.0}
\drawpath{28.0}{42.0}{32.0}{44.0}
\drawpath{28.0}{40.0}{32.0}{42.0}
\drawpath{28.0}{38.0}{32.0}{40.0}
\drawpath{28.0}{36.0}{32.0}{38.0}
\drawpath{28.0}{34.0}{32.0}{36.0}
\drawpath{28.0}{32.0}{32.0}{34.0}
\drawpath{28.0}{30.0}{32.0}{32.0}
\drawpath{28.0}{28.0}{32.0}{30.0}
\drawpath{28.0}{26.0}{32.0}{28.0}
\drawpath{28.0}{24.0}{32.0}{26.0}
\drawpath{28.0}{26.0}{32.0}{28.0}
\drawpath{28.0}{24.0}{32.0}{26.0}
\drawpath{28.0}{22.0}{32.0}{24.0}
\drawpath{28.0}{20.0}{32.0}{22.0}
\drawpath{28.0}{18.0}{32.0}{20.0}
\drawpath{28.0}{16.0}{32.0}{18.0}
\drawpath{28.0}{14.0}{32.0}{16.0}
\drawpath{28.0}{12.0}{32.0}{14.0}
\drawpath{28.0}{10.0}{32.0}{12.0}
\drawpath{28.0}{8.0}{32.0}{10.0}
\drawpath{28.0}{6.0}{32.0}{8.0}
\drawpath{28.0}{4.0}{32.0}{6.0}
\thicklines
\path(8.05,4.03)(8.05,4.03)(8.25,4.42)(8.45,4.8)(8.66,5.19)(8.87,5.57)(9.08,5.94)(9.27,6.32)(9.48,6.71)(9.69,7.07)
\path(9.69,7.07)(9.9,7.46)(10.11,7.82)(10.3,8.21)(10.51,8.57)(10.72,8.94)(10.93,9.31)(11.12,9.68)(11.33,10.05)(11.54,10.4)
\path(11.54,10.4)(11.75,10.77)(11.94,11.14)(12.16,11.5)(12.36,11.85)(12.56,12.21)(12.76,12.56)(12.97,12.92)(13.18,13.27)(13.37,13.63)
\path(13.37,13.63)(13.58,13.97)(13.79,14.32)(13.98,14.68)(14.19,15.02)(14.4,15.36)(14.59,15.71)(14.8,16.05)(15.0,16.39)(15.2,16.73)
\path(15.2,16.73)(15.41,17.07)(15.61,17.42)(15.8,17.75)(16.01,18.09)(16.22,18.42)(16.41,18.75)(16.62,19.07)(16.81,19.4)(17.01,19.73)
\path(17.01,19.73)(17.22,20.06)(17.41,20.38)(17.62,20.7)(17.81,21.04)(18.01,21.36)(18.22,21.68)(18.43,22.0)(18.62,22.3)(18.81,22.62)
\path(18.81,22.62)(19.01,22.94)(19.22,23.26)(19.41,23.56)(19.62,23.87)(19.81,24.19)(20.01,24.48)(20.22,24.79)(20.41,25.11)(20.62,25.4)
\path(20.62,25.4)(20.81,25.7)(21.01,26.01)(21.2,26.3)(21.41,26.61)(21.61,26.9)(21.8,27.2)(22.0,27.5)(22.19,27.79)(22.4,28.07)
\path(22.4,28.07)(22.59,28.37)(22.79,28.65)(22.98,28.95)(23.19,29.22)(23.37,29.52)(23.58,29.79)(23.77,30.07)(23.97,30.36)(24.16,30.63)
\path(24.16,30.63)(24.37,30.92)(24.55,31.2)(24.76,31.46)(24.94,31.75)(25.15,32.02)(25.33,32.29)(25.54,32.56)(25.73,32.83)(25.93,33.09)
\path(25.93,33.09)(26.12,33.36)(26.31,33.63)(26.51,33.88)(26.7,34.15)(26.9,34.4)(27.09,34.66)(27.29,34.93)(27.48,35.18)(27.68,35.43)
\path(27.68,35.43)(27.87,35.68)(28.05,35.93)(28.06,35.95)
\path(8.05,93.93)(8.05,93.93)(8.3,93.43)(8.58,92.94)(8.84,92.45)(9.11,91.97)(9.37,91.48)(9.63,91.01)(9.9,90.54)(10.16,90.06)
\path(10.16,90.06)(10.41,89.61)(10.68,89.15)(10.93,88.69)(11.19,88.23)(11.44,87.77)(11.69,87.33)(11.94,86.87)(12.19,86.44)(12.44,86.0)
\path(12.44,86.0)(12.69,85.55)(12.93,85.12)(13.18,84.69)(13.41,84.26)(13.66,83.84)(13.9,83.43)(14.13,83.01)(14.37,82.59)(14.61,82.19)
\path(14.61,82.19)(14.83,81.77)(15.08,81.37)(15.3,80.98)(15.54,80.58)(15.76,80.19)(15.98,79.8)(16.22,79.41)(16.44,79.02)(16.66,78.65)
\path(16.66,78.65)(16.88,78.27)(17.11,77.9)(17.33,77.52)(17.55,77.16)(17.76,76.8)(17.98,76.44)(18.19,76.08)(18.41,75.73)(18.62,75.38)
\path(18.62,75.38)(18.83,75.04)(19.04,74.69)(19.23,74.36)(19.44,74.01)(19.65,73.69)(19.86,73.36)(20.05,73.02)(20.26,72.7)(20.45,72.38)
\path(20.45,72.38)(20.66,72.06)(20.84,71.76)(21.05,71.44)(21.23,71.13)(21.43,70.83)(21.62,70.54)(21.8,70.23)(22.0,69.94)(22.19,69.66)
\path(22.19,69.66)(22.37,69.37)(22.55,69.08)(22.73,68.8)(22.91,68.52)(23.09,68.26)(23.27,67.98)(23.45,67.72)(23.63,67.45)(23.8,67.19)
\path(23.8,67.19)(23.98,66.94)(24.16,66.69)(24.33,66.44)(24.5,66.19)(24.66,65.94)(24.83,65.7)(25.0,65.48)(25.16,65.23)(25.31,65.01)
\path(25.31,65.01)(25.48,64.79)(25.65,64.55)(25.8,64.33)(25.95,64.12)(26.12,63.9)(26.26,63.7)(26.43,63.5)(26.58,63.29)(26.73,63.09)
\path(26.73,63.09)(26.87,62.9)(27.02,62.7)(27.16,62.5)(27.31,62.33)(27.45,62.13)(27.59,61.95)(27.73,61.79)(27.87,61.61)(28.01,61.43)
\path(28.01,61.43)(28.16,61.27)(28.29,61.11)(28.3,61.11)
\drawpath{28.06}{61.11}{28.06}{35.72}
\thinlines
\drawcenteredtext{23.34}{91.45}{$t$}
\drawpath{4.0}{22.0}{52.0}{22.0}
\drawcenteredtext{49.22}{26.05}{$x$}
\end{picture}
\setlength{\unitlength}{1mm}
\caption{Reflection off the attractive boundary ($\varepsilon=-1)$.
\label{fig4a}}
}

What is less easy to see is under what circumstances these solutions
satisfy the boundary conditions also at intermediate times. This requires
detailed calculations which we will perform 
for $a_n^{(1)}$ Toda theory 
in section \ref{sect:3}. We will find that a soliton-antisoliton
pair can indeed satisfy the boundary conditions \eqref{bc} if
\begin{equation}
\varepsilon^2=\frac{2}{|\aroot_i|^2}.
\end{equation}
We will choose the normalization of the roots so that $|\aroot|^2=2$
and thus we need $\varepsilon=\pm 1$.
Furthermore, in order to satisfy the boundary condition \eqref{bc} the
soliton and antisoliton must meet not at the boundary
but either behind or in front of it.
This is depicted in figures \ref{fig4a} and \ref{fig4b}.
In these figures the thin lines are the worldlines of the soliton
and antisoliton on the whole line, the fat line is the worldline
of the reflected soliton on the left half-line and the dotted lines
are the asymptotic trajectories of the reflected soliton.
The calculations in section \ref{sect:3} also show that there are
no other two-soliton solutions on the whole line which satisfy the
boundary conditions. Thus we see that a soliton on the half-line
will always be converted into its antisoliton upon reflection from
the boundary.

The distance $d(v)$ between the boundary and the center of mass of the 
soliton-antisoliton pair comes out of the calculation as
\begin{equation}\label{ddd}
d(v)=\frac{-\varepsilon\sqrt{1-v^2}}{2m_a} 
\log\left(\frac{1+\frac{m_a}{2}\sqrt{1-v^2}}
{1-\frac{m_a}{2}\sqrt{1-v^2}}\right).
\end{equation}
We see that for $\varepsilon=-1$ the scattering with the mirror soliton
takes place behind the boundary and for
$\varepsilon =1$ it takes place in front of the boundary. 
The distance between the boundary and the virtual scattering point is
smaller if the soliton is faster or lighter.

\FIGURE{
\setlength{\unitlength}{0.6mm}
\begin{picture}(62,98)
\thinlines
\drawpath{42.0}{92.0}{46.0}{94.0}
\drawpath{42.0}{90.0}{46.0}{92.0}
\drawpath{4.0}{22.0}{56.0}{22.0}
\drawcenteredtext{54.0}{18.0}{$x$}
\drawpath{42.0}{94.0}{42.0}{4.0}
\drawdotline{5.96}{4.03}{42.65}{72.58}
\path(6.0,94.0)(6.0,94.0)(6.69,92.62)(7.4,91.29)(8.11,89.98)(8.81,88.69)(9.51,87.43)(10.18,86.19)(10.86,84.98)(11.53,83.79)
\path(11.53,83.79)(12.18,82.62)(12.85,81.48)(13.51,80.37)(14.15,79.26)(14.8,78.19)(15.43,77.12)(16.05,76.08)(16.69,75.08)(17.29,74.08)
\path(17.29,74.08)(17.93,73.08)(18.54,72.12)(19.14,71.19)(19.75,70.26)(20.34,69.36)(20.94,68.44)(21.52,67.58)(22.12,66.69)(22.69,65.86)
\path(22.69,65.86)(23.27,65.01)(23.86,64.19)(24.43,63.38)(25.0,62.59)(25.55,61.79)(26.12,61.02)(26.69,60.25)(27.26,59.49)(27.79,58.74)
\path(27.79,58.74)(28.37,58.0)(28.9,57.27)(29.46,56.54)(30.01,55.83)(30.55,55.11)(31.12,54.4)(31.65,53.7)(32.2,53.02)(32.75,52.33)
\path(32.75,52.33)(33.29,51.63)(33.83,50.95)(34.36,50.27)(34.9,49.59)(35.43,48.9)(35.97,48.25)(36.54,47.56)(37.08,46.9)(37.61,46.22)
\path(37.61,46.22)(38.15,45.54)(38.68,44.86)(39.22,44.18)(39.77,43.5)(40.33,42.81)(40.86,42.13)(41.4,41.43)(41.97,40.74)(42.5,40.02)
\path(42.5,40.02)(43.06,39.31)(43.61,38.61)(44.18,37.88)(44.72,37.15)(45.29,36.4)(45.84,35.65)(46.4,34.9)(46.99,34.15)(47.54,33.36)
\path(47.54,33.36)(48.13,32.58)(48.7,31.78)(49.27,30.95)(49.86,30.13)(50.45,29.3)(51.04,28.45)(51.63,27.6)(52.24,26.71)(52.83,25.84)
\drawcenteredtext{38.0}{92.0}{$t$}
\drawpath{42.0}{88.0}{46.0}{90.0}
\drawpath{42.0}{86.0}{46.0}{88.0}
\path(6.0,4.0)(6.0,4.0)(6.69,5.3)(7.4,6.57)(8.11,7.84)(8.81,9.06)(9.51,10.27)(10.18,11.46)(10.86,12.6)(11.53,13.76)
\path(11.53,13.76)(12.18,14.85)(12.85,15.96)(13.51,17.04)(14.15,18.09)(14.8,19.12)(15.43,20.12)(16.05,21.12)(16.69,22.1)(17.29,23.04)
\path(17.29,23.04)(17.93,24.0)(18.54,24.93)(19.14,25.84)(19.75,26.71)(20.34,27.6)(20.94,28.45)(21.52,29.3)(22.12,30.13)(22.69,30.95)
\path(22.69,30.95)(23.27,31.78)(23.86,32.58)(24.43,33.36)(25.0,34.13)(25.55,34.9)(26.12,35.65)(26.69,36.4)(27.26,37.15)(27.79,37.88)
\path(27.79,37.88)(28.37,38.61)(28.9,39.31)(29.46,40.02)(30.01,40.74)(30.55,41.43)(31.12,42.13)(31.65,42.81)(32.2,43.5)(32.75,44.18)
\path(32.75,44.18)(33.29,44.86)(33.83,45.54)(34.36,46.22)(34.9,46.9)(35.43,47.56)(35.97,48.24)(36.54,48.9)(37.08,49.59)(37.61,50.27)
\path(37.61,50.27)(38.15,50.95)(38.68,51.63)(39.22,52.33)(39.77,53.02)(40.33,53.7)(40.86,54.4)(41.4,55.11)(41.97,55.83)(42.5,56.54)
\path(42.5,56.54)(43.06,57.27)(43.61,58.0)(44.18,58.74)(44.72,59.49)(45.29,60.25)(45.84,61.02)(46.4,61.79)(46.99,62.58)(47.54,63.38)
\path(47.54,63.38)(48.13,64.19)(48.7,65.01)(49.27,65.86)(49.86,66.69)(50.45,67.58)(51.04,68.44)(51.63,69.36)(52.24,70.26)(52.83,71.19)
\drawpath{42.0}{84.0}{46.0}{86.0}
\drawpath{42.0}{82.0}{46.0}{84.0}
\drawpath{42.0}{80.0}{46.0}{82.0}
\drawdotline{5.96}{93.93}{41.97}{26.51}
\drawpath{42.0}{78.0}{46.0}{80.0}
\drawpath{42.0}{76.0}{46.0}{78.0}
\drawpath{42.0}{74.0}{46.0}{76.0}
\drawpath{42.0}{72.0}{46.0}{74.0}
\drawpath{42.0}{70.0}{46.0}{72.0}
\drawpath{42.0}{68.0}{46.0}{70.0}
\drawpath{42.0}{68.0}{46.0}{70.0}
\drawpath{42.0}{66.0}{46.0}{68.0}
\drawpath{42.0}{64.0}{46.0}{66.0}
\drawpath{42.0}{62.0}{46.0}{64.0}
\drawpath{42.0}{60.0}{46.0}{62.0}
\drawpath{42.0}{58.0}{46.0}{60.0}
\drawpath{42.0}{56.0}{46.0}{58.0}
\drawpath{42.0}{54.0}{46.0}{56.0}
\drawpath{42.0}{52.0}{46.0}{54.0}
\drawpath{42.0}{50.0}{46.0}{52.0}
\drawpath{42.0}{48.0}{46.0}{50.0}
\drawpath{42.0}{46.0}{46.0}{48.0}
\drawpath{42.0}{44.0}{46.0}{46.0}
\drawpath{42.0}{42.0}{46.0}{44.0}
\drawpath{42.0}{40.0}{46.0}{42.0}
\drawpath{42.0}{38.0}{46.0}{40.0}
\drawpath{42.0}{36.0}{46.0}{38.0}
\drawpath{42.0}{34.0}{46.0}{36.0}
\drawpath{42.0}{32.0}{46.0}{34.0}
\drawpath{42.0}{30.0}{46.0}{32.0}
\drawpath{42.0}{28.0}{46.0}{30.0}
\drawpath{42.0}{26.0}{46.0}{28.0}
\drawpath{42.0}{24.0}{46.0}{26.0}
\drawpath{42.0}{22.0}{46.0}{24.0}
\drawpath{42.0}{20.0}{46.0}{22.0}
\drawpath{42.0}{18.0}{46.0}{20.0}
\drawpath{42.0}{28.0}{46.0}{30.0}
\drawpath{42.0}{26.0}{46.0}{28.0}
\drawpath{42.0}{24.0}{46.0}{26.0}
\drawpath{42.0}{22.0}{46.0}{24.0}
\drawpath{42.0}{20.0}{46.0}{22.0}
\drawpath{42.0}{18.0}{46.0}{20.0}
\drawpath{42.0}{16.0}{46.0}{18.0}
\drawpath{42.0}{14.0}{46.0}{16.0}
\drawpath{42.0}{12.0}{46.0}{14.0}
\drawpath{42.0}{10.0}{46.0}{12.0}
\drawpath{42.0}{8.0}{46.0}{10.0}
\drawpath{42.0}{6.0}{46.0}{8.0}
\drawpath{42.0}{4.0}{46.0}{6.0}
\thicklines
\path(6.0,94.0)(6.0,94.0)(6.51,93.08)(7.01,92.16)(7.51,91.23)(7.98,90.31)(8.46,89.4)(8.93,88.48)(9.38,87.55)(9.81,86.65)
\path(9.81,86.65)(10.25,85.73)(10.67,84.8)(11.09,83.9)(11.48,82.98)(11.88,82.06)(12.26,81.15)(12.61,80.23)(12.97,79.33)(13.32,78.41)
\path(13.32,78.41)(13.67,77.5)(14.0,76.58)(14.31,75.66)(14.61,74.76)(14.92,73.84)(15.19,72.94)(15.47,72.02)(15.75,71.12)(16.0,70.2)
\path(16.0,70.2)(16.23,69.3)(16.47,68.38)(16.7,67.48)(16.92,66.58)(17.12,65.66)(17.3,64.76)(17.48,63.84)(17.65,62.95)(17.81,62.04)
\path(17.81,62.04)(17.97,61.13)(18.12,60.22)(18.25,59.31)(18.37,58.41)(18.46,57.52)(18.56,56.61)(18.65,55.7)(18.73,54.79)(18.8,53.9)
\path(18.8,53.9)(18.86,53.0)(18.9,52.09)(18.95,51.2)(18.96,50.29)(18.98,49.4)(19.0,48.5)(18.98,47.59)(18.96,46.7)(18.95,45.79)
\path(18.95,45.79)(18.9,44.9)(18.87,44.0)(18.8,43.09)(18.73,42.2)(18.65,41.31)(18.56,40.4)(18.47,39.52)(18.37,38.61)(18.25,37.72)
\path(18.25,37.72)(18.12,36.83)(17.97,35.93)(17.82,35.04)(17.65,34.15)(17.48,33.25)(17.3,32.36)(17.12,31.46)(16.92,30.57)(16.7,29.68)
\path(16.7,29.68)(16.47,28.79)(16.23,27.89)(16.0,27.01)(15.75,26.12)(15.47,25.22)(15.19,24.34)(14.92,23.45)(14.61,22.55)(14.31,21.68)
\path(14.31,21.68)(14.0,20.79)(13.67,19.89)(13.32,19.01)(12.97,18.12)(12.63,17.23)(12.26,16.35)(11.88,15.47)(11.48,14.57)(11.09,13.69)
\path(11.09,13.69)(10.68,12.81)(10.25,11.93)(9.81,11.05)(9.38,10.15)(8.93,9.27)(8.47,8.39)(7.98,7.51)(7.51,6.63)(7.01,5.76)
\path(7.01,5.76)(6.51,4.88)(6.0,4.0)(6.0,4.0)
\end{picture}
\setlength{\unitlength}{1mm}
\caption{
Reflection off the repulsive boundary ($\varepsilon=1)$.
\label{fig4b}}
}

The time delay which the soliton experiences during reflection is given by
(see \eq{rt})
\begin{equation}\label{tt}
\Delta t=\frac{2\sqrt{1-v^2}}
{m_a v }\log\left(1-\varepsilon\frac{m_a}{2}\sqrt{1-v^2}\right).
\end{equation}
It is negative for the $\varepsilon=1$ boundary and
positive for the $\varepsilon=-1$ boundary.

The energy of a soliton
in the presence of an  $\varepsilon=-1$ boundary is less than
the energy of a soliton on the whole line.
Conversely the energy of a soliton
in the presence of an  $\varepsilon=1$ boundary is higher than
on the whole line. Therefore we believe that the
$\varepsilon=-1$ boundary should be called attractive and the
$\varepsilon=1$ boundary should be called repulsive.

As we will see in section \ref{sect:bb},
the $\varepsilon=-1$ boundary permits boundary breather solutions 
which are regular for all $x<0$ and which have a finite energy
on the half-line. However they have a singularity exactly at
$x=0$. We are not sure whether they have a physical interpretation
or not. The $\varepsilon=1$ boundary on the other hand does not
allow any boundary breather solutions on the left half-line. 
This again confirms the identification of the $\varepsilon=1$ boundary
as repulsive and the $\varepsilon=-1$ boundary as attractive.

The time advance
which the soliton experiences during its reflection off the
$\varepsilon=1$ boundary must be due to the fact that
the soliton turns around before it reaches the boundary as
indicated by the fat trajectory in figure \ref{fig4b}.
Note the curious fact that in the case of the Neumann boundary condition
the soliton is also experiencing a time advance but in that case
it is due to not to an early turnaround but it is due to the acceleration
by the attractive boundary. This is reminiscent of a similarly confusing
situation in 
sine-Gordon theory on the whole line. There the time advance during
soliton-soliton scattering is interpreted as being due to a repulsive
force between the solitons (the solitons reflect of each other before
they reach each other) whereas the same time advance during
soliton-antisoliton scattering is interpreted as being due to an
attractive force between them (they accelerate while they move towards
each other).

Another peculiarity is that while the boundary seems to either attract
or repel moving solitons, it does not affect stationary solitons.
All the three boundary conditions which we have studied allow
stationary solitons to sit at an arbitrary distance from the boundary
without being pulled into it or being pushed away. This is just like
solitons which attract other solitons only when they are moving relative
to each other.

While affine Toda theory in the bulk is invariant under
the shift symmetry $\bphi\rightarrow\bphi+2\pi i\weight$,
this is not true of the boundary.
If we act with the shift symmetry
on the boundary conditions (\ref{bc}) we obtain new boundary
conditions
\begin{equation}
\partial _x\bphi|_{x=0}=\sum_{i=0}^n\ A_i^{(\weight)}\,
\eta _i\,\aroot_i\ e^{\aroot_i\cdot 
\bphi/2}|_{x=0},  \label{bcn}
\end{equation}
with
\begin{equation}
A_i^{(\weight)}=\varepsilon\, e^{i\pi \aroot_i\cdot\weight}=
\pm\varepsilon.
\end{equation}
This fact that different boundary conditions are related
by imaginary translations of the fields by weights was noted in
\cite{Cor95}.
These new boundary conditions are physically equivalent to
the old ones because the solutions which satisfy these
boundary conditions are obtained from those satisfying the old
ones by a shift symmetry. We can therefore without loss of
generality concentrate on just the three boundary conditions
(\ref{bc}). The situation will be different in
real coupling Toda theory which does not have the shift symmetry.
This will be discussed below in section \ref{ireal}.

If the Coxeter number $h=\sum_{i=0}^n\eta_i$ is even (this is the case
for example for $\hat{g}=c_n^{(1)}$ and $\hat{g}=a_{2m-1}^{(1)}$) 
the two boundary conditions (\ref{bc})
with $\varepsilon=1$ and $\varepsilon=-1$ are equivalent.
They transform into each other under the shift $\bphi\rightarrow\bphi
+2\pi i{\mathbf{\rho}}$ where $\mathbf{\rho}$ 
is the sum of the fundamental coweights. Thus there is no
one-to-one correspondence between boundary conditions and boundaries.
If $h$ is even then one and the same boundary condition allows both
the attractive and the repulsive boundary. Indeed, having a solution
describing reflection of the attractive boundary one can parity
transform it ($x\rightarrow -x$) and shift it by $2\pi i\rho$ and
one will obtain another solution satisfying the same boundary 
condition but now describing reflection off the repulsive boundary.

Affine Toda theory possesses an infinite number of 
conserved charges. These charges are higher-spin
generalizations of energy and momentum and generate velocity-dependent
time- and space translation symmetries. Corrigan et.al. \cite{Bow95}
showed that boundary conditions of the form
\begin{equation}
\partial _x\bphi|_{x=0}=\sum_{i=0}^n\ A_i\,
\sqrt{\frac{2\eta _i}{|\aroot_i|^2}}\,\aroot_i\ e^{\aroot_i\cdot 
\bphi/2}|_{x=0},  \label{cbc}
\end{equation}
preserve the velocity-dependent time translation symmetries if the
coefficients $A_i$ satisfy some severe constraints. For simply laced
algebras $\hat{g}$ the $A_i$ either have to all be zero (Neumann
boundary condition) or they all have to satisfy $A_i^2=1$. 
\footnote{In most papers the equation \eqref{cbc} is given without
the factor of $\sqrt{2/|\alpha_i|^2}$. It is then implicitly
assumed that the roots are normalized so that $|\alpha_i|^2=2$. Affine
Toda theory should be independent of the choice of normalization of 
the roots. Indeed the equations of motion \eqref{phieom} are invariant
under the simultaneous rescalings
\begin{equation}
\aroot\rightarrow \frac{1}{a}\aroot,~~~~\bphi\rightarrow a\bphi,~~~~
x\rightarrow a x,~~~t\rightarrow a t.
\end{equation}
With the factor of $\sqrt{2/|\alpha_i|^2}$ included also the boundary
conditions \eqref{cbc} are invariant.}
These
possibilities also exist for the non-simply laced algebras but
the conditions are weaker and some coefficients can be allowed
to take on arbitrary values.

Comparing the integrable boundary conditions \eqref{cbc}
of Corrigan et.al with the boundary conditions \eqref{bcn}
we notice a discrepancy.
The boundary conditions \eqref{cbc}
of Corrigan et.al. are generically not satisfied by the soliton-antisoliton
pairs which we have discussed because
$\sum_{i=0}^n \sqrt{2\eta _i/|\aroot_i|^2} \aroot_i$ is generally not zero. 
One will probably have to also include
one or more stationary solitons near the boundary to soak up the 
discrepancy. How this works in general we have not yet investigated
satisfactorily.

For $\hat{g}=a_n^{(1)}$ however all Kac labels $\eta_i$ are unity and
all roots square to $2$ and thus our boundary conditions are 
integrable according to Corrigan et.al. In fact, if $n$ is even our
boundary conditions \eq{bcn} exhaust all the integrable boundary
conditions. If $n$ is odd, the coefficients $A_i^{(\weight)}$ in
our boundary conditions satisfy
the constraint $\prod_{i=0}^n A_i^{(\weight)}=1$ 
(because $\sum_{n=0}^n \aroot_i=0$).
In that case we are thus dealing with only half of all the 
possible integrable boundary conditions.

Also for $\hat{g}=c_n^{(1)}$ the integrable boundary conditions \eqref{cbc} 
reduce to the boundary conditions \eqref{bcn} because in that case
$2/|\aroot_i|^2=\eta_i$. Thus our method of images applies also to
this case. 
As was explained in \cite{Bra90}, the affine Toda theories for 
nonsimply-laced algebras can be obtained from those for simply-laced
algebras by folding. In \cite{Del94} it was explained how to apply
this folding trick to obtain the
soliton solutions of $c_n^{(1)}$ Toda theory from those of 
$a_{2n-1}^{(1)}$ Toda theory. Using this method,
the calculations which we perform
for the case of $a_n^{(1)}$ can be directly applied to the case of
$c_n^{(1)}$.

Once one has shown that the single soliton-antisoliton pair
satisfies an integrable boundary condition 
(possibly with some additional stationary solitons near the 
boundary in the case of $\hat{g}\neq a_n^{(1)}$) then one can 
deduce that a configuration
describing any number of such pairs also satisfies them. The
argument goes as follows:

\FIGURE{
\setlength{\unitlength}{0.5mm}
\begin{picture}(186,116)
\thinlines
\drawpath{4.0}{88.0}{66.0}{56.0}
\thicklines
\path(126.0,94.0)(126.0,94.0)(126.91,93.51)(127.81,93.04)(128.72,92.58)(129.61,92.11)(130.5,91.66)(131.36,91.19)(132.24,90.75)(133.1,90.3)
\thinlines
\drawpath{66.0}{84.0}{4.0}{64.0}
\thicklines
\path(24.0,110.0)(24.0,110.0)(24.8,109.72)(25.63,109.44)(26.45,109.16)(27.29,108.87)(28.09,108.62)(28.87,108.33)(29.67,108.05)(30.45,107.8)
\path(133.1,90.3)(133.94,89.87)(134.8,89.44)(135.63,89.01)(136.46,88.58)(137.27,88.16)(138.08,87.75)(138.88,87.33)(139.69,86.93)(140.47,86.52)
\path(38.45,75.0)(38.45,75.0)(38.22,74.97)(38.02,74.93)(37.79,74.9)(37.59,74.86)(37.38,74.83)(37.18,74.79)(36.97,74.75)(36.77,74.72)
\thinlines
\drawpath{24.0}{110.0}{66.0}{96.0}
\thicklines
\path(30.45,107.8)(31.2,107.54)(31.95,107.26)(32.72,107.01)(33.47,106.75)(34.2,106.48)(34.93,106.23)(35.65,105.98)(36.36,105.72)(37.04,105.47)
\path(37.11,105.48)(37.11,105.48)(37.56,105.33)(38.02,105.18)(38.47,105.04)(38.91,104.88)(39.36,104.75)(39.81,104.61)(40.24,104.47)(40.68,104.33)
\path(40.68,104.33)(41.11,104.19)(41.52,104.06)(41.95,103.94)(42.36,103.8)(42.79,103.68)(43.2,103.55)(43.61,103.41)(44.02,103.3)(44.41,103.18)
\path(44.41,103.18)(44.81,103.05)(45.22,102.94)(45.61,102.81)(46.0,102.69)(46.38,102.58)(46.75,102.47)(47.13,102.36)(47.52,102.25)(47.88,102.13)
\path(47.88,102.13)(48.25,102.04)(48.61,101.93)(48.97,101.83)(49.34,101.73)(49.7,101.62)(50.04,101.52)(50.4,101.43)(50.74,101.33)(51.08,101.23)
\path(51.08,101.23)(51.41,101.15)(51.75,101.05)(52.08,100.97)(52.4,100.87)(52.72,100.79)(53.04,100.7)(53.36,100.62)(53.68,100.54)(54.0,100.45)
\path(54.0,100.45)(54.29,100.38)(54.61,100.3)(54.9,100.23)(55.2,100.16)(55.5,100.08)(55.77,100.01)(56.06,99.94)(56.34,99.87)(56.63,99.81)
\path(56.63,99.81)(56.9,99.75)(57.18,99.69)(57.45,99.62)(57.72,99.56)(57.97,99.51)(58.24,99.44)(58.49,99.4)(58.75,99.34)(59.0,99.29)
\path(59.0,99.29)(59.24,99.23)(59.47,99.19)(59.72,99.15)(59.95,99.09)(60.18,99.05)(60.41,99.01)(60.63,98.98)(60.86,98.94)(61.08,98.9)
\path(61.08,98.9)(61.29,98.87)(61.5,98.83)(61.7,98.8)(61.91,98.76)(62.11,98.73)(62.31,98.7)(62.5,98.68)(62.7,98.66)(62.88,98.63)
\path(62.88,98.63)(63.06,98.62)(63.25,98.59)(63.43,98.58)(63.59,98.55)(63.77,98.54)(63.93,98.52)(64.09,98.51)(64.26,98.51)(64.41,98.5)
\path(36.77,74.72)(36.59,74.69)(36.38,74.66)(36.2,74.62)(36.0,74.58)(35.81,74.55)(35.63,74.54)(35.43,74.5)(35.25,74.47)(35.08,74.44)
\path(35.08,74.44)(34.9,74.43)(34.72,74.4)(34.54,74.36)(34.38,74.33)(34.2,74.3)(34.04,74.29)(33.86,74.26)(33.7,74.23)(33.54,74.22)
\path(64.41,98.5)(64.56,98.48)(64.72,98.48)(64.87,98.48)(65.01,98.48)(65.15,98.48)(65.29,98.48)(65.41,98.48)(65.55,98.48)(65.66,98.48)
\path(65.66,98.48)(65.8,98.5)(65.91,98.51)(65.91,98.51)
\path(39.15,75.37)(39.15,75.37)(39.56,75.5)(39.97,75.62)(40.36,75.76)(40.77,75.88)(41.18,76.01)(41.56,76.15)(41.97,76.27)(42.36,76.4)
\path(42.36,76.4)(42.75,76.51)(43.13,76.65)(43.52,76.76)(43.9,76.88)(44.27,77.0)(44.65,77.12)(45.02,77.23)(45.38,77.34)(45.75,77.45)
\path(45.75,77.45)(46.11,77.56)(46.47,77.68)(46.83,77.79)(47.18,77.9)(47.52,78.0)(47.88,78.11)(48.22,78.2)(48.56,78.3)(48.9,78.41)
\path(48.9,78.41)(49.24,78.51)(49.56,78.61)(49.9,78.7)(50.22,78.8)(50.54,78.9)(50.86,78.98)(51.18,79.08)(51.5,79.16)(51.81,79.26)
\path(51.81,79.26)(52.11,79.33)(52.43,79.43)(52.72,79.51)(53.02,79.59)(53.31,79.68)(53.61,79.76)(53.9,79.83)(54.2,79.91)(54.47,79.98)
\path(54.47,79.98)(54.75,80.06)(55.04,80.13)(55.31,80.2)(55.59,80.27)(55.86,80.34)(56.13,80.41)(56.38,80.48)(56.65,80.55)(56.9,80.62)
\path(56.9,80.62)(57.16,80.68)(57.41,80.73)(57.66,80.8)(57.9,80.86)(58.15,80.91)(58.4,80.97)(58.63,81.02)(58.86,81.08)(59.09,81.12)
\path(59.09,81.12)(59.33,81.18)(59.54,81.23)(59.77,81.27)(60.0,81.33)(60.2,81.37)(60.43,81.41)(60.63,81.45)(60.84,81.5)(61.04,81.54)
\path(33.54,74.22)(33.38,74.18)(33.22,74.16)(33.06,74.15)(32.9,74.11)(32.75,74.08)(32.61,74.08)(32.47,74.05)(32.31,74.04)(32.18,74.01)
\thinlines
\drawpath{66.0}{40.0}{24.0}{18.0}
\thicklines
\path(32.18,74.01)(32.04,74.0)(31.87,73.97)(31.76,73.94)(31.62,73.93)(31.5,73.91)(31.36,73.9)(31.21,73.87)(31.11,73.86)(30.96,73.83)
\path(30.96,73.83)(30.86,73.83)(30.75,73.8)(30.62,73.8)(30.51,73.79)(30.37,73.76)(30.29,73.75)(30.18,73.75)(30.04,73.73)(29.95,73.72)
\path(29.95,73.72)(29.86,73.69)(29.76,73.69)(29.63,73.68)(29.54,73.68)(29.45,73.66)(29.37,73.65)(29.29,73.65)(29.2,73.62)(29.11,73.62)
\path(29.11,73.62)(29.02,73.62)(28.95,73.61)(28.86,73.61)(28.79,73.58)(28.7,73.58)(28.62,73.58)(28.54,73.58)(28.5,73.58)(28.43,73.58)
\path(28.43,73.58)(28.36,73.58)(28.29,73.55)(28.25,73.55)(28.18,73.55)(28.12,73.55)(28.05,73.55)(28.02,73.55)(27.95,73.55)(27.93,73.55)
\path(27.93,73.55)(27.87,73.55)(27.84,73.55)(27.79,73.55)(27.76,73.55)(27.7,73.55)(27.69,73.55)(27.62,73.55)(27.61,73.58)(27.59,73.58)
\path(27.59,73.58)(27.54,73.58)(27.54,73.58)(27.52,73.58)(27.5,73.58)(27.45,73.61)(27.45,73.61)(27.45,73.61)(27.44,73.62)(27.43,73.62)
\path(27.43,73.62)(27.42,73.65)(27.42,73.65)(27.42,73.65)
\path(27.42,73.65)(27.42,73.65)(27.38,73.66)(27.38,73.68)(27.37,73.69)(27.37,73.72)(27.37,73.73)(27.36,73.75)(27.36,73.75)(27.34,73.76)
\path(27.34,73.76)(27.29,73.79)(27.29,73.79)(27.27,73.8)(27.21,73.8)(27.2,73.83)(27.18,73.83)(27.12,73.83)(27.1,73.83)(27.04,73.86)
\path(27.04,73.86)(27.02,73.86)(26.95,73.86)(26.93,73.86)(26.87,73.86)(26.8,73.87)(26.77,73.87)(26.7,73.87)(26.63,73.87)(26.59,73.87)
\path(26.59,73.87)(26.53,73.87)(26.45,73.87)(26.37,73.86)(26.29,73.86)(26.25,73.86)(26.17,73.86)(26.09,73.86)(26.01,73.83)(25.93,73.83)
\path(25.93,73.83)(25.84,73.83)(25.75,73.83)(25.63,73.8)(25.54,73.8)(25.45,73.79)(25.36,73.79)(25.27,73.76)(25.17,73.75)(25.04,73.75)
\path(25.04,73.75)(24.95,73.73)(24.85,73.72)(24.7,73.69)(24.61,73.68)(24.5,73.66)(24.37,73.65)(24.26,73.62)(24.12,73.61)(24.0,73.58)
\path(24.0,73.58)(23.86,73.58)(23.71,73.55)(23.61,73.51)(23.45,73.5)(23.3,73.48)(23.19,73.44)(23.04,73.43)(22.87,73.4)(22.75,73.37)
\path(22.75,73.37)(22.6,73.36)(22.44,73.33)(22.29,73.29)(22.12,73.26)(21.95,73.23)(21.79,73.19)(21.62,73.18)(21.45,73.15)(21.29,73.11)
\path(21.29,73.11)(21.12,73.08)(20.95,73.04)(20.77,73.01)(20.59,72.97)(20.37,72.93)(20.2,72.9)(20.03,72.86)(19.85,72.83)(19.63,72.79)
\path(19.63,72.79)(19.45,72.75)(19.26,72.69)(19.04,72.66)(18.86,72.62)(18.62,72.58)(18.45,72.54)(18.21,72.5)(18.02,72.44)(17.79,72.4)
\path(17.79,72.4)(17.59,72.36)(17.36,72.3)(17.12,72.25)(16.93,72.22)(16.7,72.16)(16.45,72.11)(16.21,72.05)(16.0,72.0)(15.77,71.94)
\path(15.77,71.94)(15.52,71.9)(15.27,71.86)(15.27,71.86)
\path(15.05,77.93)(15.05,77.93)(15.3,77.83)(15.52,77.72)(15.77,77.62)(16.0,77.54)(16.2,77.43)(16.45,77.33)(16.67,77.25)(16.87,77.16)
\path(16.87,77.16)(17.11,77.08)(17.3,76.98)(17.54,76.9)(17.76,76.8)(17.95,76.72)(18.17,76.65)(18.37,76.55)(18.55,76.47)(18.78,76.4)
\path(18.78,76.4)(18.95,76.3)(19.17,76.23)(19.37,76.16)(19.54,76.08)(19.75,76.01)(19.93,75.93)(20.11,75.86)(20.29,75.79)(20.45,75.72)
\path(20.45,75.72)(20.62,75.65)(20.79,75.58)(20.96,75.51)(21.17,75.44)(21.3,75.37)(21.5,75.33)(21.63,75.26)(21.79,75.19)(21.95,75.15)
\path(21.95,75.15)(22.12,75.08)(22.29,75.01)(22.45,74.97)(22.6,74.91)(22.71,74.86)(22.87,74.8)(23.03,74.75)(23.17,74.69)(23.29,74.65)
\path(23.29,74.65)(23.45,74.61)(23.55,74.55)(23.7,74.5)(23.84,74.47)(23.95,74.41)(24.09,74.37)(24.2,74.33)(24.34,74.29)(24.45,74.25)
\path(24.45,74.25)(24.54,74.22)(24.68,74.18)(24.79,74.12)(24.87,74.11)(25.01,74.05)(25.11,74.04)(25.2,74.0)(25.29,73.97)(25.38,73.94)
\path(25.38,73.94)(25.51,73.91)(25.6,73.87)(25.69,73.86)(25.78,73.83)(25.86,73.8)(25.95,73.79)(26.03,73.76)(26.11,73.75)(26.19,73.72)
\path(26.19,73.72)(26.26,73.69)(26.3,73.68)(26.37,73.68)(26.45,73.65)(26.54,73.65)(26.6,73.62)(26.63,73.61)(26.7,73.61)(26.77,73.61)
\path(26.77,73.61)(26.8,73.58)(26.87,73.58)(26.93,73.58)(26.95,73.58)(27.02,73.58)(27.04,73.55)(27.1,73.55)(27.12,73.55)(27.18,73.55)
\path(27.18,73.55)(27.2,73.55)(27.21,73.58)(27.27,73.58)(27.29,73.58)(27.29,73.58)(27.34,73.58)(27.36,73.61)(27.37,73.61)(27.37,73.62)
\path(27.37,73.62)(27.37,73.62)(27.38,73.65)(27.42,73.65)
\path(61.04,81.54)(61.25,81.58)(61.45,81.61)(61.65,81.65)(61.84,81.68)(62.02,81.72)(62.2,81.75)(62.4,81.77)(62.58,81.8)(62.75,81.83)
\path(62.75,81.83)(62.93,81.86)(63.09,81.88)(63.27,81.91)(63.43,81.93)(63.59,81.94)(63.75,81.97)(63.91,81.98)(64.08,82.01)(64.23,82.01)
\path(64.23,82.01)(64.37,82.04)(64.51,82.05)(64.66,82.05)(64.8,82.06)(64.94,82.08)(65.06,82.08)(65.19,82.08)(65.33,82.09)(65.44,82.09)
\path(37.22,70.47)(37.02,70.58)(36.79,70.68)(36.59,70.76)(36.38,70.86)(36.18,70.97)(35.97,71.05)(35.77,71.15)(35.58,71.23)(35.38,71.33)
\path(35.38,71.33)(35.18,71.41)(35.0,71.5)(34.81,71.58)(34.61,71.66)(34.43,71.75)(34.25,71.83)(34.06,71.9)(33.88,71.97)(33.7,72.05)
\path(33.7,72.05)(33.54,72.12)(33.36,72.19)(33.2,72.26)(33.02,72.33)(32.86,72.4)(32.7,72.47)(32.54,72.54)(32.38,72.58)(32.22,72.65)
\path(32.22,72.65)(32.06,72.72)(31.92,72.76)(31.77,72.83)(31.61,72.87)(31.45,72.94)(31.3,73.0)(31.18,73.04)(31.04,73.08)(30.92,73.15)
\path(30.92,73.15)(30.77,73.18)(30.62,73.23)(30.52,73.26)(30.37,73.3)(30.27,73.36)(30.12,73.4)(30.02,73.43)(29.88,73.47)(29.79,73.5)
\path(29.79,73.5)(29.68,73.54)(29.54,73.58)(29.45,73.61)(29.35,73.62)(29.25,73.65)(29.12,73.68)(29.04,73.69)(28.95,73.73)(28.85,73.75)
\path(28.85,73.75)(28.76,73.76)(28.67,73.79)(28.55,73.8)(28.46,73.83)(28.38,73.83)(28.3,73.86)(28.25,73.86)(28.17,73.87)(28.09,73.87)
\path(28.09,73.87)(28.02,73.87)(27.95,73.9)(27.87,73.9)(27.79,73.9)(27.75,73.9)(27.69,73.9)(27.62,73.9)(27.54,73.9)(27.52,73.87)
\path(27.52,73.87)(27.45,73.87)(27.42,73.87)(27.42,73.87)
\drawpath{4.0}{84.0}{16.0}{78.0}
\drawpath{16.0}{72.0}{4.0}{68.0}
\path(65.44,82.09)(65.58,82.09)(65.69,82.11)(65.69,82.11)
\drawcenteredtext{36.0}{6.0}{\Large{a)}}
\thinlines
\drawpath{178.0}{112.0}{178.0}{20.0}
\drawpath{178.0}{20.0}{182.0}{22.0}
\drawpath{178.0}{22.0}{182.0}{24.0}
\drawpath{178.0}{24.0}{182.0}{26.0}
\drawpath{178.0}{26.0}{182.0}{28.0}
\drawpath{178.0}{28.0}{182.0}{30.0}
\drawpath{178.0}{30.0}{182.0}{32.0}
\drawpath{178.0}{32.0}{182.0}{34.0}
\drawpath{178.0}{34.0}{182.0}{36.0}
\drawpath{178.0}{86.0}{182.0}{88.0}
\drawpath{178.0}{88.0}{182.0}{90.0}
\drawpath{178.0}{90.0}{182.0}{92.0}
\drawpath{178.0}{92.0}{182.0}{94.0}
\drawpath{178.0}{94.0}{182.0}{96.0}
\drawpath{178.0}{96.0}{182.0}{98.0}
\drawpath{178.0}{98.0}{182.0}{100.0}
\drawpath{178.0}{100.0}{182.0}{102.0}
\drawpath{178.0}{102.0}{182.0}{104.0}
\drawpath{178.0}{104.0}{182.0}{106.0}
\drawpath{178.0}{106.0}{182.0}{108.0}
\drawpath{178.0}{108.0}{182.0}{110.0}
\drawpath{178.0}{110.0}{182.0}{112.0}
\thicklines
\drawcenteredtext{152.0}{6.0}{\Large{b)}}
\thinlines
\drawpath{178.0}{68.0}{126.0}{94.0}
\drawpath{178.0}{52.0}{126.0}{26.0}
\drawpath{178.0}{66.0}{124.0}{84.0}
\drawpath{178.0}{54.0}{124.0}{36.0}
\thicklines
\path(140.47,86.52)(141.25,86.12)(142.02,85.73)(142.8,85.36)(143.55,84.97)(144.3,84.59)(145.03,84.22)(145.77,83.86)(146.5,83.5)(147.21,83.13)
\path(147.21,83.13)(147.91,82.77)(148.61,82.44)(149.3,82.08)(150.0,81.76)(150.66,81.41)(151.33,81.08)(152.0,80.76)(152.64,80.44)(153.28,80.12)
\path(153.28,80.12)(153.92,79.83)(154.55,79.51)(155.17,79.22)(155.78,78.93)(156.38,78.62)(156.99,78.34)(157.58,78.06)(158.16,77.79)(158.72,77.51)
\path(158.72,77.51)(159.28,77.26)(159.85,76.98)(160.39,76.73)(160.94,76.48)(161.47,76.23)(161.99,76.0)(162.5,75.76)(163.02,75.51)(163.52,75.3)
\path(163.52,75.3)(164.0,75.06)(164.49,74.86)(164.97,74.63)(165.44,74.43)(165.89,74.23)(166.35,74.02)(166.78,73.83)(167.22,73.65)(167.66,73.45)
\path(167.66,73.45)(168.08,73.27)(168.49,73.11)(168.88,72.94)(169.28,72.76)(169.67,72.61)(170.05,72.44)(170.42,72.3)(170.78,72.16)(171.14,72.01)
\path(171.14,72.01)(171.5,71.87)(171.83,71.73)(172.16,71.62)(172.49,71.5)(172.8,71.37)(173.11,71.26)(173.41,71.16)(173.71,71.05)(173.99,70.95)
\path(173.99,70.95)(174.27,70.86)(174.53,70.76)(174.8,70.69)(175.05,70.61)(175.28,70.54)(175.52,70.47)(175.75,70.4)(175.97,70.33)(176.19,70.29)
\path(176.19,70.29)(176.38,70.23)(176.58,70.19)(176.77,70.15)(176.96,70.11)(177.13,70.08)(177.28,70.05)(177.44,70.02)(177.6,70.01)(177.74,70.0)
\path(177.74,70.0)(177.86,70.0)(178.0,70.0)(178.0,70.0)
\path(124.0,84.0)(124.0,84.0)(124.94,83.68)(125.9,83.36)(126.83,83.05)(127.76,82.73)(128.69,82.43)(129.6,82.12)(130.5,81.83)(131.41,81.54)
\path(131.41,81.54)(132.28,81.23)(133.17,80.94)(134.05,80.66)(134.91,80.38)(135.77,80.11)(136.61,79.83)(137.44,79.55)(138.27,79.27)(139.1,79.01)
\path(139.1,79.01)(139.91,78.75)(140.72,78.48)(141.52,78.23)(142.3,77.98)(143.08,77.73)(143.85,77.48)(144.61,77.23)(145.36,77.0)(146.11,76.76)
\path(146.11,76.76)(146.85,76.51)(147.58,76.29)(148.3,76.05)(149.0,75.83)(149.72,75.61)(150.41,75.38)(151.1,75.18)(151.77,74.95)(152.44,74.76)
\path(152.44,74.76)(153.11,74.55)(153.77,74.34)(154.41,74.15)(155.05,73.94)(155.66,73.76)(156.28,73.55)(156.91,73.37)(157.5,73.19)(158.1,73.01)
\path(158.1,73.01)(158.69,72.83)(159.27,72.66)(159.83,72.48)(160.39,72.31)(160.94,72.16)(161.49,72.0)(162.02,71.83)(162.55,71.68)(163.08,71.52)
\path(163.08,71.52)(163.58,71.37)(164.08,71.23)(164.58,71.08)(165.07,70.94)(165.55,70.81)(166.0,70.68)(166.47,70.55)(166.92,70.43)(167.36,70.3)
\path(167.36,70.3)(167.8,70.19)(168.22,70.06)(168.64,69.95)(169.05,69.83)(169.46,69.73)(169.85,69.62)(170.24,69.52)(170.61,69.44)(170.97,69.33)
\path(170.97,69.33)(171.33,69.25)(171.69,69.16)(172.03,69.08)(172.36,69.0)(172.69,68.91)(173.0,68.83)(173.32,68.76)(173.61,68.69)(173.91,68.62)
\path(173.91,68.62)(174.19,68.56)(174.47,68.51)(174.74,68.45)(175.0,68.4)(175.25,68.36)(175.49,68.3)(175.72,68.26)(175.94,68.23)(176.16,68.19)
\path(176.16,68.19)(176.36,68.16)(176.57,68.12)(176.77,68.09)(176.94,68.06)(177.11,68.05)(177.28,68.04)(177.44,68.01)(177.6,68.01)(177.74,68.0)
\path(177.74,68.0)(177.86,68.0)(178.0,68.0)(178.0,68.0)
\path(126.0,26.0)(126.0,26.0)(126.91,26.46)(127.81,26.95)(128.72,27.4)(129.61,27.87)(130.5,28.34)(131.36,28.79)(132.24,29.23)(133.1,29.68)
\path(133.1,29.68)(133.94,30.12)(134.8,30.54)(135.63,30.97)(136.46,31.4)(137.27,31.82)(138.08,32.24)(138.88,32.65)(139.69,33.06)(140.47,33.45)
\path(140.47,33.45)(141.25,33.86)(142.02,34.25)(142.8,34.63)(143.55,35.02)(144.3,35.38)(145.03,35.77)(145.77,36.13)(146.5,36.5)(147.21,36.84)
\path(147.21,36.84)(147.91,37.2)(148.61,37.54)(149.3,37.9)(150.0,38.24)(150.66,38.56)(151.33,38.9)(152.0,39.22)(152.64,39.54)(153.28,39.84)
\path(153.28,39.84)(153.92,40.15)(154.55,40.47)(155.17,40.77)(155.78,41.06)(156.38,41.34)(156.99,41.63)(157.58,41.91)(158.16,42.2)(158.72,42.47)
\path(158.72,42.47)(159.28,42.72)(159.85,43.0)(160.39,43.25)(160.94,43.5)(161.47,43.75)(161.99,43.99)(162.5,44.22)(163.02,44.47)(163.52,44.68)
\thinlines
\drawpath{66.0}{112.0}{66.0}{32.0}
\thicklines
\path(37.11,70.41)(37.11,70.41)(37.56,70.16)(38.02,69.93)(38.47,69.68)(38.91,69.44)(39.36,69.2)(39.81,68.98)(40.24,68.73)(40.68,68.51)
\path(40.68,68.51)(41.11,68.29)(41.52,68.05)(41.95,67.83)(42.38,67.62)(42.79,67.41)(43.2,67.19)(43.61,66.98)(44.02,66.76)(44.43,66.55)
\path(44.43,66.55)(44.83,66.36)(45.22,66.16)(45.61,65.95)(46.0,65.76)(46.38,65.55)(46.77,65.37)(47.15,65.18)(47.52,64.98)(47.9,64.8)
\path(47.9,64.8)(48.27,64.62)(48.63,64.44)(49.0,64.26)(49.36,64.08)(49.72,63.91)(50.06,63.75)(50.41,63.58)(50.77,63.4)(51.11,63.25)
\path(51.11,63.25)(51.45,63.09)(51.77,62.93)(52.11,62.77)(52.45,62.61)(52.77,62.47)(53.09,62.31)(53.4,62.16)(53.72,62.02)(54.04,61.88)
\path(54.04,61.88)(54.34,61.75)(54.65,61.61)(54.95,61.47)(55.25,61.34)(55.54,61.22)(55.84,61.09)(56.13,60.97)(56.41,60.84)(56.7,60.72)
\path(56.7,60.72)(56.97,60.61)(57.25,60.49)(57.52,60.38)(57.79,60.27)(58.06,60.15)(58.31,60.06)(58.58,59.95)(58.83,59.84)(59.08,59.75)
\path(59.08,59.75)(59.33,59.65)(59.58,59.56)(59.81,59.47)(60.06,59.38)(60.29,59.29)(60.52,59.22)(60.75,59.13)(60.97,59.06)(61.2,58.97)
\path(61.2,58.97)(61.4,58.9)(61.63,58.84)(61.84,58.77)(62.04,58.7)(62.25,58.63)(62.45,58.58)(62.65,58.52)(62.84,58.45)(63.02,58.4)
\path(63.02,58.4)(63.22,58.36)(63.4,58.31)(63.59,58.25)(63.75,58.22)(63.93,58.16)(64.11,58.13)(64.27,58.09)(64.44,58.06)(64.59,58.02)
\path(64.59,58.02)(64.76,58.0)(64.91,57.97)(65.05,57.95)(65.2,57.93)(65.34,57.9)(65.48,57.88)(65.62,57.86)(65.76,57.86)(65.88,57.84)
\thinlines
\drawpath{66.0}{32.0}{66.0}{20.0}
\thicklines
\path(24.0,18.0)(24.0,18.0)(24.8,18.43)(25.63,18.87)(26.45,19.29)(27.29,19.71)(28.09,20.17)(28.87,20.59)(29.67,21.01)(30.45,21.44)
\path(30.45,21.44)(31.2,21.84)(31.95,22.26)(32.72,22.68)(33.47,23.05)(34.2,23.46)(34.93,23.87)(35.65,24.27)(36.36,24.68)(37.04,25.05)
\path(65.88,57.84)(66.01,57.84)(66.13,57.84)(66.15,57.84)
\path(36.9,25.03)(36.9,25.03)(37.36,25.28)(37.81,25.53)(38.27,25.78)(38.74,26.03)(39.18,26.28)(39.63,26.52)(40.08,26.76)(40.52,27.0)
\path(40.52,27.0)(40.95,27.22)(41.4,27.46)(41.83,27.7)(42.25,27.92)(42.68,28.14)(43.09,28.37)(43.5,28.59)(43.93,28.8)(44.33,29.03)
\path(44.33,29.03)(44.74,29.23)(45.13,29.45)(45.54,29.65)(45.93,29.87)(46.33,30.06)(46.7,30.28)(47.09,30.46)(47.47,30.67)(47.86,30.87)
\path(47.86,30.87)(48.22,31.05)(48.59,31.25)(48.97,31.43)(49.33,31.62)(49.68,31.79)(50.04,31.97)(50.4,32.15)(50.75,32.33)(51.09,32.5)
\path(51.09,32.5)(51.43,32.66)(51.77,32.84)(52.11,33.0)(52.43,33.16)(52.77,33.33)(53.09,33.49)(53.4,33.63)(53.72,33.79)(54.04,33.95)
\path(54.04,33.95)(54.34,34.09)(54.65,34.24)(54.95,34.38)(55.25,34.52)(55.54,34.65)(55.84,34.79)(56.13,34.93)(56.41,35.06)(56.7,35.18)
\path(56.7,35.18)(56.97,35.31)(57.25,35.43)(57.52,35.56)(57.79,35.68)(58.04,35.79)(58.31,35.9)(58.56,36.02)(58.81,36.13)(59.06,36.24)
\path(59.06,36.24)(59.31,36.34)(59.56,36.45)(59.79,36.54)(60.02,36.65)(60.27,36.74)(60.49,36.83)(60.72,36.91)(60.93,37.0)(61.15,37.09)
\path(61.15,37.09)(61.36,37.18)(61.58,37.25)(61.79,37.34)(61.99,37.4)(62.18,37.47)(62.38,37.54)(62.58,37.61)(62.77,37.68)(62.95,37.75)
\thinlines
\drawpath{66.0}{110.0}{70.0}{112.0}
\drawpath{66.0}{106.0}{70.0}{108.0}
\drawpath{66.0}{108.0}{70.0}{110.0}
\drawpath{66.0}{104.0}{70.0}{106.0}
\drawpath{66.0}{102.0}{70.0}{104.0}
\drawpath{66.0}{100.0}{70.0}{102.0}
\drawpath{66.0}{98.0}{70.0}{100.0}
\drawpath{66.0}{96.0}{70.0}{98.0}
\drawpath{66.0}{94.0}{70.0}{96.0}
\drawpath{66.0}{92.0}{70.0}{94.0}
\drawpath{66.0}{90.0}{70.0}{92.0}
\drawpath{66.0}{88.0}{70.0}{90.0}
\drawpath{66.0}{86.0}{70.0}{88.0}
\drawpath{66.0}{84.0}{70.0}{86.0}
\drawpath{66.0}{82.0}{70.0}{84.0}
\drawpath{66.0}{80.0}{70.0}{82.0}
\drawpath{66.0}{78.0}{70.0}{80.0}
\drawpath{66.0}{76.0}{70.0}{78.0}
\drawpath{66.0}{74.0}{70.0}{76.0}
\drawpath{66.0}{72.0}{70.0}{74.0}
\drawpath{66.0}{70.0}{70.0}{72.0}
\drawpath{66.0}{68.0}{70.0}{70.0}
\drawpath{66.0}{66.0}{70.0}{68.0}
\drawpath{66.0}{64.0}{70.0}{66.0}
\drawpath{66.0}{62.0}{70.0}{64.0}
\drawpath{66.0}{60.0}{70.0}{62.0}
\drawpath{66.0}{58.0}{70.0}{60.0}
\drawpath{66.0}{56.0}{70.0}{58.0}
\drawpath{66.0}{54.0}{70.0}{56.0}
\drawpath{66.0}{52.0}{70.0}{54.0}
\drawpath{66.0}{50.0}{70.0}{52.0}
\drawpath{66.0}{48.0}{70.0}{50.0}
\drawpath{66.0}{46.0}{70.0}{48.0}
\drawpath{66.0}{44.0}{70.0}{46.0}
\drawpath{66.0}{42.0}{70.0}{44.0}
\drawpath{66.0}{40.0}{70.0}{42.0}
\drawpath{66.0}{38.0}{70.0}{40.0}
\drawpath{66.0}{36.0}{70.0}{38.0}
\drawpath{66.0}{34.0}{70.0}{36.0}
\drawpath{66.0}{34.0}{70.0}{36.0}
\drawpath{66.0}{32.0}{70.0}{34.0}
\drawpath{66.0}{30.0}{70.0}{32.0}
\drawpath{66.0}{28.0}{70.0}{30.0}
\drawpath{66.0}{26.0}{70.0}{28.0}
\drawpath{66.0}{24.0}{70.0}{26.0}
\drawpath{66.0}{22.0}{70.0}{24.0}
\drawpath{66.0}{20.0}{70.0}{22.0}
\thicklines
\path(163.52,44.68)(164.0,44.91)(164.49,45.13)(164.97,45.34)(165.44,45.56)(165.89,45.75)(166.35,45.95)(166.78,46.15)(167.22,46.34)(167.66,46.52)
\path(167.66,46.52)(168.08,46.7)(168.49,46.88)(168.88,47.04)(169.28,47.22)(169.67,47.38)(170.05,47.54)(170.42,47.68)(170.78,47.83)(171.14,47.97)
\path(171.14,47.97)(171.5,48.11)(171.83,48.25)(172.16,48.36)(172.49,48.49)(172.8,48.61)(173.11,48.72)(173.41,48.83)(173.71,48.93)(173.99,49.02)
\path(173.99,49.02)(174.27,49.13)(174.53,49.22)(174.8,49.29)(175.05,49.38)(175.28,49.45)(175.52,49.52)(175.75,49.59)(175.97,49.65)(176.19,49.7)
\path(176.19,49.7)(176.38,49.75)(176.58,49.79)(176.77,49.84)(176.96,49.88)(177.13,49.9)(177.28,49.93)(177.44,49.95)(177.6,49.97)(177.74,49.99)
\path(177.74,49.99)(177.86,49.99)(178.0,50.0)(178.0,50.0)
\path(124.0,36.0)(124.0,36.0)(124.94,36.31)(125.9,36.63)(126.83,36.93)(127.76,37.25)(128.69,37.54)(129.6,37.86)(130.5,38.15)(131.41,38.45)
\path(131.41,38.45)(132.28,38.75)(133.17,39.02)(134.05,39.31)(134.91,39.59)(135.77,39.88)(136.61,40.15)(137.44,40.43)(138.27,40.7)(139.1,40.97)
\path(139.1,40.97)(139.91,41.24)(140.72,41.5)(141.52,41.75)(142.3,42.0)(143.08,42.25)(143.85,42.5)(144.61,42.75)(145.36,43.0)(146.11,43.22)
\path(146.11,43.22)(146.85,43.47)(147.58,43.7)(148.3,43.93)(149.0,44.15)(149.72,44.38)(150.41,44.59)(151.1,44.81)(151.77,45.02)(152.44,45.22)
\path(152.44,45.22)(153.11,45.43)(153.77,45.63)(154.41,45.84)(155.05,46.04)(155.66,46.22)(156.28,46.43)(156.91,46.61)(157.5,46.79)(158.1,46.97)
\path(158.1,46.97)(158.69,47.15)(159.27,47.33)(159.83,47.5)(160.39,47.66)(160.94,47.83)(161.49,47.99)(162.02,48.15)(162.55,48.31)(163.08,48.45)
\path(163.08,48.45)(163.58,48.61)(164.08,48.75)(164.58,48.9)(165.07,49.04)(165.55,49.16)(166.0,49.31)(166.47,49.43)(166.92,49.56)(167.36,49.68)
\path(167.36,49.68)(167.8,49.79)(168.22,49.91)(168.64,50.02)(169.05,50.15)(169.46,50.25)(169.85,50.36)(170.24,50.45)(170.61,50.54)(170.97,50.65)
\path(170.97,50.65)(171.33,50.74)(171.69,50.83)(172.03,50.9)(172.36,50.99)(172.69,51.06)(173.0,51.15)(173.32,51.22)(173.61,51.29)(173.91,51.34)
\path(173.91,51.34)(174.19,51.41)(174.47,51.47)(174.74,51.52)(175.0,51.59)(175.25,51.63)(175.49,51.68)(175.72,51.72)(175.94,51.75)(176.16,51.79)
\path(176.16,51.79)(176.36,51.83)(176.57,51.86)(176.77,51.88)(176.94,51.91)(177.11,51.93)(177.28,51.95)(177.44,51.97)(177.6,51.97)(177.74,51.99)
\path(177.74,51.99)(177.86,51.99)(178.0,52.0)(178.0,52.0)
\Thicklines
\drawpath{178.0}{68.0}{178.0}{52.0}
\thicklines
\drawpath{178.0}{70.0}{178.0}{68.0}
\thinlines
\drawpath{178.0}{84.0}{182.0}{86.0}
\drawpath{178.0}{82.0}{182.0}{84.0}
\drawpath{178.0}{80.0}{182.0}{82.0}
\drawpath{178.0}{78.0}{182.0}{80.0}
\drawpath{178.0}{76.0}{182.0}{78.0}
\drawpath{178.0}{74.0}{182.0}{76.0}
\drawpath{178.0}{72.0}{182.0}{74.0}
\drawpath{178.0}{70.0}{182.0}{72.0}
\drawpath{178.0}{68.0}{182.0}{70.0}
\drawpath{178.0}{66.0}{182.0}{68.0}
\drawpath{178.0}{64.0}{182.0}{66.0}
\drawpath{178.0}{62.0}{182.0}{64.0}
\drawpath{178.0}{60.0}{182.0}{62.0}
\drawpath{178.0}{58.0}{182.0}{60.0}
\drawpath{178.0}{56.0}{182.0}{58.0}
\drawpath{178.0}{54.0}{182.0}{56.0}
\drawpath{178.0}{54.0}{182.0}{56.0}
\drawpath{178.0}{52.0}{182.0}{54.0}
\drawpath{178.0}{50.0}{182.0}{52.0}
\drawpath{178.0}{48.0}{182.0}{50.0}
\drawpath{178.0}{44.0}{182.0}{46.0}
\drawpath{178.0}{46.0}{182.0}{48.0}
\drawpath{178.0}{42.0}{182.0}{44.0}
\drawpath{178.0}{40.0}{182.0}{42.0}
\drawpath{178.0}{38.0}{182.0}{40.0}
\drawpath{178.0}{36.0}{182.0}{38.0}
\thicklines
\path(62.95,37.75)(63.13,37.81)(63.31,37.86)(63.49,37.93)(63.65,37.97)(63.83,38.02)(63.99,38.08)(64.15,38.13)(64.3,38.16)(64.45,38.2)
\path(64.45,38.2)(64.61,38.25)(64.76,38.29)(64.9,38.31)(65.04,38.34)(65.18,38.38)(65.3,38.4)(65.44,38.43)(65.55,38.45)(65.68,38.47)
\path(65.68,38.47)(65.8,38.49)(65.91,38.5)(65.91,38.5)
\drawpath{178.0}{52.0}{178.0}{50.0}
\end{picture}
\setlength{\unitlength}{1mm}
\caption{a) Part of a multisoliton solution describing reflection
off the attractive boundary in which the individual reflection processes
are separated in time\label{figtwo}.
b) This multisoliton configuration in which the two reflection
processes are superposed is obtained from the solution in
a) by a higher spin time translation symmetry. Thus it
too is a solution satisfying the boundary condition.}
}

If we have a
configuration with several soliton-antisoliton pairs with real, nonzero
and pairwise different velocities then we can use the velocity-dependent
time translation symmetry 
transformations to separate the soliton pairs arbitrarily far so that
at any time no more than one pair is close to the boundary. Thus the
problem of showing that the configuration with many soliton-antisoliton
pairs satisfies the boundary condition is reduced to showing that each
pair individually does so. This is illustrated in figure \ref{figtwo}.
Configurations to which this argument
doesn't apply directly because some velocities are zero or complex
can be viewed as limits or analytic continuations of the generic 
situation.

\subsection{Real-coupling Toda theory\label{ireal}}

We have so far discussed the solutions to the Toda equations of motion
and the boundary conditions under the assumption that the field
$\bphi$ takes complex values. However the energy functional
corresponding to the Toda equations of motion on the whole line is 
\begin{equation}
E[\bphi]=\frac 1{\beta ^2}\int\limits_{-\infty }^\infty dx\left(
\frac 12\dot{\bphi}^2+\frac 12\bphi^{\prime
}{}^2+\sum_{i=0}^n\eta _i\left( e^{\aroot_i\cdot \bphi%
}-1\right) \right) ,  \label{E}
\end{equation}
where $\beta $ is an arbitrary normalization constant. For $\bphi$
complex, this energy functional is complex and neither its real nor its
imaginary part are bounded from below. It would thus be very difficult to
make sense of the corresponding quantum theory. Therefore before we can
quantize affine Toda theory we have to restrict the allowed values for $%
\bphi$.

The simplest possibility is to let $\bphi$ and $\beta$ be real. Then
the energy is manifestly positive definite
and quantization is straightforward. This leads to the
so-called real coupling Toda theories. Another much less
straightforward possibility is called imaginary coupling
Toda theory and consists in restricting $\bphi$ and $\beta$
to be purely imaginary, i.e., one lets $\tilde{\bphi}=-i\bphi$
and $\tilde{\beta}=-i\beta$ be real.

When one restricts to real fields
none of the soliton or breather solutions survive (except in the
case of sine-Gordon, which we do not treat). Indeed there
is only one nonsingular classical solution which is real. It is
the trivial solution $\bphi=0$. This is the vacuum around which one
can then quantize the theory. The particles are the small fluctuations
around this vacuum. Their scattering matrices have been determined exactly
\cite{Bra90,Del92}.

Two interesting things happen once one restricts to
the half-line. First of all, while there are no nontrivial real solutions 
to the affine
Toda equations which are nonsingular on the whole line, there may be
real solutions whose singularities all lie on the right half-
line, out of harms way. Secondly the shift symmetry transformation
$\bphi\rightarrow\bphi+2\pi i\weight$
is no longer allowed because it would map to a complex field.
Therefore the boundary conditions in \eq{bcn} are no longer
equivalent. Instead each choice of the signs $A_i$ defines
a different real coupling Toda theory on the half-line.

The task therefore arises to find for each boundary condition
those real solutions which are nonsingular on the left half-line.
This is particularly important because
the trivial vacuum solution $\bphi=0$
does not satisfy most of these boundary conditions and
an alternative vacuum solution must be found before one
can start to quantize these theories.

This problem has been addressed already by Bowcock in \cite{Bow96}
for $a_n^{(1)}$ Toda theory on the half-line. We
repeat his analysis in section \ref{sect:real} in our formalism.
We find that if one puts a single stationary soliton (and its mirror
antisoliton)
at just the right distance (given by \eq{t43}) from the $\varepsilon=-1$
boundary and adds a constant $2\pi i\weight$ 
where $\weight$ is the topological charge of the soliton
one obtains a real nonsingular solution. One obtains the identical
solution if one replaces $\weight$ by $-\weight$ because soliton
and antisoliton enter symmetrically in the solution. 

While we can not
prove it, we believe that this is the only
nonsingular  real solution. No nonsingular real solutions
can be obtained from the solutions for the $\varepsilon=1$ boundary.

In the approach of \cite{Bow96} it was not so easy to determine which 
particular boundary conditions the real solutions satisfy. For us
this is trivial because we can view the stationary real soliton as the
limit of a moving complex soliton as the velocity goes to zero. For a
moving soliton one can perform the analysis at a time when
the soliton is far away. At the boundary the field then takes the constant
value $\bphi=2i\pi\weight$. It therefore satisfies the boundary
condition
\begin{equation}\label{e113}
\partial _x\bphi|_{x=0}=\sum_{i=0}^n\ A_i^{(\weight)}\,
\eta _i\,\aroot_i\ e^{\aroot_i\cdot 
\bphi/2}|_{x=0},  
\end{equation}
with
\begin{equation}
A_i^{(\weight)}=-e^{i\pi \aroot_i\cdot\weight}.
\end{equation}
Because the different boundary conditions are not continuously
connected, the
boundary condition satisfied by the solution can not change
as the velocity is taken to zero. Thus also the real stationary
soliton solution satisfies this boundary condition.

The topological charges $\weight$ of single solitons in $a_n^{(1)}$
Toda theory are known to be weights in the fundamental representations 
of $a_n$. As $\weight$ runs through all the weights in the fundamental
representations $A_i^{(\weight)}$ runs through all possible sign
combinations with the property that
\begin{equation}\label{ec}
\prod_{i=0}^n A_i^{(\weight)}=
(-1)^{n+1},
\end{equation}
except for the one with $\weight=0$ which corresponds to the
trivial solution $\bphi=0$.
This implies that if all the weights in the fundamental representation
were realized as topological charges of solitons, then we would
have found a vacuum solution for every real coupling $a_n^{(1)}$
Toda theory with a boundary condition which satisfies the constraint
\eqref{ec}. Furthermore every sign combination is obtained exactly
twice as $A_i^{(\weight)}=A_i^{(-\weight)}$. These two correspond
to the same solution however and thus we find exactly one vacuum
solution to every boundary condition.

However 
not for every weight in the fundamental representations has a classical
soliton solution been found which has this weight as its topological
charge. This has been a puzzle because in the quantum theory of imaginary
coupling Toda theory 
the solitons need to fill out the entire representations. This puzzle has
often been interpreted as an indication that imaginary coupling
Toda theory is sick. Now however it is hitting us even in
real coupling Toda theory. If solitons exist for all weights in the
fundamental representations then we get a vacuum solution for all
boundary conditions of the form  \eqref{e113} with the constraint
\eqref{ec}. If solitons do not exist for all weights, then some of
those boundary conditions don't have a vacuum. This seems a bit unnatural.
It seems that the semiclassical analysis, which is used to make
the correspondence between classical solutions and quantum states
is breaking down. It is my belief that the same, as yet unknown, mechanism
which provides the missing quantum soliton states in the imaginary
coupling Toda theory will also provide the missing quantum vacuum states in
the real coupling Toda theory on the half-line with the boundary conditions
satisfying the constraint \eqref{ec}.

Even more dubious is the situation for those real coupling Toda theories
on the half-line for which the boundary conditions do not satisfy \eqref{ec}.
For these we don't find any vacuum solutions even if we work with the
full set of solitons. We thus seem to have no handle on the quantum
theory for these. This agrees with an observation made by Corrigan et.al.
during their work on \cite{Cor95}. They tried to derive the classical
reflection factor for $a_2^{(1)}$ with boundary condition specified by
$A_0=A_1=-1$ and $A_2=1$ which does not satisfy the condition \eqref{ec}
and found a bizzare-looking result [private communication].

There is another puzzling fact about the classical soliton solutions of
affine Toda theory which now finds its way into real coupling Toda theory.
The classical soliton solution contains a certain angular parameter
$\zeta$ which can take on any value between $0$ and $2\pi$. This is
the parameter which determines the asymptotic value of the solution and
thus the topological charge of the soliton. However there is only
a discrete set of topological charges. The way this works is that the
interval $[0,2\pi]$ is split into several intervals. The topological
charge stays constant for $\zeta$ within one of these intervals and
changes discontinuously as $\zeta$ crosses into the next interval.
Thus there is actually a family of solutions for each soliton with
a specific topological charge, parametrized by a continuous parameter
ranging over an interval. It is not well understood in the semiclassical
approximation what the effect of such a restricted zero-mode is. But
because there are so many things in the semiclassical approximation for
imaginary coupling Toda theory which are not well understood, not much
attention has been paid to this particular problem. 

\EPSFIGURE{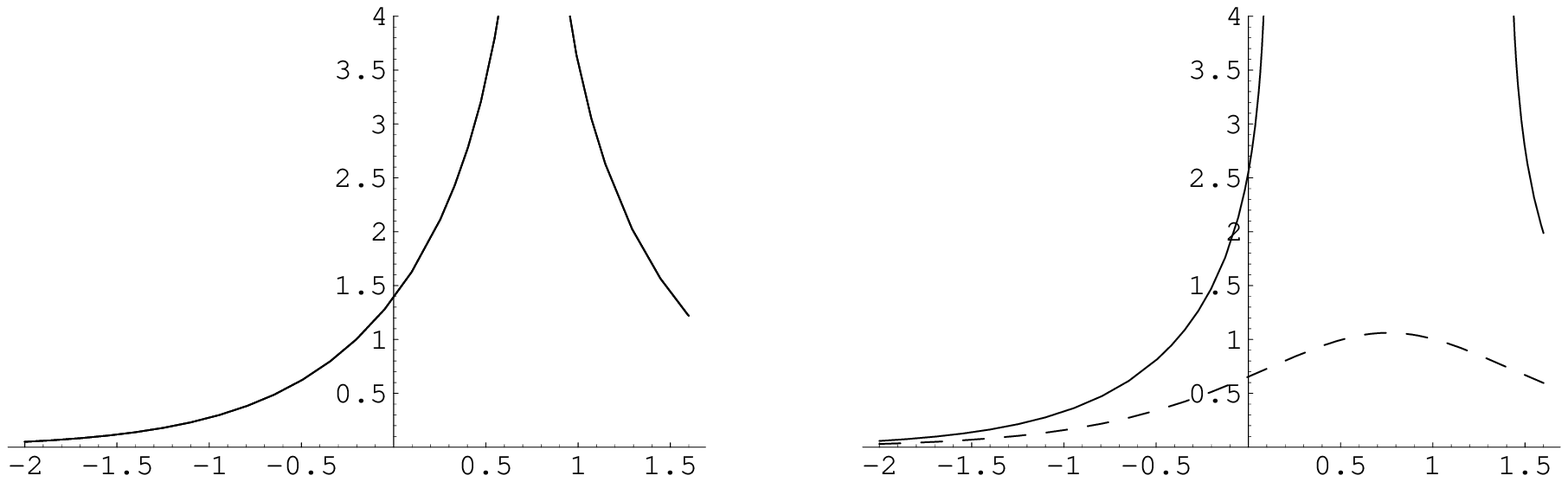,width=15cm}
{Plot of the vacuum solution of $a_2^{(1)}$ Toda theory with
$A_0=-1,A_1=A_2=1$ at two different values of the modulus parameter
$\zeta$. The left plot is at $\zeta=0$, the right plot is at $\zeta=0.5$.
\label{figv}}

Now the same continuous parameter appears in the vacuum solutions
of real coupling Toda theory with a boundary. To give an idea of what
these vacuum solutions look like, we have plotted the vacuum solution
for $a_2^{(1)}$ affine Toda theory with the boundary condition
specified by $A_0=-1$, $A_1=A_2=1$ in figure \ref{figv} for two different
values of the parameter $\zeta$. The field $\bphi$ of $a_2^{(1)}$ Toda theory 
is a two-component vector. We have plotted $\phi_1=\weight_1\cdot\bphi$ with
a solid line and $\phi_2=\weight_2\cdot\bphi$ with a dashed line. For
$\zeta=0$ the two components coincide and both have a singularity on
the right half-line. For $\zeta=0.5$ the component
$\phi_2$ is no longer singular.
Any value of $\zeta$ between $-\pi/6$ and $\pi/6$ gives a different
solution to these boundary conditions, all with the same energy.
It will be interesting to study what, if any, consequences this vacuum
degeneracy has for the quantum theory.

This concludes our general discussion and we now turn to concrete 
calculations.

\section{Review of classical solutions on the whole line\label{sect:review}}

This section reviews well known facts about the classical solutions
of affine Toda theory which we will use.

\subsection{Hirota's tau functions}

Applying a method developed by Hirota \cite{Hir} and adapted to affine Toda
theories by Hollowood \cite{Hol92}, one parameterizes the $n$-component field 
$\bphi$ by the $n+1$ tau functions $\tau _i$ as 
\begin{equation}
\bphi(x)-\bphi(-\infty)=
-\sum_{i=0}^n\frac{2\aroot_i}{\aroot_i^2}%
\,\ln \tau _i.  \label{tau}
\end{equation}
For finite energy solutions $\bphi(-\infty)$ will have to lie in one
of the minima of the Toda potential, i.e., $\bphi(-\infty)=2\pi i \weight$
where $\weight$ can be any coweight of the underlying finite dimensional
Lie algebra $g$. Which minima is chosen is irrelevant because the
constant shift $\bphi\rightarrow\bphi+2\pi i \weight$ is a symmetry
of affine Toda theory for any coweight $\weight$.

Substituting the parameterization (\ref{tau}) 
into the equations of motion (\ref{phieom}) gives 
\begin{equation}
\sum_{i=0}^n\,\aroot_i\,\frac 1{\tau _i^2}\,\left( \frac 2{\aroot_i^2}
\left( \ddot{\tau _i}\tau _i-\dot{\tau}_i^2-\tau _i^{\prime
\prime }\tau _i+{\tau ^{\prime }}_i^2\right) -\eta _i\prod_{j=0}^n\,\tau
_j^{I_{ij}}\right) =0,  \label{taueom}
\end{equation}
where $I_{ij}$ is the incidence matrix $I_{ij}=-2\aroot_i\cdot 
\aroot_j/\aroot_j^2+2\delta _{ij}$. Having introduced $%
n+1$ tau functions we can choose one of them to satisfy the equation 
\begin{equation}
\ddot{\tau _0}\tau _0-\dot{\tau _0}^2-\tau _0^{\prime \prime }\tau _0+{\tau
^{\prime }}_0^2=\frac{\aroot_0^2}2\left( \prod_{j=0}^n\,\tau
_j^{I_{0j}}-\tau _0^2\right)  \label{tauoeom}
\end{equation}
Multiplying eq.~(\ref{taueom}) by fundamental weights $\weight_i$
and using eq.~(\ref{tauoeom}) one obtains the following set of
equations of motion 
\begin{equation}
\ddot{\tau _i}\tau _i-\dot{\tau _i}^2-\tau _i^{\prime \prime }\tau _i+{\tau
^{\prime }}_i^2=\frac{\aroot_i^2}2\eta _i\left(
\prod_{j=0}^n\,\tau _j^{I_{ij}}-\tau _i^2\right) ,~~~~i=1,\dots ,n
\label{tauieom}
\end{equation}
which thus take the same form as eq.~(\ref{tauoeom}).

\subsection{The soliton solutions for $a_n^{(1)}$}

The parameterization of $\bphi$ in terms of Hirota $\tau $ functions
turns out to be particularly useful in the case of $\hat{g}=a_n^{(1)}$ where
the equations of motion (\ref{tauieom}) are quadratic in the $\tau $ once
one substitutes the particular form of the incidence matrix $I_{ij}$. 
\begin{equation}
\ddot{\tau _i}\tau _i-\dot{\tau _i}^2-\tau _i^{\prime \prime }\tau _i+{\tau
^{\prime }}_i^2=\tau _{i-1}\tau _{i+1}-\tau _i^2,~~~~i=0,\dots ,n
\label{tauieoma}
\end{equation}
We have defined $\tau _{-1}\equiv \tau _n$ and $\tau _{n+1}\equiv \tau _0$.

The single soliton solutions to these equations are 
\begin{equation}
\tau _j^{(a)}=1+e^\Omega \omega ^{ja},~~~~~~a=1,\dots ,n  \label{sol}
\end{equation}
with 
\begin{equation}
\Omega =\sigma (x-vt)-\xi ,\qquad \qquad \omega =e^{\frac{2\pi i}{n+1}}
\label{e27}
\end{equation}
and 
\begin{equation}
\sigma ^2(1-v^2)=4\sin ^2\frac{a\pi }{n+1}\equiv m_a^2  \label{smass}
\end{equation}
Here $a=1,\dots ,n$ labels the species of soliton, $v$ is the velocity, $%
1/\sigma $ the width of the soliton and relation (\ref{smass}) expresses the
Lorentz contraction. We choose $\sigma $ positive. 
$\xi=\rho+i\zeta$ is an arbitrary complex number. Its real part $\rho$
determines the center of mass position\footnote{%
There is a subtlety here. While the mass of a soliton solution turns out to
always be real, this is not true of the mass density or the center of mass
position. They are both complex. However the real part of the center of mass
position at  $t=0$ equals the real part of $\xi$.}, while its imaginary part
$\zeta$ determines the asymptotic values of the soliton solution, and 
therefore its
topological charge \cite{Hol92,McG93}. 
For a fixed topological charge the parameter $\zeta$ can
still range over an interval. The topological charge is defined
as $(\bphi(\infty)-\bphi(-\infty))/2\pi i$ and turns out to always be a
weight of the $a$th
fundamental representation of $a_n$. The fact that the
asymptotic values of $\bphi$ are purely imaginary means that these
solutions will play a role only in the imaginary coupling Toda theory.

The Hirota Ansatz gives the multi-soliton solutions by a nonlinear
superposition. For two solitons this takes the form 
\begin{equation}  \label{twosolitons}
\tau_j= 1+ e^{\Omega_1}\omega^{ja_1} +e^{\Omega_2}\omega^{ja_2}
+X_{12}e^{\Omega_1+\Omega_2}\omega^{j(a_1+a_2)}
\end{equation}
with the ``interaction'' function 
\begin{equation}  \label{intfunc}
X_{12}= -\frac{(\sigma_1-\sigma_2)^2-(\sigma_1
v_1-\sigma_2v_2)^2-m^2_{a_1-a_2}} {(\sigma_1+\sigma_2)^2-(\sigma_1v_1+%
\sigma_2v_2)^2-m^2_{a_1+a_2}}.
\end{equation}

It is sometimes useful to use the rapidity variables $\theta_p$ defined by $%
v_p=\tanh(\theta_p)$. Then $\sigma_p=m_{a_p}\,\cosh(\theta_p)$ and $%
\sigma_p\,v_p=m_{a_p}\,\sinh(\theta_p)$. The interaction function $X_{pq}$
depends only on the relative rapidity $\theta=\theta_p-\theta_q$ 
\begin{equation}  \label{gammadef}
X_{pq}(\theta)=\frac {\sinh\left(\frac{\theta}{2}+\frac{i\pi(a_p-a_q)}{2n+2}%
\right) \sinh\left(\frac{\theta}{2}-\frac{i\pi(a_p-a_q)}{2n+2}\right)}
{\sinh\left(\frac{\theta}{2}+\frac{i\pi(a_p+a_q)}{2n+2}\right) \sinh\left(%
\frac{\theta}{2}-\frac{i\pi(a_p+a_q)}{2n+2}\right)}.
\end{equation}

At the far past the two solitons described by the solution (\ref{twosolitons}%
) are far apart and have their original shape. They then interact and deform
at intermediate times and reemerge in the far future with their original
shape regained but at shifted positions.
We can determine the asymptotic trajectories of the right-moving soliton
by taking the limits of the solution as $t\rightarrow
\pm \infty $ while keeping $x-vt$ fixed 
\begin{eqnarray}
\lim_{\stackrel{t\rightarrow -\infty }{x-vt \text{ fixed}}}
\tau _j &=&1+e^{\Omega _1}\omega ^{ja_1},\label{t1} \\
\lim_{\stackrel{t\rightarrow \infty }{x-vt \text{ fixed}}}
\tau _j &=&\left( 1+X_{12}e^{\Omega _1}\omega
^{ja_1}\right) e^{\Omega _2}\omega ^{ja_2}.
\end{eqnarray}
Similarly we can look at the left-moving soliton
\begin{eqnarray}
\lim_{\stackrel{t\rightarrow -\infty }{x+vt \text{ fixed}}}
\tau_j &=&\left( 1+X_{12}e^{\Omega _2}\omega
^{ja_2}\right) e^{\Omega _1}\omega ^{ja_1},\\
\lim_{\stackrel{t\rightarrow \infty }{x+vt \text{ fixed}}}
\tau _j &=&1+e^{\Omega _2}\omega ^{ja_2}.\label{t4}
\end{eqnarray}
We have plotted these trajectories for the case $\sigma_1=\sigma_2$
and $v_1=-v_2=v>0$ in figure \ref{figt}.

\FIGURE{
\setlength{\unitlength}{1mm}
\begin{picture}(182,82)(12,0)
\thinlines
\drawpath{162.0}{16.0}{30.0}{16.0}
\drawpath{38.0}{6.0}{38.0}{78.0}
\drawpath{130.0}{10.0}{30.0}{60.0}
\drawpath{66.0}{10.0}{162.0}{58.0}
\thicklines
\path(34.0,8.0)(34.0,8.0)(34.99,8.51)(35.97,9.02)(36.95,9.53)(37.93,10.03)(38.9,10.53)(39.86,11.02)(40.81,11.51)(41.75,11.98)
\path(41.75,11.98)(42.7,12.45)(43.63,12.94)(44.56,13.39)(45.47,13.86)(46.38,14.31)(47.29,14.77)(48.18,15.2)(49.06,15.65)(49.95,16.08)
\thinlines
\drawpath{162.0}{8.0}{34.0}{72.0}
\drawpath{114.0}{48.0}{34.0}{8.0}
\thicklines
\path(49.95,16.08)(50.83,16.5)(51.7,16.93)(52.56,17.36)(53.4,17.77)(54.25,18.18)(55.08,18.58)(55.91,18.97)(56.75,19.36)(57.55,19.75)
\path(57.55,19.75)(58.37,20.13)(59.16,20.52)(59.97,20.88)(60.76,21.25)(61.54,21.61)(62.3,21.97)(63.06,22.31)(63.83,22.66)(64.58,23.0)
\path(64.58,23.0)(65.33,23.34)(66.06,23.68)(66.8,24.0)(67.51,24.31)(68.23,24.63)(68.94,24.93)(69.63,25.25)(70.33,25.54)(71.02,25.84)
\path(71.02,25.84)(71.69,26.13)(72.37,26.4)(73.04,26.68)(73.69,26.95)(74.34,27.22)(74.98,27.49)(75.62,27.75)(76.26,28.0)(76.87,28.25)
\path(76.87,28.25)(77.5,28.49)(78.09,28.72)(78.7,28.95)(79.3,29.18)(79.87,29.4)(80.45,29.61)(81.02,29.83)(81.59,30.04)(82.16,30.24)
\path(82.16,30.24)(82.7,30.43)(83.25,30.63)(83.77,30.81)(84.3,30.99)(84.83,31.15)(85.34,31.33)(85.86,31.5)(86.34,31.65)(86.84,31.81)
\path(86.84,31.81)(87.33,31.95)(87.8,32.09)(88.27,32.24)(88.73,32.36)(89.19,32.5)(89.65,32.61)(90.08,32.74)(90.52,32.84)(90.94,32.95)
\path(90.94,32.95)(91.37,33.06)(91.79,33.15)(92.19,33.24)(92.58,33.33)(92.98,33.4)(93.37,33.49)(93.75,33.56)(94.12,33.61)(94.48,33.68)
\path(94.48,33.68)(94.83,33.72)(95.18,33.77)(95.51,33.83)(95.86,33.86)(96.19,33.9)(96.5,33.93)(96.81,33.95)(97.12,33.97)(97.41,33.97)
\path(97.41,33.97)(97.7,33.99)(97.98,34.0)(98.0,34.0)
\path(162.0,58.0)(162.0,58.0)(161.0,57.48)(160.0,56.97)(159.02,56.45)(158.05,55.95)(157.08,55.45)(156.11,54.98)(155.16,54.48)(154.22,54.01)
\path(154.22,54.01)(153.28,53.54)(152.36,53.08)(151.42,52.61)(150.5,52.16)(149.6,51.7)(148.69,51.26)(147.8,50.83)(146.91,50.38)(146.03,49.95)
\path(146.03,49.95)(145.16,49.54)(144.28,49.12)(143.44,48.72)(142.58,48.3)(141.74,47.91)(140.89,47.51)(140.07,47.12)(139.25,46.75)(138.42,46.37)
\path(138.42,46.37)(137.61,46.0)(136.82,45.63)(136.02,45.27)(135.24,44.9)(134.44,44.56)(133.67,44.22)(132.91,43.88)(132.16,43.54)(131.41,43.22)
\path(131.41,43.22)(130.66,42.9)(129.91,42.59)(129.19,42.27)(128.47,41.97)(127.75,41.68)(127.05,41.38)(126.35,41.09)(125.64,40.81)(124.96,40.54)
\path(124.96,40.54)(124.28,40.27)(123.61,40.0)(122.94,39.74)(122.28,39.49)(121.63,39.24)(121.0,39.0)(120.36,38.75)(119.72,38.52)(119.11,38.29)
\path(119.11,38.29)(118.49,38.08)(117.88,37.86)(117.27,37.65)(116.69,37.45)(116.11,37.25)(115.52,37.06)(114.95,36.88)(114.38,36.68)(113.83,36.52)
\path(113.83,36.52)(113.27,36.34)(112.73,36.18)(112.2,36.02)(111.68,35.86)(111.16,35.72)(110.63,35.58)(110.12,35.45)(109.63,35.31)(109.13,35.18)
\path(109.13,35.18)(108.66,35.06)(108.18,34.95)(107.7,34.84)(107.25,34.75)(106.79,34.65)(106.33,34.56)(105.9,34.47)(105.45,34.38)(105.04,34.31)
\path(105.04,34.31)(104.61,34.25)(104.19,34.18)(103.8,34.11)(103.4,34.06)(103.01,34.02)(102.62,33.97)(102.23,33.95)(101.87,33.91)(101.51,33.88)
\path(101.51,33.88)(101.16,33.88)(100.8,33.86)(100.47,33.84)(100.12,33.84)(99.8,33.86)(99.48,33.86)(99.16,33.88)(98.87,33.9)(98.56,33.93)
\path(98.56,33.93)(98.27,33.95)(98.0,34.0)(98.0,34.0)
\path(30.0,60.0)(30.0,60.0)(31.06,59.44)(32.13,58.88)(33.2,58.33)(34.25,57.8)(35.29,57.26)(36.33,56.73)(37.36,56.22)(38.38,55.7)
\path(38.38,55.7)(39.38,55.19)(40.38,54.69)(41.38,54.19)(42.38,53.7)(43.36,53.22)(44.33,52.73)(45.29,52.26)(46.25,51.8)(47.2,51.33)
\path(47.2,51.33)(48.13,50.88)(49.06,50.44)(50.0,50.0)(50.9,49.55)(51.81,49.12)(52.72,48.69)(53.61,48.27)(54.5,47.87)(55.37,47.45)
\path(55.37,47.45)(56.23,47.05)(57.09,46.66)(57.94,46.27)(58.79,45.88)(59.62,45.52)(60.45,45.15)(61.27,44.77)(62.08,44.41)(62.88,44.06)
\path(62.88,44.06)(63.69,43.72)(64.48,43.38)(65.26,43.04)(66.02,42.72)(66.79,42.4)(67.55,42.08)(68.3,41.77)(69.04,41.45)(69.76,41.15)
\path(69.76,41.15)(70.48,40.86)(71.2,40.58)(71.91,40.29)(72.62,40.02)(73.3,39.75)(73.98,39.5)(74.66,39.24)(75.33,38.99)(76.0,38.74)
\path(76.0,38.74)(76.65,38.5)(77.29,38.27)(77.93,38.04)(78.55,37.81)(79.18,37.61)(79.79,37.4)(80.38,37.2)(80.98,37.0)(81.58,36.81)
\path(81.58,36.81)(82.16,36.61)(82.73,36.43)(83.29,36.27)(83.84,36.09)(84.4,35.93)(84.94,35.79)(85.47,35.63)(85.98,35.5)(86.51,35.36)
\path(86.51,35.36)(87.01,35.22)(87.51,35.09)(88.01,34.97)(88.48,34.86)(88.97,34.75)(89.44,34.65)(89.9,34.56)(90.34,34.47)(90.79,34.4)
\path(90.79,34.4)(91.23,34.31)(91.66,34.25)(92.08,34.18)(92.48,34.11)(92.88,34.06)(93.29,34.02)(93.68,33.97)(94.05,33.95)(94.43,33.91)
\path(94.43,33.91)(94.79,33.9)(95.15,33.88)(95.5,33.86)(95.83,33.86)(96.16,33.86)(96.48,33.86)(96.8,33.88)(97.12,33.9)(97.41,33.93)
\path(97.41,33.93)(97.7,33.95)(97.98,34.0)(98.0,34.0)
\path(162.0,8.0)(162.0,8.0)(161.0,8.51)(160.0,9.02)(159.02,9.53)(158.05,10.03)(157.08,10.53)(156.11,11.02)(155.16,11.51)(154.22,11.98)
\path(154.22,11.98)(153.28,12.45)(152.36,12.94)(151.42,13.39)(150.5,13.86)(149.6,14.31)(148.69,14.77)(147.8,15.2)(146.91,15.65)(146.03,16.08)
\path(146.03,16.08)(145.16,16.5)(144.28,16.93)(143.44,17.36)(142.58,17.77)(141.74,18.18)(140.89,18.58)(140.07,18.97)(139.25,19.36)(138.42,19.75)
\path(138.42,19.75)(137.61,20.13)(136.82,20.52)(136.02,20.88)(135.24,21.25)(134.44,21.61)(133.67,21.97)(132.91,22.31)(132.16,22.66)(131.41,23.0)
\path(131.41,23.0)(130.66,23.34)(129.91,23.68)(129.19,24.0)(128.47,24.31)(127.75,24.63)(127.05,24.93)(126.35,25.25)(125.64,25.54)(124.96,25.84)
\path(124.96,25.84)(124.28,26.13)(123.61,26.4)(122.94,26.68)(122.28,26.95)(121.63,27.22)(121.0,27.49)(120.36,27.75)(119.72,28.0)(119.11,28.25)
\path(119.11,28.25)(118.49,28.49)(117.88,28.72)(117.27,28.95)(116.69,29.18)(116.11,29.4)(115.52,29.61)(114.95,29.83)(114.38,30.04)(113.83,30.24)
\path(113.83,30.24)(113.27,30.43)(112.73,30.63)(112.2,30.81)(111.68,30.99)(111.16,31.15)(110.63,31.33)(110.12,31.5)(109.63,31.65)(109.13,31.81)
\path(109.13,31.81)(108.66,31.95)(108.18,32.09)(107.7,32.24)(107.25,32.36)(106.79,32.5)(106.33,32.61)(105.9,32.74)(105.45,32.84)(105.04,32.95)
\path(105.04,32.95)(104.61,33.06)(104.19,33.15)(103.8,33.24)(103.4,33.33)(103.01,33.4)(102.62,33.49)(102.23,33.56)(101.87,33.61)(101.51,33.68)
\path(101.51,33.68)(101.16,33.72)(100.8,33.77)(100.47,33.83)(100.12,33.86)(99.8,33.9)(99.48,33.93)(99.16,33.95)(98.87,33.97)(98.56,33.97)
\path(98.56,33.97)(98.27,33.99)(98.0,34.0)(98.0,34.0)
\thinlines
\drawdotline{98.0}{34.0}{98.0}{10.0}
\drawcenteredtext{50.0}{12.0}{$\frac{\xi_1}{\sigma}$}
\drawcenteredtext{78.0}{12.0}{$\frac{\xi_1-\log X}{\sigma}$}
\drawcenteredtext{148.0}{12.0}{$\frac{\xi_2-\log X}{\sigma}$}
\drawcenteredtext{118.0}{12.0}{$\frac{\xi_2}{\sigma}$}
\drawcenteredtext{98.0}{6.0}{$\frac{\xi_1+\xi_2-\log X}{2\sigma}$}
\drawrighttext{34.0}{56.0}{$\frac{\xi_2}{\sigma v}$}
\drawrighttext{34.0}{70.0}{$\frac{\xi_2-\log X}{\sigma v}$}
\drawcenteredtext{160.0}{20.0}{$x$}
\drawlefttext{42.0}{76.0}{$t$}
\drawpath{36.0}{76.0}{38.0}{78.0}
\drawpath{38.0}{78.0}{40.0}{76.0}
\drawpath{160.0}{14.0}{162.0}{16.0}
\drawpath{162.0}{16.0}{160.0}{18.0}
\drawrightbrace{39.0}{63.0}{8.0}
\thicklines
\drawlefttext{42.0}{62.0}{$-\Delta t=-\frac{\log X}{\sigma v}$}
\end{picture}
\setlength{\unitlength}{1mm}
\caption{Details of a two-soliton scattering process. For simplicity we 
plotted two solitons of equal mass in their center of mass frame. 
The thin lines are the asymptotic trajectories as read off from equations
(\ref{t1})-(\ref{t4}). The thick lines are fictitious interpolations 
between them.
The drawing takes into account that $\log X$ is always negative.
\label{figt}}
}
\setlength{\unitlength}{0.6mm}
We read off that asymptotically the only observable effect of the
interaction between the two solitons is that they have 
experienced a time-delay $\Delta t$ given
by
\begin{equation}
\Delta t=\frac{\log X_{12}}{\sigma v}.  \label{std}
\end{equation}
This time delay is always negative (a time advance really) because 
$\log X_{12}$ is always
negative \cite{Fri94}.

The solitons satisfy a classical exclusion principle: one can not have two
or more solitons of the same species traveling with the same velocity. If
one tries to construct such a solution one finds that $X_{12}$ in eq.~(\ref
{intfunc}) vanishes and thus the $\tau_j$ in eq.~(\ref{twosolitons})
collapse to a one-soliton solution with a shifted position.

An $N$-soliton solution is given by 
\begin{equation}  \label{N}
\tau_j=\sum_{\mu_1=0}^1\cdots\sum_{\mu_N=0}^1\prod_{p=1}^N
\left(\left(\prod_{q=1}^{p-1}\left(X_{pq}\right)^{\mu_q}\right)
e^{\Omega_p}\omega^{j a_p}\right)^{\mu_p}.
\end{equation}
Let us again look at the asymptotic positions of the solitons. We choose the
labeling so that $v_1\geq v_2\geq \cdots \geq v_N$. We then look at the $r$%
-th soliton by taking the limits $t\rightarrow \pm \infty $ while keeping $%
x-v_rt$ fixed. We assume for simplicity that there is no other soliton with
velocity $v_r$. If there were such a soliton, it would have no effect on the
position of the $r$-th soliton. 
\begin{eqnarray}
\lim_{t\rightarrow -\infty }\tau _j &=&\left(
1+\prod_{p=1}^{r-1}X_{pr}e^{\Omega _r}\omega ^{ja_r}\right)
\prod_{p=1}^{r-1}\left( \prod_{q=1}^{p-1}X_{pq}\right) e^{\Omega _p}\omega
^{ja_p},  \nonumber \\
\lim_{t\rightarrow \infty }\tau _j &=&\left(
1+\prod_{p=r+1}^NX_{pr}e^{\Omega _r}\omega ^{ja_r}\right)
\prod_{p=r+1}^N\left( \prod_{q=p+1}^NX_{pq}\right) e^{\Omega _p}\omega
^{ja_p}.
\end{eqnarray}
Thus we see that at $t=\infty $ the soliton is shifted with respect to the
expected position by the amount 
\begin{equation}
\sigma \,\Delta x=\sum_{p=1}^{r-1}\log X_{pr}-\sum_{r+1}^N\log X_{pr}.
\end{equation}
Thus each time a soliton $p$ scatters with another soliton $q$ it is shifted
by $\log X_{pq}$ if soliton $q$ comes from the left and it is shifted by the
negative of $\log X_{pq}$ if it comes from the right. Because $\log X_{pq}$
is always negative \cite{Fri94} this can be interpreted as an attractive
force between the solitons.

\subsection{The energy in terms of tau functions}

An expression for the energy in terms of the $\tau $ functions can be
obtained by substituting eq.~(\ref{tau}) into the energy functional (\ref{E}%
), giving 
\begin{equation}
E[\tau ]=\frac 1{\beta ^2}\int\limits_{-\infty }^\infty dx\sum_{i=0}^n\left(
\frac 1{\aroot_i^2}\sum_{j=0}^na_{ij}\frac{\dot{\tau}_i\dot{\tau}%
_j+\tau _i^{\prime }\tau _j^{\prime }}{\tau _i\tau _j}+\eta
_i\prod_{j=0}^n\tau _j^{-a_{ij}}-\eta _i\right) ,  \label{et}
\end{equation}
where $a_{ij}=2\aroot_i\cdot \aroot_j/\aroot%
_j^2$ is the Cartan matrix of $\hat{g}$.

In \cite{Oli93} it was observed that for soliton solutions the energy
density could be rewritten as a total derivative so that the energy is given
as a surface term, which in terms of $\tau $ functions is 
\begin{equation}
E[\tau ]=\left[ -\frac 2{\beta ^2}\sum_{i=0}^n\frac 2{\alpha _i^2}\frac{\tau
_i^{\prime }}{\tau _i}\right] _{x=-\infty }^\infty .  \label{es}
\end{equation}
Evaluating this for the soliton solutions of $a_n^{(1)}$ Toda theory in eq.~(%
\ref{sol}) one gets 
\begin{equation}
E_{\text{{\small soliton}}}=-\frac 2{\beta ^2}\sum_{i=0}^n\sigma =-\frac{%
2(n+1)}{\beta ^2}\sigma .  \label{esol}
\end{equation}
This energy is real and positive for imaginary coupling constant $\beta =i%
\tilde{\beta}$. The rest mass of a soliton of type $a$ is thus\cite{Hol92} 
\begin{equation}
M_a=\frac{2(n+1)}{\tilde{\beta}^2}\,m_a.  \label{msol}
\end{equation}
The energy of multi-soliton solutions is just the sum of the energy of the
individual solitons \cite{Oli93}.

\section{Classical solutions on the half-line}
\label{sect:3}

Among the solutions of the equations of motion (%
\ref{tauieom}) reviewed in the previous section we will now
identify those which satisfy the boundary conditions (\ref
{bc}) at $x=0$. Except for the following first two general
sections we will again restrict ourselves to $\hat{g}=a_n^{(1)}$.
We will also not pay attention to the rather special case
of $n=1$ (the sine-Gordon model) because this has
been well studied already \cite{Sal94}.

\subsection{The boundary conditions\label{subsect:bc}}

The Hirota formalism is extremely well suited for dealing with boundary
conditions of the form (\ref{bc}) because they too lead to equations
quadratic in the $\tau$. This has first been observed and exploited by
Bowcock in \cite{Bow96}.

Substituting the Hirota Ansatz (\ref{tau}) into the boundary condition (\ref
{bc}) and taking the inner products with the fundamental weights
$\weight_i$ leads to the set of $n+1$ equations 
\begin{equation}
\frac{\tau _i^{\prime }}{\tau _i}+\varepsilon\,\eta _i\,\frac{
\alpha_i^2}{2}
\prod_{j=0}^n\tau _j^{-\frac{a_{ij}}2}=\eta_i\,C~~~~\text{at }
x=0,~~~i=0,\dots ,n,
\label{taubc1}
\end{equation}
where $C$ is some constant. Taking the square of eq.~(\ref{taubc1}) and
multiplying by $\tau _i^2$ gives 
\begin{equation}
\varepsilon^2\,\prod_{j=0}^n\tau _j^{I_{ij}}=
\left(\frac{2}{\alpha _i^2\eta _i}\right)^2
\left(\eta_i C\tau_i-\tau _i^{\prime }\right) ^2~~~~\text{at }x=0.
\label{taubc2}
\end{equation}
Upon specializing to $a_n^{(1)}$ Toda theory one obtains 
\begin{equation}
\varepsilon^2\tau _{i-1}\tau _{i+1}-\left( C\tau _i-\tau _i^{\prime }\right)
^2=0,~~~\;i=0,\dots n,  \label{taubc}
\end{equation}
again valid only at $x=0$. We have defined $\tau _{-1}\equiv \tau _n$ and $%
\tau _{n+1}\equiv \tau _0$.

In addition to this condition on the tau functions the boundary condition
also restricts the asymptotic value of the solution at $x=-\infty$. If
the solution describes $m$ incoming solitons with topological charges
$\weight_k$ then we need
\begin{equation}
\bphi(-\infty)=2\pi i\sum_{k=1}^m \weight_k\,+4\pi i\weight,
\end{equation}
where $\weight$ can be any coweight of $g$. As already explained in the
introduction this ensures that at $t=-\infty$, when all solitons are
infinitely far from the boundary, the field takes the constant value
$4\pi i\weight$ at $x=0$ which satisfies the boundary condition.
Because the full Lagrangian including the boundary potential is
invariant under constant shifts $\bphi\rightarrow\bphi+4\pi i\weight$
for any coweight $\weight$, all configurations which differ only by
such a shift are physically equivalent. 

In affine Toda theory on the whole line two configurations
which differ by $2\pi i\weight$ are equivalent. But on the half
line, unless $\weight$ is equal to twice some other coweight,
only one of these combinations can satisfy the boundary condition
and the other one is projected out of the theory.

\subsection{The energy}

It was observed by Bowcock \cite{Bow96} that the constant $C$ occurring 
above is proportional to the energy of the solution. This is seen as
follows.
The energy functional for Toda theory on the half-line consists of a
contribution from the bulk and a contribution from the boundary, $E[\bphi%
]=E_{\text{{\small bulk}}}[\bphi]+E_{\text{{\small boundary}}%
}[\bphi_0]$, where the bulk contribution is similar to eq.~(\ref{E}%
) or eq.~(\ref{et}) except that the integration runs only from $-\infty $ to 
$0$. The boundary contribution is 
\begin{equation}
E_{\text{{\small boundary}}}[\bphi_0]=-\varepsilon\frac 2{\beta^2}
\sum_{i=0}^n\,\eta _i\,e^{\aroot_i\cdot \bphi(0)/2}.
\end{equation}
Again, for soliton solutions the bulk contribution is given by a surface
term and, putting the two together, one has 
\begin{equation}
E[\tau ]=-\frac 2{\beta ^2}\sum_{i=0}^n\left[ 
\frac{2}{\alpha_i^2}\frac{\tau _i^{\prime }}{\tau
_i}+\varepsilon\,\eta _i\,e^{\aroot%
_i\cdot \bphi/2}\right] ^{x=0},  \label{f}
\end{equation}
(the contribution from $x=-\infty $ is zero). Comparing this to eq.~(\ref
{taubc1}) one finds that
\begin{equation}
E=-\frac{2}{\beta^2}\tilde{h}\,C,~~~\text{where }
\tilde{h}=\left(\sum_{i=0}^n\frac{2\eta_i}{\alpha_i^2}\right).
\end{equation}
From here on we will specialize to $a_n^{(1)}$. Then $\tilde{h}=n+1$.

\subsection{Restricting two-soliton solutions to the half-line \label%
{subsect:double}}

We will now look for two-soliton solutions which satisfy the boundary
conditions (\ref{taubc}) and which can thus be restricted to the
half-line to describe the reflection of a single soliton off a boundary.
It follows from integrability that the reflection of a soliton off the
boundary has to be completely elastic. The mass and the velocity of the
soliton have to be the same before and after the reflection. In terms of the
two-soliton solution on the whole line this means that the right-moving and
the left-moving solitons have to have the same but opposite velocities and
have to lie in either the same or in conjugate representations. Such a
two-soliton solution is given by eq.~(\ref{twosolitons}) with 
\begin{equation}
\Omega _1=\sigma (x-vt)-\xi _1,~~~~\Omega _2=\sigma (x+vt)-\xi _2,
\end{equation}
We first treat the case
\begin{equation}\label{c38}
a_2=n+1-a_1,
\end{equation}
i.e., the reflected soliton is in the conjugate representation. We introduce
the abbreviations $a_1=a$, $X_{12}=X$ and $(\xi _1+\xi _2)/2=\xi $ and find
from eq.~(\ref{twosolitons}) that 
\begin{eqnarray}
\tau _i|_{x=0} &=&1+Xe^{-2\xi }+\omega ^{ia}e^{-\sigma vt-\xi _1}+\omega
^{-ia}e^{\sigma vt-\xi _2}, \\
\tau _i^{\prime }|_{x=0} 
&=&\sigma \left( 2Xe^{-2\xi }+\omega ^{ia}e^{-\sigma
vt-\xi _1}+\omega ^{-ia}e^{\sigma vt-\xi _2}\right) .
\end{eqnarray}

Substituting these expressions into the boundary conditions (\ref{taubc})
one obtains the equations 
\begin{eqnarray}
0\,=\,B_0 &+&B_1\left( \omega ^{ia}e^{-\sigma vt-\xi _1}+\omega
^{-ia}e^{\sigma vt-\xi _2}\right)  \label{Beqn} \\
&&+B_2\left( \omega ^{2ia}e^{2(-\sigma vt-\xi _1)}+\omega ^{-2ia}e^{2(\sigma
vt-\xi _2)}\right) ,~~~~\forall i  \nonumber
\end{eqnarray}
where 
\begin{eqnarray}
B_2 &=&\varepsilon^2-\left( C-\sigma \right) ^2, \\
B_1 &=&\left( \varepsilon^2(\omega ^a+\omega ^{-a})
-2C\left( C-\sigma \right) \right) 
\nonumber \\
&&+e^{-2\xi }X\left(\varepsilon^2( \omega ^a+\omega ^{-a})
-2\left( C-2\sigma \right) \left(
C-\sigma \right) \right) \\
B_0 &=&\allowbreak \left( \varepsilon^2
-C^2\right) +e^{-2\xi }\left( \varepsilon^2(\omega ^{2a}+\omega^{-2a})
-2(C-\sigma )^2\right)  \nonumber \\
&&+2e^{-2\xi }X\left( \varepsilon^2-C^2+2\sigma C\right) \\
&&+e^{-4\xi }X^2\left( \varepsilon^2-C^2+4\sigma \left( C-\sigma \right) 
\right) 
\nonumber
\end{eqnarray}

We can easily calculate the constant $C$ 
by making use of the fact that it is proportional
to the energy and thus is time-independent. This allows us to
choose to evaluate it at $t=-\infty$ where at $x=0$ we have
$\tau_i^\prime/\tau_i=\sigma$ and $\bphi=0$ and thus
\begin{equation}\label{solC}
C=\sigma+\varepsilon.
\end{equation}

Because the equations (\ref{Beqn}) have to hold for all times, all the $B_p$
have to be zero if $v\neq 0$. Even if $v=0$ the terms still have to vanish
separately unless $\omega ^{2a}=1$. We will deal with this case
separately later. $B_2=0$ automatically because of \eq{solC}.
The equation $B_1=0$ becomes
\begin{equation}\label{xx}
e^{2\xi }=X\frac{\left( 1-\varepsilon \frac{m_a^2}{2\sigma }\right) }
{\left(1+\varepsilon \frac{m_a^2}{2\sigma }\right) }
\end{equation}
where we used that $\omega^a+\omega^{-a}-2=-m_a^2$ and that
$m_a^2<2\sigma$ unless $a=(n+1)/2$ (the case $\omega^{2a}=1$).
We are left with one more equation $B_0=0$ and no further parameters which
could be adjusted. Luckily one finds 
that this equation is satisfied automatically.

Using that the expression (\ref{intfunc}) for $X$ simplifies to 
\begin{equation}
X=\left( 1+\frac{m_a^2}{2\sigma }\right) \left( 1-\frac{m_a^2}{2\sigma }%
\right), \label{gamma}
\end{equation}
we find
\begin{equation}\label{e45}
e^{2\xi}=\left\{
\begin{array}{l}
\left(
1-\varepsilon \,\frac{m_a^2}{2\sigma }\right)^2
\text{for }\varepsilon=\pm 1\\
\left(1-\varepsilon \,\frac{m_a^2}{2\sigma }\right)
\left(1+\varepsilon \,\frac{m_a^2}{2\sigma }\right)~~~
\text{for }\varepsilon=0
\end{array}
\right.
\end{equation}
This can be expressed in terms of the rapidity $\theta $ of the soliton by
noting that 
\begin{equation}\label{e50}
1\pm \frac{m_a^2}{2\sigma }=\frac{\cosh \theta \pm \sin \frac{a\pi }{n+1}}{%
\cosh \theta }.
\end{equation}
From this expression it is immediately obvious that $1\pm \frac{m_a^2}{%
2\sigma }$ is always real and positive for real $\theta $ (real velocity)
and therefore $\xi $ is real. However for imaginary $\theta$ (imaginary
velocity), which will be relevant for breather solutions, 
$1-\frac{m_a^2}{2\sigma }$ can be negative and thus for $\varepsilon=0$
the imaginary part of $2 \xi$ can be equal to  $\pi$.

\FIGURE{
\setlength{\unitlength}{0.7mm}
\begin{picture}(200,106)
\thicklines
\drawcenteredtext{160.0}{100.0}{\small{$\varepsilon=1$ boundary}}
\drawlefttext{164.0}{76.0}{$\frac{-\xi_1+\log X}{\sigma v}$}
\drawlefttext{164.0}{26.0}{$\frac{\xi_2-\log X}{\sigma v}$}
\Thicklines
\drawpath{162.0}{90.0}{10.0}{14.0}
\drawpath{162.0}{12.0}{10.0}{88.0}
\thinlines
\drawpath{160.0}{24.0}{160.0}{10.0}
\drawpath{160.0}{22.0}{164.0}{24.0}
\drawpath{160.0}{20.0}{164.0}{22.0}
\drawpath{160.0}{18.0}{164.0}{20.0}
\drawpath{160.0}{16.0}{164.0}{18.0}
\drawpath{160.0}{14.0}{164.0}{16.0}
\drawpath{160.0}{12.0}{164.0}{14.0}
\drawpath{160.0}{10.0}{164.0}{12.0}
\drawlefttext{164.0}{12.0}{$\frac{\xi_2}{\sigma v}$}
\drawlefttext{164.0}{90.0}{$\frac{-\xi_1}{\sigma v}$}
\drawcenteredtext{38.0}{20.0}{$x=0$}
\drawpath{98.0}{24.0}{98.0}{78.0}
\drawcenteredtext{160.0}{6.0}{$x=0$}
\drawcenteredtext{98.0}{20.0}{$d=\frac{\xi_1+\xi_2-\log X}{2\sigma}$}
\drawpath{98.0}{66.0}{102.0}{68.0}
\drawpath{98.0}{64.0}{102.0}{66.0}
\drawpath{98.0}{62.0}{102.0}{64.0}
\drawpath{98.0}{60.0}{102.0}{62.0}
\drawpath{98.0}{58.0}{102.0}{60.0}
\drawpath{98.0}{56.0}{102.0}{58.0}
\drawpath{98.0}{54.0}{102.0}{56.0}
\drawpath{98.0}{52.0}{102.0}{54.0}
\drawpath{98.0}{50.0}{102.0}{52.0}
\drawpath{98.0}{48.0}{102.0}{50.0}
\drawpath{98.0}{46.0}{102.0}{48.0}
\drawpath{98.0}{44.0}{102.0}{46.0}
\drawpath{98.0}{42.0}{102.0}{44.0}
\drawpath{98.0}{40.0}{102.0}{42.0}
\drawpath{98.0}{38.0}{102.0}{40.0}
\drawpath{98.0}{36.0}{102.0}{38.0}
\drawpath{98.0}{68.0}{102.0}{70.0}
\drawpath{98.0}{70.0}{102.0}{72.0}
\drawpath{98.0}{72.0}{102.0}{74.0}
\drawpath{98.0}{74.0}{102.0}{76.0}
\drawpath{98.0}{76.0}{102.0}{78.0}
\drawcenteredtext{98.0}{86.0}{\small{$\varepsilon=0$ boundary}}
\drawpath{170.0}{34.0}{168.0}{36.0}
\drawcenteredtext{168.0}{38.0}{$x$}
\drawpath{168.0}{32.0}{170.0}{34.0}
\drawpath{162.0}{34.0}{170.0}{34.0}
\drawpath{160.0}{24.0}{160.0}{96.0}
\drawpath{160.0}{96.0}{158.0}{94.0}
\drawpath{160.0}{96.0}{162.0}{94.0}
\thicklines
\drawcenteredtext{38.0}{100.0}{\small{$\varepsilon=-1$ boundary}}
\thinlines
\drawpath{38.0}{90.0}{42.0}{92.0}
\drawpath{38.0}{88.0}{42.0}{90.0}
\drawpath{38.0}{86.0}{42.0}{88.0}
\drawpath{38.0}{84.0}{42.0}{86.0}
\drawpath{38.0}{82.0}{42.0}{84.0}
\drawpath{38.0}{80.0}{42.0}{82.0}
\drawpath{38.0}{78.0}{42.0}{80.0}
\drawpath{38.0}{76.0}{42.0}{78.0}
\drawpath{38.0}{74.0}{42.0}{76.0}
\drawpath{38.0}{72.0}{42.0}{74.0}
\drawpath{38.0}{70.0}{42.0}{72.0}
\drawpath{38.0}{68.0}{42.0}{70.0}
\drawpath{38.0}{66.0}{42.0}{68.0}
\drawpath{38.0}{64.0}{42.0}{66.0}
\drawpath{38.0}{62.0}{42.0}{64.0}
\drawpath{38.0}{60.0}{42.0}{62.0}
\drawpath{38.0}{58.0}{42.0}{60.0}
\drawpath{38.0}{56.0}{42.0}{58.0}
\drawpath{38.0}{54.0}{42.0}{56.0}
\drawpath{38.0}{52.0}{42.0}{54.0}
\drawpath{38.0}{50.0}{42.0}{52.0}
\drawpath{38.0}{48.0}{42.0}{50.0}
\drawpath{38.0}{46.0}{42.0}{48.0}
\drawpath{38.0}{44.0}{42.0}{46.0}
\drawpath{38.0}{42.0}{42.0}{44.0}
\drawpath{38.0}{40.0}{42.0}{42.0}
\drawpath{38.0}{38.0}{42.0}{40.0}
\drawpath{38.0}{36.0}{42.0}{38.0}
\drawpath{38.0}{34.0}{42.0}{36.0}
\drawpath{38.0}{32.0}{42.0}{34.0}
\drawpath{38.0}{30.0}{42.0}{32.0}
\drawpath{38.0}{28.0}{42.0}{30.0}
\drawpath{38.0}{26.0}{42.0}{28.0}
\drawpath{38.0}{24.0}{42.0}{26.0}
\drawpath{160.0}{90.0}{164.0}{92.0}
\drawpath{160.0}{88.0}{164.0}{90.0}
\drawpath{160.0}{86.0}{164.0}{88.0}
\drawpath{160.0}{84.0}{164.0}{86.0}
\drawpath{160.0}{82.0}{164.0}{84.0}
\drawpath{160.0}{80.0}{164.0}{82.0}
\drawpath{160.0}{78.0}{164.0}{80.0}
\drawpath{160.0}{76.0}{164.0}{78.0}
\drawpath{160.0}{74.0}{164.0}{76.0}
\drawpath{160.0}{72.0}{164.0}{74.0}
\drawpath{160.0}{70.0}{164.0}{72.0}
\drawpath{160.0}{68.0}{164.0}{70.0}
\drawpath{160.0}{66.0}{164.0}{68.0}
\drawpath{160.0}{64.0}{164.0}{66.0}
\drawpath{160.0}{62.0}{164.0}{64.0}
\drawpath{160.0}{60.0}{164.0}{62.0}
\drawpath{160.0}{58.0}{164.0}{60.0}
\drawpath{160.0}{56.0}{164.0}{58.0}
\drawpath{160.0}{54.0}{164.0}{56.0}
\drawpath{160.0}{52.0}{164.0}{54.0}
\drawpath{160.0}{50.0}{164.0}{52.0}
\drawpath{160.0}{48.0}{164.0}{50.0}
\drawpath{162.0}{34.0}{30.0}{34.0}
\drawpath{38.0}{24.0}{38.0}{96.0}
\drawpath{66.0}{28.0}{162.0}{76.0}
\thicklines
\path(34.0,26.0)(34.0,26.0)(34.99,26.51)(35.97,27.02)(36.95,27.52)(37.93,28.02)(38.9,28.52)(39.86,29.02)(40.81,29.51)(41.75,29.97)
\path(41.75,29.97)(42.7,30.45)(43.63,30.93)(44.56,31.38)(45.47,31.86)(46.38,32.31)(47.29,32.77)(48.18,33.2)(49.06,33.65)(49.95,34.08)
\thinlines
\drawpath{162.0}{26.0}{34.0}{90.0}
\thicklines
\path(49.95,34.08)(50.83,34.5)(51.7,34.93)(52.55,35.36)(53.4,35.77)(54.25,36.18)(55.08,36.58)(55.91,36.97)(56.75,37.36)(57.55,37.75)
\path(57.55,37.75)(58.36,38.12)(59.16,38.52)(59.97,38.87)(60.75,39.25)(61.54,39.61)(62.3,39.97)(63.05,40.3)(63.83,40.65)(64.58,41.0)
\path(64.58,41.0)(65.33,41.34)(66.05,41.68)(66.8,42.0)(67.51,42.3)(68.23,42.62)(68.94,42.93)(69.62,43.25)(70.33,43.54)(71.01,43.84)
\path(71.01,43.84)(71.69,44.12)(72.37,44.4)(73.04,44.68)(73.69,44.95)(74.33,45.22)(74.98,45.48)(75.62,45.75)(76.26,46.0)(76.87,46.25)
\path(76.87,46.25)(77.5,46.48)(78.08,46.72)(78.69,46.95)(79.3,47.18)(79.87,47.4)(80.44,47.61)(81.01,47.83)(81.58,48.04)(82.16,48.23)
\path(82.16,48.23)(82.69,48.43)(83.25,48.62)(83.76,48.8)(84.3,48.98)(84.83,49.15)(85.33,49.33)(85.86,49.5)(86.33,49.65)(86.83,49.8)
\path(86.83,49.8)(87.33,49.95)(87.8,50.09)(88.26,50.24)(88.73,50.36)(89.19,50.5)(89.65,50.61)(90.08,50.74)(90.51,50.84)(90.94,50.95)
\path(90.94,50.95)(91.37,51.06)(91.79,51.15)(92.19,51.24)(92.58,51.33)(92.98,51.4)(93.37,51.49)(93.75,51.56)(94.12,51.61)(94.48,51.68)
\path(94.48,51.68)(94.83,51.72)(95.18,51.77)(95.51,51.83)(95.86,51.86)(96.19,51.9)(96.5,51.93)(96.8,51.95)(97.12,51.97)(97.41,51.97)
\path(97.41,51.97)(97.69,51.99)(97.98,52.0)(98.0,52.0)
\path(162.0,76.0)(162.0,76.0)(161.0,75.47)(160.0,74.97)(159.02,74.44)(158.05,73.94)(157.08,73.44)(156.11,72.97)(155.16,72.47)(154.22,72.0)
\path(154.22,72.0)(153.27,71.54)(152.36,71.08)(151.41,70.61)(150.5,70.16)(149.6,69.69)(148.69,69.25)(147.8,68.83)(146.91,68.38)(146.02,67.94)
\path(146.02,67.94)(145.16,67.54)(144.27,67.11)(143.44,66.72)(142.58,66.3)(141.74,65.91)(140.88,65.5)(140.07,65.11)(139.25,64.75)(138.41,64.36)
\path(138.41,64.36)(137.61,64.0)(136.82,63.63)(136.02,63.27)(135.24,62.9)(134.44,62.55)(133.66,62.22)(132.91,61.88)(132.16,61.54)(131.41,61.22)
\path(131.41,61.22)(130.66,60.9)(129.91,60.58)(129.19,60.27)(128.47,59.97)(127.75,59.68)(127.05,59.38)(126.35,59.08)(125.63,58.8)(124.95,58.54)
\path(124.95,58.54)(124.27,58.27)(123.61,58.0)(122.94,57.74)(122.27,57.49)(121.63,57.24)(121.0,57.0)(120.36,56.75)(119.72,56.52)(119.11,56.29)
\path(119.11,56.29)(118.48,56.08)(117.88,55.86)(117.26,55.65)(116.69,55.44)(116.11,55.25)(115.51,55.05)(114.94,54.88)(114.37,54.68)(113.83,54.52)
\path(113.83,54.52)(113.26,54.33)(112.73,54.18)(112.19,54.02)(111.68,53.86)(111.16,53.72)(110.62,53.58)(110.12,53.45)(109.62,53.31)(109.12,53.18)
\path(109.12,53.18)(108.66,53.06)(108.18,52.95)(107.69,52.84)(107.25,52.75)(106.79,52.65)(106.33,52.56)(105.9,52.47)(105.44,52.38)(105.04,52.31)
\path(105.04,52.31)(104.61,52.25)(104.19,52.18)(103.8,52.11)(103.4,52.06)(103.01,52.02)(102.62,51.97)(102.23,51.95)(101.87,51.9)(101.51,51.88)
\path(101.51,51.88)(101.16,51.88)(100.8,51.86)(100.47,51.84)(100.12,51.84)(99.8,51.86)(99.48,51.86)(99.16,51.88)(98.87,51.9)(98.55,51.93)
\path(98.55,51.93)(98.26,51.95)(98.0,52.0)(98.0,52.0)
\path(30.0,78.0)(30.0,78.0)(31.05,77.44)(32.13,76.88)(33.2,76.33)(34.25,75.8)(35.29,75.25)(36.33,74.72)(37.36,74.22)(38.38,73.69)
\path(38.38,73.69)(39.38,73.19)(40.38,72.69)(41.38,72.19)(42.38,71.69)(43.36,71.22)(44.33,70.72)(45.29,70.25)(46.25,69.8)(47.2,69.33)
\path(47.2,69.33)(48.13,68.88)(49.06,68.44)(50.0,68.0)(50.9,67.55)(51.81,67.11)(52.72,66.69)(53.61,66.27)(54.5,65.86)(55.36,65.44)
\path(55.36,65.44)(56.22,65.05)(57.08,64.66)(57.94,64.27)(58.79,63.88)(59.61,63.52)(60.44,63.15)(61.27,62.77)(62.08,62.41)(62.88,62.05)
\path(62.88,62.05)(63.69,61.72)(64.48,61.38)(65.26,61.04)(66.01,60.72)(66.79,60.4)(67.55,60.08)(68.3,59.77)(69.04,59.44)(69.76,59.15)
\path(69.76,59.15)(70.48,58.86)(71.19,58.58)(71.91,58.29)(72.62,58.02)(73.3,57.75)(73.98,57.5)(74.66,57.24)(75.33,56.99)(76.0,56.74)
\path(76.0,56.74)(76.65,56.5)(77.29,56.27)(77.93,56.04)(78.55,55.8)(79.18,55.61)(79.79,55.4)(80.37,55.19)(80.98,55.0)(81.58,54.8)
\path(81.58,54.8)(82.16,54.61)(82.73,54.43)(83.29,54.27)(83.83,54.08)(84.4,53.93)(84.94,53.79)(85.47,53.63)(85.98,53.5)(86.51,53.36)
\path(86.51,53.36)(87.01,53.22)(87.51,53.09)(88.01,52.97)(88.48,52.86)(88.97,52.75)(89.44,52.65)(89.9,52.56)(90.33,52.47)(90.79,52.4)
\path(90.79,52.4)(91.23,52.31)(91.66,52.25)(92.08,52.18)(92.48,52.11)(92.87,52.06)(93.29,52.02)(93.68,51.97)(94.05,51.95)(94.43,51.9)
\path(94.43,51.9)(94.79,51.9)(95.15,51.88)(95.5,51.86)(95.83,51.86)(96.16,51.86)(96.48,51.86)(96.8,51.88)(97.12,51.9)(97.41,51.93)
\path(97.41,51.93)(97.69,51.95)(97.98,52.0)(98.0,52.0)
\path(162.0,26.0)(162.0,26.0)(161.0,26.51)(160.0,27.02)(159.02,27.52)(158.05,28.02)(157.08,28.52)(156.11,29.02)(155.16,29.51)(154.22,29.97)
\path(154.22,29.97)(153.27,30.45)(152.36,30.93)(151.41,31.38)(150.5,31.86)(149.6,32.31)(148.69,32.77)(147.8,33.2)(146.91,33.65)(146.02,34.08)
\path(146.02,34.08)(145.16,34.5)(144.27,34.93)(143.44,35.36)(142.58,35.77)(141.74,36.18)(140.88,36.58)(140.07,36.97)(139.25,37.36)(138.41,37.75)
\path(138.41,37.75)(137.61,38.12)(136.82,38.52)(136.02,38.87)(135.24,39.25)(134.44,39.61)(133.66,39.97)(132.91,40.3)(132.16,40.65)(131.41,41.0)
\path(131.41,41.0)(130.66,41.34)(129.91,41.68)(129.19,42.0)(128.47,42.3)(127.75,42.62)(127.05,42.93)(126.35,43.25)(125.63,43.54)(124.95,43.84)
\path(124.95,43.84)(124.27,44.12)(123.61,44.4)(122.94,44.68)(122.27,44.95)(121.63,45.22)(121.0,45.48)(120.36,45.75)(119.72,46.0)(119.11,46.25)
\path(119.11,46.25)(118.48,46.48)(117.88,46.72)(117.26,46.95)(116.69,47.18)(116.11,47.4)(115.51,47.61)(114.94,47.83)(114.37,48.04)(113.83,48.23)
\path(113.83,48.23)(113.26,48.43)(112.73,48.62)(112.19,48.8)(111.68,48.98)(111.16,49.15)(110.62,49.33)(110.12,49.5)(109.62,49.65)(109.12,49.8)
\path(109.12,49.8)(108.66,49.95)(108.18,50.09)(107.69,50.24)(107.25,50.36)(106.79,50.5)(106.33,50.61)(105.9,50.74)(105.44,50.84)(105.04,50.95)
\path(105.04,50.95)(104.61,51.06)(104.19,51.15)(103.8,51.24)(103.4,51.33)(103.01,51.4)(102.62,51.49)(102.23,51.56)(101.87,51.61)(101.51,51.68)
\path(101.51,51.68)(101.16,51.72)(100.8,51.77)(100.47,51.83)(100.12,51.86)(99.8,51.9)(99.48,51.93)(99.16,51.95)(98.87,51.97)(98.55,51.97)
\path(98.55,51.97)(98.26,51.99)(98.0,52.0)(98.0,52.0)
\thinlines
\drawcenteredtext{50.0}{30.0}{$\frac{\xi_1}{\sigma}$}
\drawcenteredtext{78.0}{30.0}{$\frac{\xi_1-\log X}{\sigma}$}
\drawcenteredtext{148.0}{30.0}{$\frac{\xi_2-\log X}{\sigma}$}
\drawcenteredtext{118.0}{30.0}{$\frac{\xi_2}{\sigma}$}
\drawpath{160.0}{46.0}{164.0}{48.0}
\drawrighttext{34.0}{74.0}{$\frac{\xi_2}{\sigma v}$}
\drawrighttext{34.0}{88.0}{$\frac{\xi_2-\log X}{\sigma v}$}
\drawpath{160.0}{44.0}{164.0}{46.0}
\drawlefttext{42.0}{94.0}{$t$}
\drawpath{36.0}{94.0}{38.0}{96.0}
\drawpath{38.0}{96.0}{40.0}{94.0}
\drawpath{160.0}{42.0}{164.0}{44.0}
\drawpath{160.0}{40.0}{164.0}{42.0}
\drawpath{160.0}{38.0}{164.0}{40.0}
\drawpath{160.0}{36.0}{164.0}{38.0}
\drawpath{160.0}{34.0}{164.0}{36.0}
\drawpath{160.0}{32.0}{164.0}{34.0}
\drawpath{160.0}{30.0}{164.0}{32.0}
\drawpath{160.0}{28.0}{164.0}{30.0}
\drawpath{162.0}{26.0}{166.0}{28.0}
\drawpath{160.0}{24.0}{164.0}{26.0}
\drawpath{78.0}{48.0}{78.0}{48.0}
\drawpath{78.0}{48.0}{78.0}{48.0}
\end{picture}
\setlength{\unitlength}{1mm}
\caption{Diagram of the soliton-antisoliton solution. Depending on the value of $d=
\xi_1+\xi_2-\log X$ the solution can describe the reflection of a soliton on the
left half-line of three different boundaries. We have combined all three cases in this
one diagram. The thick lines are the asymptotic trajectories of the 
reflected soliton. The time delay can be read of from the distance between the
points at which these lines intersect the boundary.
\label{figb}}
}

The displacement $d$ of the center of
mass of the two-soliton system is (see figure \ref{figb})
\begin{equation}
d=\frac{1}{2\sigma}\left(2\xi-\log X\right) =
\frac{1}{2\sigma} \log \left( \frac{1-\varepsilon\frac{m_a^2}{2\sigma}}
{1+\varepsilon\frac{m_a^2}{2\sigma}}\right).  \label{d}
\end{equation}
Thus $\varepsilon $ 
determines whether the virtual scattering takes place behind the boundary
$(\varepsilon=-1)$, exactly at the boundary $(\varepsilon=0)$,
or in front of it $(\varepsilon=+1)$. This is the result for $d$ which
was reported in the introduction in \eq{ddd}.

The time-delay $\Delta t$ which the soliton on the half-line is experiencing
at the boundary is given by $\Delta t=2\xi/\sigma v$ 
(see figure \ref{figb}).
\begin{equation}
\Delta t=\left\{
\begin{array}{l}
\frac{2}{\sigma v}\log \left(
1-\varepsilon \,\frac{m_a^2}{2\sigma }\right)~~~
\text{for }\varepsilon=\pm 1\\
\frac{1}{\sigma v}\log\left(\left(1+\frac{m_a^2}{2\sigma }\right)
\left(1-\frac{m_a^2}{2\sigma }\right)\right)~~~
\text{for }\varepsilon=0
\end{array}
\right.
\label{rt}
\end{equation}

If we choose $a_2=a_1$ instead of \eq{c38} and repeat the analysis,
we find that this does not lead to solutions. Thus a soliton always
turns into an antisoliton upon reflection from the boundary, it can
never come back as a soliton.

The energy of the soliton on the left half-line plus the boundary is
\begin{equation}\label{se}
E=\frac{2(n+1)}{\tilde{\beta}^2}\left(\sigma+\varepsilon)\right).
\end{equation}
Although the energy can be negative it is always higher than the 
energy of the vacuum solution $\bphi=0$.

\subsection{Restricting single solitons to the half-line \label%
{subsect:single}}

When the soliton-antisoliton pair solution is made up of solitons
of the self-conjugate type $a=(n+1)/2$ and one sets the
velocity to zero, then the interaction function $X$ vanishes and the
solution reduces to a single soliton solution. This is exactly
the case $\omega^{2a}=1$ which in the previous section we promised 
to treat separately. Of course this can happen only for $n$ odd.
But it shows that at least in those cases there is the possibility
of having a single soliton solution which satisfies the boundary
condition. We will now study whether there are others.

It is clear that a single \emph{moving }soliton can not satisfy integrable
boundary conditions because at some time this soliton would move off the
half-line and would thus violate energy conservation. We thus need to
only consider stationary solitons. For these 
the values of the tau functions $\tau _j$ and their derivatives $\tau
_j^{\prime }$ at $x=0$ are 
\begin{equation}
\tau _j=1+e^{-\xi }\omega ^{ja},\hspace{1cm}\tau _j^{\prime }=\sigma e^{-\xi
}\omega ^{ja}\hspace{1cm}\text{at }x=0\text{.}  \label{s}
\end{equation}
Substituting this into the boundary conditions (\ref{taubc}) leads to the
equations 
\begin{multline}  \label{bcs} 
0=\left(\varepsilon^2-C^2\right) +
e^{-\xi }\omega ^{ja}\left(\varepsilon^2( \omega ^a+\omega
^{-a})+2C\left( \sigma -C\right) \right)\\
+e^{-2\xi }\omega ^{2ja}\left( \varepsilon^2-C^2-
\sigma\left( \sigma -2C\right)\right) \hspace{%
0.2in}\forall \,\,j=0,\cdots ,n. 
\end{multline}
This has a solution only if $\omega ^{2a}=1$. Assume $\omega
^{2a}\neq 1$. Then (\ref{bcs}) can be satisfied for all $j$ only if each of
the three terms vanish separately. From the first we find 
$C=\pm \varepsilon^2$, the
third then implies $\sigma =2C=\pm 2 \varepsilon^2$ (which is already
possible only for $\epsilon=0$), and substituting this into the second
implies $\omega ^a=-1$ and thus $\omega ^{2a}=1$ after all.

$\omega^{2a}=1$ requires $a=\frac{n+1}2$ and this is excluded if $n$ is
even. However if $n$ is odd and $\varepsilon^2=1$, 
then a single stationary soliton of type $a=%
\frac{n+1}2$ can live on the half plane. It is the collapsed soliton-
antisoliton solution. For such a soliton $\sigma
=m_a=2\sin \frac \pi 2=2.$ Eq. (\ref{bcs}) is then satisfied with $C=1$.
The fact that there is no restriction on the
separation between
the soliton and the antisoliton translates into the fact that 
there is no restriction on the location $\xi $ of the single soliton.

Besides this solution which is obtained from the collapse of a stationary
self-conjugate soliton-antisoliton pair there are no single 
soliton solutions which satisfy the boundary condition. We have
not investigated this rigorously but we believe strongly
that there are also no $2m+1$-soliton solutions 
on the whole line which satisfy the boundary condition
other than the one obtained from a $2m+2$-soliton solution with
one stationary self-conjugate pair.

\subsection{$m$ reflected solitons from $2m$-soliton solutions}
\label{subsect:multi}

We have seen that a pair of a right-moving and a left-moving soliton with
the same velocity satisfy one of the integrable boundary conditions provided
they transform in conjugate representations, their topological charges add
to zero and the displacement $d$
of their center of mass is related to their velocity by eq.~(\ref{d}). The
Hirota method allows us to build a solution which describes any number $m$
of such pairs. We have argued in the introduction that such a 
$2m$ soliton solution also satisfies
integrable boundary conditions at $x=0$. 

Thus the general solution describing $m$ solitons on the half-line with
velocities $v_1,\dots v_m$ is obtained 
(up to a shift in $\phi$) from the general $2m$-soliton
solution given by eq.~(\ref{N}) on the whole line with the solitons arranged
in pairs $\{r,\bar{r}=2m+1-r\}$ such that $a_{\bar{r}}=n+1-a_r$ and 
\begin{eqnarray}
\Omega _r &=&\sigma _r(x-v_rt)-\xi _r-\sum_{\stackrel{1\leq
p\leq m}{p\neq r}}\log X_{pr},~~~~1\leq r\leq m,  \nonumber  \label{or} \\
\Omega _{\bar{r}} &=&\sigma _r(x+v_rt)-\xi_{\bar{r}}-\sum_{\stackrel{%
1\leq p\leq m}{p\neq r}}\log X_{p{\bar{r}}}.
\end{eqnarray}
The reason why we included the $\log X$ terms in eq.~(\ref{or}) becomes
clear when one isolates soliton pair $r$ by taking the limit 
$\xi_p\rightarrow-\infty$ and $\xi_{\bar{p}}\rightarrow\infty$
for $p\neq r$ 
\begin{eqnarray}
\tau_j&\rightarrow&\left(\prod_{\stackrel{1\leq p\leq m}{p\neq r}}
\left(\prod_{\stackrel{p<q\leq m}{q\neq r}}X_{pq}\right)
e^{\Omega_p}\omega^{j a_p}\right) \left(1+\prod_{\stackrel{1\leq p\leq m}{%
p\neq r}}X_{pr}e^{\Omega_r}\omega^{j a_r} \right.  \nonumber \\
&&\left.+\prod_{\stackrel{1\leq p\leq m}{p\neq r}} X_{p\bar{r}}e^{\Omega_{%
\bar{r}}}\omega^{-j a_r}+ X_{r\bar{r}} \prod_{\stackrel{1\leq p\leq m}{p\neq
r}}X_{pr}X_{p\bar{r}} e^{\Omega_r+\Omega_{\bar{r}}}\right).
\end{eqnarray}
For every soliton pair we need that $\exp(\xi_r+\xi_{\bar{r}})$
is given by \eq{e45}

\subsection{Boundary breathers\label{sect:bb}}

If one continues a solution describing a soliton of type $a$, topological
charge $\weight$ and velocity $v$ and an antisoliton of type $\bar{%
a}=n+1-a$, topological charge $-\weight$ and velocity $-v$ to
imaginary values of the velocity $v=iu$, one obtains a breather
solution. It is described by the $\tau $-functions 
\begin{equation}
\tau _j=1+e^{\sigma (x-iut)-\xi _1}\omega ^{aj}+e^{\sigma (x+iut)-\xi
_2}\omega ^{-aj}+X\,e^{2\sigma x-\xi _1-\xi _2}  \label{bb}
\end{equation}
with (see eqs. (\ref{smass}) and (\ref{gamma}))
\begin{equation}
\sigma =\frac{m_a}{\sqrt{1+u^2}},~~~~~X=\cos ^2\frac{a\pi }{n+1}-u^2\sin ^2%
\frac{a\pi }{n+1}.
\end{equation}
These are well-defined solutions on the whole line with real energy and
momentum
provided the velocity and the separation of the two solitons stay within
certain bounds. These breather solutions of $a_n^{(1)}$ Toda theory have
been studied in \cite{Har94}. 

The results of the previous sections imply that such a breather solution
satisfies the boundary condition
provided $\exp(\xi _1+\xi _2)$ is given by \eq{e45}.
We will call these solutions \textit{boundary breathers} because in contrast
to ordinary breathers they can not move but rather are stuck near the
boundary. They should thus not be interpreted as particles but as
excitations of the boundary. 

We need to make sure that the continuation to imaginary velocities
does not lead to singularities in the solution on the left half-line.
Singularities on the right half-line are irrelevant, but singularities
on the left half-line mean that the solution has to be discarded.

From the expression (\ref{tau}) for the field $\bphi$ in terms of the 
tau functions it is clear that $\bphi$ will have a singularity
wherever any one of the $\tau_j$ has a zero. Let us therefore study the
zeros of the expression (\ref{bb}) for the $\tau_j$. This is most easily
done by splitting it into its real and imaginary parts. For this
purpose we introduce the notation
\begin{equation}
\xi=\frac{1}{2}(\xi_1+\xi_2)=\rho+i\zeta,~~~~~
\xi_-=\frac{1}{2}(\xi_1-\xi_2)=\rho_-+i\zeta_-.
\end{equation}

We will first treat the cases $\varepsilon=\pm 1$ and deal with the
Neumann boundary later.

\subsubsection{Boundary breathers for the $\varepsilon=\pm1$
boundary \label{sect:bbe}}

From the expression (\ref{e45}) for $\xi$ we see that
$\zeta=\pi n$ with $n\in\integer$. 
The real and imaginary parts of $\tau_j$ are then
\begin{align}
\real{\tau_j}&=1+2(-1)^n e^{\sigma x -\rho}\cosh{\rho_-}\cos\left(\sigma
u t+\zeta_--\frac{2\pi a j}{n+1}\right)+X e^{2\sigma x -2 \rho},\label{rp}\\
\imaginary{\tau_j}&=2(-1)^n e^{\sigma x-\rho}\sinh{\rho_-}\sin\left(\sigma
u t+\zeta_--\frac{2\pi a j}{n+1}\right).\label{ip}
\end{align}
We see that $\imaginary{\tau_j}$ vanishes whenever 
$\sin(\sigma u t+\zeta_--\frac{2\pi a j}{n+1})
=0$. 
The real part of $\tau_j$ is equal to $1$ at $x=-\infty$. If it
is to have no zero on the left half-line then it must stay positive
there. Upon substituting $\cos(\sigma u t+\zeta_--\frac{2\pi a j}{n+1})
=\pm 1$ into \eq{rp} this gives the conditions
\begin{equation}\label{e334}
1\pm2e^{\sigma x-\rho}\cosh{\rho_-}+X e^{2\sigma x-2\rho}>0,~~~~~
\forall x<0.
\end{equation}
Clearly it is the condition with the minus sign which is the
more stringent one. We substitute the values for $X$ and $e^{-\xi}$
from eqs. (\ref{gamma}) and (\ref{e45}), and multiply by
$\lvert 1-\varepsilon\frac{m_a^2}{2\sigma}\rvert e^{-\sigma x}$.
\begin{equation}
\sign\left(1-\varepsilon\frac{m_a^2}{2\sigma}\right)
\left(\left(1-\varepsilon\frac{m_a^2}{2\sigma}\right)e^{-\sigma x}+
\left(1+\varepsilon\frac{m_a^2}{2\sigma}\right)e^{\sigma x}\right)>
2\cosh{\rho_-}.
\end{equation}
Clearly this requires $1-\varepsilon\frac{m_a^2}{2\sigma}>0$. Then
the left hand side has its only minimum at $\tanh\sigma x=
-\varepsilon\frac{m_a^2}{2\sigma}$. This is on the left half
line for $\varepsilon=1$ and on the right half-line for 
$\varepsilon=-1$. At $x=0$ the left hand side is equal to $2$.
We deduce that there is no non-singular boundary breather solution
for the $\varepsilon=1$ boundary. For $\varepsilon=-1$ the
solution with $\rho_-=0$ has no singularity on the left half-line
but is singular exactly at $x=0$ at the times when
$\sin(\sigma u t+\zeta_--\frac{2\pi a j}{n+1})
=0$.

\subsubsection{Boundary breathers for the Neumann boundary
\label{bbnb}}

We now repeat the analysis for the Neumann boundary ($\varepsilon=0$).
Let us first deal with the case where $e^{2\xi}$ as given in
\eq{e45} is positive so that 
$\zeta=\pi n$ with $n\in\integer$. Then the condition for
absence of singularities is again given by \eq{e334}. But now
$X e^{-2\rho}=1$ (see \eq{xx}) and $e^{-\rho}=1/\sqrt{1-m_a^4/4\sigma^2}$
and the condition becomes
\begin{equation}
\cosh \sigma x>\frac{\cosh\rho_-}{\sqrt{1-\frac{m_a^4}{4\sigma^2}}},~~~~~
\forall x<0.
\end{equation}
This is clearly not satisfied for small negative $x$.

Next we consider the case where $e^{2\xi}$ is negative and
thus $\zeta=\pi n+\pi/2$. This happens if the velocity is big enough
\begin{equation}\label{mu}
u^2>\tan^2\frac{a\pi}{n+1}.
\end{equation}
The expressions (\ref{rp}) and (\ref{ip}) for
the real and imaginary parts of $\tau_j$ change, cosines and sines
get interchanged as well as hyperbolic cosines and hyperbolic sines.
One finally ends up with the condition
\begin{equation}
\cosh \sigma x>\frac{\lvert\cosh\rho_-\rvert}{\sqrt{\lvert
1-\frac{m_a^4}{4\sigma^2}\rvert}},~~~~~
\forall x<0.
\end{equation}
This is indeed true provided $\lvert\rho_-\rvert$, 
the separation between the 
solitons, does not get too big,
\begin{equation}\label{mr}
\lvert \rho_-\rvert<\arcsinh{\sqrt
{\lvert 1-\frac{m_a^4}{4\sigma^2}\rvert}}.
\end{equation}
Thus we conclude that there are nonsingular boundary breather solutions
for the Neumann boundary conditions. They have a minimal frequency 
determined by \eq{mu} and a maximal size determined by \eq{mr}.

Given the discussion in the previous section it
is clear that one can obtain a solution satisfying the boundary
conditions describing several boundary breather excitations
by giving several pairs of solitons imaginary velocities.
Unfortunately the investigation which of these configurations 
are non-singular on the
left half-line and thus constitute valid solutions
becomes exceedingly difficult, but we consider it unlikely
that any of these is nonsingular.

On the whole line there are also breathers with non-zero topological charge
(called type A breathers in \cite{Har94} and excited solitons in \cite
{Hol93,Gan96}). These do not satisfy any of our boundary conditions and
do not lead to boundary breathers.

\section{Real coupling Toda theory on the half-line\label{sect:real}}

Among the solutions for Toda theory on the half-line determined in section 
\ref{sect:3} we can now search for those solutions which are real 
and nonsingular on 
the half-line and which thus play a role in real coupling Toda theory.

The condition that the field $\bphi$ must be real is equivalent to
the requirement that $\tau _j/\tau _0$ must be real and positive for all $j$: 
\begin{equation}
\bphi=\sum_{j=1}^n\alpha _j\log \frac{\tau _j}{\tau _0}\in \Bbb{R}^n
\hspace{0.5cm}\Leftrightarrow \hspace{0.5cm}\frac{\tau _j}{\tau _0}>0
\hspace{0.5cm}\forall j=1,\cdots ,n.
\end{equation}
This must hold for all times and for all $x<0$.

Even without studying these conditions in detail we know that only
very particular solutions can satisfy them. We use the fact that
the soliton and breather solutions of affine Toda theory are never
real on the whole line (again we exclude the rather special case
of sine-Gordon). The only way we can have a real solution
on the left half-line is by hiding the non-real part on the right
half-line. This tells us immediately that there can be no real
solutions
which describe solitons or breathers with a real part to
their velocity.
Because if a solution describes an object which is moving,
then at the far past or the far future it
will move completely onto the left half-line and its non-real part
is no longer hidden. 

Let us study the case of a single stationary soliton on the left
half-line.
Its $\tau $ functions and their real and imaginary parts can be
obtained from the corresponding formulas (\ref{bb}), (\ref{rp}) and
(\ref{ip}) for the boundary breathers by setting the velocity $u$
to zero. 
One finds that in order for $\tau _j/\tau _0$ to be real for all $j$ 
one needs that $\rho_-=0$. Solutions with $\rho_-=0$ are always
singular. We need to ensure that all singularities can be safely
hidden on the right half-line. Thus as in section \ref{sect:bb}
we need that all $\tau_j$ stay positive on the left half-line.
This leads to the condition
\begin{equation}
\left(1-\varepsilon\frac{m_a}{2}\right) e^{-\sigma x}+
\left(1+\varepsilon\frac{m_a}{2}\right) e^{\sigma x}>
2\cos\left(\zeta_--\frac{2\pi a j}{n+1}\right),~~~~
\forall x<0, \forall j.
\end{equation}
The right hand side is always strictly smaller than $2$ for some $j$. 
The left hand side is
equal to $2$ at $x=0$ an for $\varepsilon=-1$ it has its minimum
on the right half-line. Thus the condition is always satisfied for 
$\varepsilon=-1$. However it is never satisfied for $\varepsilon=1$.

We conclude that a stationary soliton solution of type $a$ for the 
$\varepsilon=-1$ boundary can be made real by setting
\begin{equation}\label{t43}
\rho_1=\rho_2=\rho=\log\left(1+\frac{m_a}{2}\right)=
\log\left(1+\sin\frac{a\pi}{n+1}\right).
\end{equation}
The $\zeta_-$ parameter is free.
The energy of this soliton solution is (see \eq{se})
\begin{equation}
E=-2\left( n+1\right) \left( 
1+m_a\right) /\beta ^2.
\end{equation}
It is negative.

These results agree with those of Bowcock in \cite{Bow96}. The advantage
of our approach is that it is easy to determine which particular
boundary condition a solution satisfies. This was explained already
in section \ref{ireal}. Like Bowcock we believe that these are the
only non-singular real solutions, even though we are not able to
prove this. Solutions with more stationary solitons
seem to always have singularities on the left half-line.

\section{Semi-classical calculations\label{sect:semiclassical}}

There is a well known relation \cite{Jac75} between the classical time delay 
$\Delta t(\theta )$ which a soliton experiences during scattering and the
semiclassical phase shift $\delta (\theta )$ which is defined as the leading 
$\hbar $ term in the logarithm of the corresponding quantum transmission
amplitude $A(\theta )$, 
\begin{equation}
A(\theta )=\exp \frac{2i}\hbar \left( \delta (\theta )+\text{O}(\hbar
)\right) .  \label{defdelta}
\end{equation}
The relation is 
\begin{equation}
\delta (\theta )=\frac 12n_B\pi -\frac 12\int_0^\theta d\theta ^{\prime }%
\frac{dE(\theta ^{\prime })}{d\theta ^{\prime }}\Delta t(\theta ^{\prime }),
\label{expdelta}
\end{equation}
Here $n_B$ is a constant, not necessarily an integer. The largest integer
smaller than $n_B$ gives the number of bound states in the direct channel. $%
\theta $ is the relative rapidity between the two particles or between
the particle and the reflecting boundary. It is taken to
be real and positive.

The relationship (\ref{expdelta}) has provided a nontrivial check on the
soliton S-matrices which have been determined by quantum group methods. We
will repeat this calculation below for the soliton S-matrices of $a_2^{(1)}$
affine Toda theory and then extend them to a study of the semiclassical
limit of the quantum reflection matrices for this model which have recently
been determined by Georg Gandenberger\cite{Gan98}.

\subsection{Useful formulas}

In this subsection we will give some formulas which are needed to extract
the semiclassical phase shift $\delta(\theta)$ from the S-matrices $%
S(\theta) $ or the reflection matrices $K(\theta)$ in a form which allows
comparison to the classical time delays.

The scalar factors of the quantum S-matrices and reflection matrices are
often written in terms of infinite products of gamma functions. These
infinite products are generically built out of building blocks of the form 
\begin{equation}  \label{G}
G(\mu,\gamma,a,b)=F(\mu,\gamma,a,b)F(\mu,\gamma,1-\gamma-a,1-\gamma-b),
\end{equation}
where 
\begin{equation}  \label{productrep}
F(\mu,\gamma,a,b)=\prod_{j=1}^\infty\frac{ \Gamma(\mu+\gamma
j+a)\Gamma(-\mu+\gamma j+b)} {\Gamma(-\mu+\gamma j+a)\Gamma(\mu+\gamma j+b)}
\end{equation}
In applications $\mu$ will be proportional to the rapidity and can take
complex values. $\gamma$ will be related to the coupling constant and takes
real positive values. $a$ and $b$ are real. These building blocks have the $%
\gamma\rightarrow 1/\gamma$ duality property 
\begin{equation}  \label{gduality}
G(\mu,\gamma,a,b)=G\left(\frac{\mu}{\gamma},\frac{1}{\gamma},\frac{-a}{\gamma%
}, \frac{-b}{\gamma}\right)
\end{equation}
which follows from 
\begin{equation}  \label{duality}
F(\mu,\gamma,a,b)=F\left(\frac{\mu}{\gamma},\frac{1}{\gamma},\frac{a-1+\gamma%
}{\gamma}, \frac{b-1+\gamma}{\gamma}\right)
\end{equation}
This can be proven by expressing the gamma functions on the left-hand side
as infinite products by using the Weierstrass product formula, then dividing
all numerators and denominators by $\gamma$ and finally reexpressing the
result in terms of gamma functions to arrive at the right-hand side. Further
useful properties of these functions to be used below are 
\begin{eqnarray}
G(-\mu,\gamma,a,b)&=&G(\mu,\gamma,b,a), \\
G\left(\frac{1}{2}-\mu,\gamma,a,b\right)&=& G\left(\mu,\gamma,b+\frac{1}{2}%
,a+\frac{1}{2}\right) \frac{\sin(\frac{\pi}{\gamma}(\mu+a-\frac{1}{2}))} {%
\sin(\frac{\pi}{\gamma}(\mu+b-\frac{1}{2}))}.  \label{crossG}
\end{eqnarray}

Using the integral representation of the gamma function (eq. 8.341.3 in \cite
{Gra80}) 
\begin{equation}
\log \,\Gamma (z)=\int\limits_0^\infty \frac{dt}t\,\left( \frac{%
e^{-zt}-e^{-t}}{1-e^{-t}}+(z-1)e^{-t}\right) ,~~~(\text{Re}(z)>0),
\end{equation}
one obtains an integral representation of $F$ \cite[appendix B]{Gan96} 
\begin{eqnarray}
&&\log \,F(\mu ,\gamma ,a,b)=  \label{integralrep} \\
&&~~~~~~~-\int\limits_0^\infty \frac{dt}t\frac{\sinh \left( \mu t\right) }{%
2\sinh \left( \frac 12\gamma t\right) \sinh \left( \frac 12t\right) }%
e^{-\frac 12\gamma t+\frac 12t}\left( e^{-at}-e^{-bt}\right) ,  \nonumber
\end{eqnarray}
provided the arguments of all gamma functions in $F$ have a positive real
part. This leads to the integral representation for $G$ 
\begin{eqnarray}
&&\log \,G(\mu ,\gamma ,a,b)=  \label{gintrep} \\
&&~~~~-\int\limits_{-\infty }^\infty \frac{dt}te^{\mu t}\frac{\sinh \left(
\frac 12(a+b+\gamma -1)t\right) \sinh \left( \frac 12(a-b)t\right) }{\sinh
\left( \frac 12\gamma t\right) \sinh \left( \frac 12t\right) },  \nonumber
\end{eqnarray}
where we used the symmetry of the integrand under $t\rightarrow -t$ to let
the integration run over the whole real axis. This integral exists for 
\begin{equation}
\left| \mbox{Re}(\mu )\right| <\min (a+\gamma ,b+\gamma ,1-a,1-b).
\end{equation}

The duality properties (\ref{duality}) and (\ref{gduality}) can be derived
directly from the integral representations by changing the integration
variable to $t^\prime=\gamma t$.

For comparison with \cite{Joh} we also give the relation between the
function $G$ and the ``regularized'' quantum dilogarithm $S_q$. Actually we
need the ``square root'' of $S_q$ 
\begin{equation}
S_q^{1/2}(\omega )=\exp \left( \frac 18\int_{-\infty }^\infty \frac{dx}x%
\frac{(\omega )^{ix}}{\sinh (\pi x)\sinh (\xi x)}\right) ,~~~\text{where }%
q=e^{i\xi }.
\end{equation}
(The notation $S_q(\omega )$ for the quantum dilogarithm is not very
fortunate because $S_q$ does not depend on $q=\exp (\xi )$ but on $\xi $
itself.) We have 
\begin{eqnarray}
G(\mu ,\gamma ,a,b) &=&\frac{S_q^{1/2}\left( e^{i\pi (2\mu +2b+\gamma
+1)}\right) S_q^{1/2}\left( e^{-i\pi (2\mu +2b+\gamma +1)}\right) }{%
S_q^{1/2}\left( e^{i\pi (-2\mu +2b+\gamma +1)}\right) S_q^{1/2}\left(
e^{-i\pi (-2\mu +2b+\gamma +1)}\right) } \\
&&\times \frac{S_q^{1/2}\left( e^{i\pi (-2\mu +2a+\gamma +1)}\right)
S_q^{1/2}\left( e^{-i\pi (-2\mu +2a+\gamma +1)}\right) }{S_q^{1/2}\left(
e^{i\pi (2\mu +2a+\gamma +1)}\right) S_q^{1/2}\left( e^{-i\pi (2\mu
+2a+\gamma +1)}\right) },  \nonumber
\end{eqnarray}
where now $q=\exp (i\pi \gamma )$.

The arguments of the blocks $F$ which make up the scalar factors of soliton
scattering and reflection matrices are generically such that the equivalence
between the infinite product representation eq.~(\ref{productrep}) and the
integral representation eq.~(\ref{integralrep}) can not be shown to hold
(the arguments of the gamma functions can have a negative real part). We
believe that in those cases the integral formula rather than the infinite
product of gamma functions should be used. It has the advantage that it is
easy to see for what values of $\mu$ it is well defined and it is easy to
take the semiclassical limit.

It will turn out that for taking the semiclassical limit of the quantum
phase factors we will need to know how to extract the leading term in $%
1/\gamma $ 
\begin{equation}
\log \,G(\mu ,\gamma ,a,b)=\mp \,\frac i{2\pi \gamma }\int\limits_0^{2\pi
i\mu }\,d\theta \,\log \frac{\sinh \left( \frac \theta 2+i\pi a\right) \sinh
\left( \frac \theta 2-i\pi a\right) }{\sinh \left( \frac \theta 2+i\pi
b\right) \sinh \left( \frac \theta 2-i\pi b\right) }+\mbox{O}(1),
\label{leading}
\end{equation}
The minus sign holds for $\mbox{Im}(\mu )>0$ and the plus sign for $\mbox{Im}%
(\mu )<0$. This expression has been obtained by taking the leading term of
the integrand in eq.~(\ref{gintrep}), integrating it by closing the contour
and picking up the residues from the poles at $2\pi im$ for $m$ a positive
or negative integer (depending on where the contour is closed). These
residues were then summed by using 
\begin{equation}
\sum_{m=1}^\infty (-1)^{m+1}\frac{e^{(\phi +r)m}}{m^2}=\int_0^\phi d\theta
\log \left( 1+e^{\theta +r}\right) .
\end{equation}

\subsection{Semiclassical limit of $a_2^{(1)}$ S-matrices}

As a warm-up exercise we will compare the semiclassical limit of the
scattering matrices for $a_2^{(1)}$ solitons with the time delay given in
eq.~(\ref{std}). The $a_n^{(1)}$ soliton scattering matrices were first
determined in \cite{Hol93} and also a comparison with the classical time
delay was carried out. We will use the formula for the $a_2^{(1)}$ Toda
soliton S-matrix given in \cite{Gan96}.

The Lie algebra $a_2=sl(3)$ has two fundamental representations which are
conjugate to each other and each has dimension 3. Therefore the $a_2^{(1)}$
affine Toda theory contains two soliton multiplets, each consisting of three
solitons. We denote the solitons in the first multiplet by $|i,\theta
\rangle ,\,i\in \{1,2,3\}$. The solitons in the second multiplet are the
antiparticles of those in the first and are denoted by $|\bar{i},\theta
\rangle ,\,\bar{i}\in \{1,2,3\}$. The S-matrix maps the incoming states to
the outgoing states, e.g., 
\begin{equation}
\left( |i,\theta ^{\prime }\rangle \otimes |j,\theta \rangle \right) _{\text{%
out}}=S_{kl}^{ij}(\theta -\theta ^{\prime })\left( |k,\theta \rangle \otimes
|l,\theta ^{\prime }\rangle \right) _{\text{in}}.
\end{equation}

Below we will study the semiclassical limits for the scattering of a) two
identical solitons, b) two different solitons, and c) a soliton and its
antisoliton. 
Note that these calculations provide a check only on the part of the quantum
amplitude which goes as $\exp(1/\hbar)$. So, for example, they can not check
the sign of the amplitude.

\sssection{a) Scattering of two identical solitons}

In our notation the amplitude for the scattering of two identical solitons
is 
\begin{equation}
S_{ii}^{ii}(\theta )=-G\left( \frac{3\lambda \theta }{2\pi i},3\lambda
,-\lambda ,0\right) =-G\left( \frac \theta {2\pi i},\frac 1{3\lambda},\frac
13,0\right) ,
\end{equation}
where the last expression is obtained with the help of the duality property (%
\ref{gduality}). The parameter $\lambda $ is related to the coupling
constant $\tilde{\beta} $ and $\hbar $ by 
\begin{equation}
\lambda =\frac{4\pi }{\hbar \tilde{\beta} ^2}-1  \label{lambda}
\end{equation}
Thus the classical limit $\hbar \rightarrow 0$ corresponds to $\lambda
\rightarrow \infty $. Using the formula (\ref{leading}) for the
semiclassical limit of $G$ with Im$(\mu )=-\theta /2\pi <0$ leads to 
\begin{eqnarray}
\delta _{ii}^{ii}(\theta ) &=&\frac 3{\tilde{\beta} ^2}
\int\limits_0^\theta d\theta
^{\prime }\,\log \frac{\sinh \left( \frac{\theta ^{\prime }}2+\frac{i\pi }%
3\right) \sinh \left( \frac{\theta ^{\prime }}2-\frac{i\pi }3\right) }{\sinh
\left( \frac{\theta ^{\prime }}2\right) ^2}  \nonumber \\
&=&-\frac 3{\tilde{\beta} ^2}
\int\limits_0^\theta d\theta ^{\prime }\,\log X_{12}(\theta ^{\prime }),
\end{eqnarray}
where $\gamma _{12}$ is given by eq.~(\ref{gammadef}) with $a_1=a_2$.
Comparing this to the general formula (\ref{expdelta}) we find that $n_B=0$,
i.e., there are no bound states in the direct channel of the identical
particle process, and the time delay is 
\begin{equation}
\Delta t(\theta )=\frac{\log X_{12}}{\sigma v},
\end{equation}
where we used that $E(\theta )=2M\cosh (\theta /2)$ in the center of mass
frame and thus $dE/d\theta =6\sigma v/\tilde{\beta} ^2$.
We find this to be in agreement with eq.~(\ref{std}).

\sssection{b) Transmission of two different solitons}

The amplitude for the transmission of two different solitons is given by 
\cite{Hol93,Gan96} 
\begin{equation}
S_{ij}^{ji}(\theta )=\frac{\sin \left( \frac{3\lambda \theta }{2i}\right) }{%
\sin \left( \pi \lambda -\frac{3\lambda \theta }{2i}\right) }%
\,S_{ii}^{ii}(\theta ),
\end{equation}
and thus the semiclassical phase shift is 
\begin{equation}
\delta _{ij}^{ji}(\theta )=\frac 12\pi 
\frac{4\pi }{\tilde{\beta} ^2}+\delta
_{ii}^{ii}(\theta ),
\end{equation}
i.e., the number of bound states (breathers) is given by the largest integer
smaller than $4\pi /\tilde{\beta}^2$ to this order in $\hbar $. This is in
agreement with the number of bound state poles in $S_{ij}^{ji}$. The time
delay is the same as for the identical particle process because $X_{12}$ is
independent of which solitons of the fundamental soliton multiplet are
taking part in the scattering process.

This is a general feature that the time delay depends only on the scalar
prefactor of the S-matrix and not on the R-matrix part. It is therefore very
easy to generalize these semiclassical calculations to arbitrary solitons of
the $a_n^{(1)}$ Toda theory.

\sssection{c) Transmission of soliton and its antisoliton}

The amplitude $S_{i\bar{i}}^{\bar{i}i}$ for the transmission of a soliton $i$
through its own antisoliton $\bar{i}$ is obtained by crossing symmetry from
the amplitude $S_{ii}^{ii}$ for the identical soliton process 
\begin{eqnarray}
S_{i\bar{i}}^{\bar{i}i}(\theta ) &=&S_{ii}^{ii}(i\pi -\theta )=-G\left(
\frac 12-\frac \theta {2\pi i},\frac 1{3\lambda },\frac 13,0\right) 
\nonumber \\
&=&-G\left( \frac \theta {2\pi i},\frac 1{3\lambda },\frac 12,\frac
56\right) \frac{\sin \left( 3\lambda \pi \left( \frac \theta {2\pi i}-\frac
16\right) \right) }{\sin \left( 3\lambda \pi \left( \frac \theta {2\pi
i}-\frac 12\right) \right) },
\end{eqnarray}
where we used eq.~(\ref{crossG}). From this we obtain the semiclassical
phase shift 
\begin{eqnarray}
\delta _{i\bar{i}}^{\bar{i}i}(\theta ) &=&\frac 12\pi 
\frac{4\pi }{\tilde{\beta}^2}%
+\frac 3{\tilde{\beta}^2}
\int\limits_0^\theta d\theta ^{\prime }\,\log \frac{\sinh
\left( \frac{\theta ^{\prime }}2+\frac{i\pi }2\right) \sinh \left( \frac{%
\theta ^{\prime }}2-\frac{i\pi }2\right) }{\sinh \left( \frac{\theta
^{\prime }}2+\frac{5i\pi }6\right) \sinh \left( \frac{\theta ^{\prime }}2-%
\frac{5i\pi }6\right) }  \nonumber  \label{pa} \\
&=&\frac 12\pi \frac{4\pi }{\tilde{\beta}^2}-
\frac 3{\tilde{\beta}^2}\int\limits_0^\theta
d\theta ^{\prime }\,\log X_{12}(\theta ^{\prime }),
\end{eqnarray}
with $X_{12}$ given by eq.~(\ref{gammadef}) with $a_1=1,a_2=2$.

\subsection{Semiclassical limit of $a_2^{(1)}$ reflection matrices}

Recently Georg Gandenberger has determined the quantum reflection matrix
describing the reflection of the solitons of $a_2^{(1)}$ affine Toda theory
off an integrable boundary \cite{Gan98}. He finds that the requirements of
boundary unitarity, boundary crossing symmetry and the boundary bootstrap
have only a small number of families of solutions. Among these
there are three multiplet-changing families of solutions,
denoted in \cite{Gan98} as $K^{(+)}$, $K^{(-)}$, and $K^{(d)}$. 
Gandenberger conjectures that
$K^{(+)}$ describes the reflection of the Toda solitons off a boundary with
von Neumann condition and that $K^{(d)}$ corresponds to the attractive
boundary (i.e., $\varepsilon=-1$). We will now check these conjectures
semiclassically.

\subsubsection{Reflection matrix for the Neumann boundary}

We introduce the shorthand $\mu=\frac{3\lambda\theta}{2\pi i}$. The quantum
reflection matrix is 
\begin{equation}  \label{rm}
K^{(+)} =\left( 
\begin{array}{ccc}
-\frac{\sin(\pi(\mu-\frac{\lambda}{4}))}{\sin(\pi\frac{\lambda}{2})}
h^2(\lambda) & 
e^{i\pi(\frac{\mu}{3}-\frac{\lambda}{4})} & 
e^{-i\pi(\frac{\mu}{3}-\frac{\lambda}{4})}h(\lambda) \\ 
e^{-i\pi(\frac{\mu}{3}-\frac{\lambda}{4})} & 
-\frac{\sin(\pi(\mu-\frac{\lambda%
}{4}))}{\sin(\pi\frac{\lambda}{2})} \frac{1}{h^2(\lambda)} & 
-e^{i\pi(\frac{%
\mu}{3}-\frac{\lambda}{4})}\frac{1}{h(\lambda)} \\ 
e^{i\pi(\frac{\mu}{3}-\frac{\lambda}{4})}h(\lambda) & 
-e^{-i\pi(\frac{\mu}{3}-%
\frac{\lambda}{4})}\frac{1}{h(\lambda)} & 
-\frac{\sin(\pi(\mu-\frac{\lambda}{4%
}))}{\sin(\pi\frac{\lambda}{2})}
\end{array}
\right) A^{(+)}(\mu)
\end{equation}
where the scalar factor $A^{(+)}(\mu)$ is given by 
\begin{eqnarray}\label{e94}
A^{(+)}(\mu)&=& 
\frac{\sin(\pi\frac{\lambda}{2})}{\sin(\pi(\mu-\frac{3}{4}\lambda))%
} \frac {\sin\left(\frac{\theta}{2i}+\frac{\pi}{12}\right)} {\sin\left(\frac{%
\theta}{2i}-\frac{\pi}{12}\right)} \frac {\sin\left(\frac{\theta}{2i}+\frac{%
5\pi}{12}\right)} {\sin\left(\frac{\theta}{2i}-\frac{5\pi}{12}\right)} 
\nonumber \\
&&\times G\left(\mu,3\lambda, -\frac{7}{4}\lambda,-\frac{1}{4}\lambda\right)
G\left(2\mu,6\lambda, -\frac{3}{2}\lambda,-\frac{5}{2}\lambda\right).
\end{eqnarray}
and $h(\lambda)$ is an as yet undetermined function. Using the duality
property (\ref{gduality}) of $G$ the scalar factor can be rewritten as 
\begin{eqnarray}
A^{(+)}(\mu)&=& 
\frac{\sin(\pi\frac{\lambda}{2})}{\sin(\pi(\mu-\frac{3}{4}\lambda))%
} \frac {\sin\left(\frac{\theta}{2i}+\frac{\pi}{12}\right)} {\sin\left(\frac{%
\theta}{2i}-\frac{\pi}{12}\right)} \frac {\sin\left(\frac{\theta}{2i}+\frac{%
5\pi}{12}\right)} {\sin\left(\frac{\theta}{2i}-\frac{5\pi}{12}\right)} 
\nonumber \\
&&\times G\left(\frac{\theta}{2\pi i},\frac{1}{3\lambda}, \frac{7}{12},\frac{%
1}{12}\right) G\left(\frac{\theta}{2\pi i},\frac{1}{6\lambda}, \frac{3}{12},%
\frac{5}{12}\right).
\end{eqnarray}
In the semiclassical limit only the diagonal entries of the reflection
matrix eq.~(\ref{rm}) survive and we obtain the semiclassical phase shifts 
\footnote{%
We believe that semiclassically all solitons in the same multiplet should
behave identically and that therefore the function $h(\lambda)=1$
semiclassically.} 
\begin{eqnarray}\label{e96}
\delta _i^{\bar{i}}(\theta ) &=&\frac{1}{2}\pi 
\frac{2\pi }{\tilde{\beta}^2}
+\frac{3}{\tilde{\beta}^2}\int\limits_0^\theta 
d\theta ^{\prime }\log \left( \frac{\sinh
\left( \frac{\theta ^{\prime }}2+\frac 7{12}i\pi \right) \sinh \left( \frac{%
\theta ^{\prime }}2-\frac 7{12}i\pi \right) }{\sinh \left( \frac{\theta
^{\prime }}2+\frac 1{12}i\pi \right) \sinh \left( \frac{\theta ^{\prime }}%
2-\frac 1{12}i\pi \right) }\right)  \nonumber \\
&&+\frac 6{
\tilde{\beta} ^2}\int\limits_0^\theta d\theta ^{\prime }\log \left( \frac{%
\sinh \left( \frac{\theta ^{\prime }}2+\frac 3{12}i\pi \right) \sinh \left( 
\frac{\theta ^{\prime }}2-\frac 3{12}i\pi \right) }{\sinh \left( \frac{%
\theta ^{\prime }}2+\frac 5{12}i\pi \right) \sinh \left( \frac{\theta
^{\prime }}2-\frac 5{12}i\pi \right) }\right) .
\end{eqnarray}
Using trigonometric identities this can be transformed into
\begin{eqnarray}
\delta _i^{\bar{i}}(\theta )&=&
\frac 12\pi \frac{2\pi }{\tilde{\beta} ^2}+
\frac{3}{\tilde{\beta}^2}\int\limits_0^\theta\,d\theta^\prime
\log\frac{\cosh \theta^\prime+\sin \frac{\pi}{3}}
{\cosh \theta^\prime-\sin \frac{\pi}{3}}\nonumber\\
&&+\frac{6}{\tilde{\beta}^2}\int\limits_0^\theta\,d\theta^\prime
\log\frac{\cosh \theta^\prime}
{\cosh \theta^\prime+\sin \frac{\pi}{3}},
\end{eqnarray}
which, with the help of \eq{e50} and $dE/d\theta=6\sigma v/\tilde{\beta}^2$,
becomes
\begin{equation}
\delta _i^{\bar{i}}(\theta )=
\frac 12\pi \frac{2\pi }{\tilde{\beta}^2}+\frac{1}{2}
\int\limits_0^\theta\,d\theta^\prime 
\frac{dE(\theta^\prime)}{d\theta^\prime} \ \frac{-1}{\sigma v}
\log\left(\left(1+\frac{m_a^2}{2\sigma^\prime}\right)
\left(1-\frac{m_a^2}{2\sigma^\prime}\right)\right).
\end{equation}
Comparing with \eq{expdelta} we can read off 
that $n_B=2\pi/\tilde{\beta^2}$ and that $\Delta t$ 
agrees with the time delay calculated in \eq{rt} for $\varepsilon=0$.
We thus confirm Gandenberger's conjecture semiclassically.
The bound states arise from the quantization of the boundary
breather which we found in section \ref{bbnb}.

\subsubsection{Reflection matrix for the $\varepsilon=-1$ boundary}

The reflection matrix which Gandenberger conjectures for the attractive
boundary $(\varepsilon=-1)$ is diagonal
\begin{equation}
K^{(d)}(\theta)=A^{(d)}(\mu)\left(
\begin{array}{ccc}
1&0&0\\
0&d(\lambda)&0\\
0&0&\frac{1}{d(\lambda)}
\end{array}
\right),
\end{equation}
where
\begin{equation}\label{ap}
A^{(d)}(\mu)=\frac{\sin\left(\frac{\theta}{2i}-\frac{\pi}{4}\right)}
{\sin\left(\frac{\theta}{2i}+\frac{\pi}{4}\right)}
G(2\mu,6\lambda,-\frac{3}{2}\lambda,-\frac{5}{2}\lambda).
\end{equation}
We will again assume $d(\lambda)=1$.
We notice that the $G$ in \eqref{ap} appeared already as a part of the
factor $A^{(+)}$ in \eq{e94}. We thus quickly find
\begin{eqnarray}
\delta _i^{\bar{i}}(\theta ) &=&\frac 6{
\tilde{\beta} ^2}\int\limits_0^\theta d\theta ^{\prime }\log \left( \frac{%
\sinh \left( \frac{\theta ^{\prime }}2+\frac 3{12}i\pi \right) \sinh \left( 
\frac{\theta ^{\prime }}2-\frac 3{12}i\pi \right) }{\sinh \left( \frac{%
\theta ^{\prime }}2+\frac 5{12}i\pi \right) \sinh \left( \frac{\theta
^{\prime }}2-\frac 5{12}i\pi \right) }\right)\nonumber\\
&=&\frac{6}{\tilde{\beta}^2}\int\limits_0^\theta\,d\theta^\prime
\log\frac{\cosh \theta^\prime}
{\cosh \theta^\prime+\sin \frac{\pi}{3}}\\
&=&\frac{1}{2}
\int\limits_0^\theta\,d\theta^\prime 
\frac{dE(\theta^\prime)}{d\theta^\prime} \ \frac{-2}{\sigma v}
\log\left(1+\frac{m_a^2}{2\sigma^\prime}\right).
\end{eqnarray}
Thus the reflection matrix $K^{(d)}$ reproduces the semiclassical
time delay which we calculated in \eq{rt} for $\varepsilon=-1$.
There are no bound states. This answers our question whether the
classical boundary breather solutions discussed in section
\ref{sect:bbe} which were singular at $x=0$
are physical. They are not, otherwise there should be bound
states in this reflection matrix.

\section{Summary and open problems\label{sect:discussion}}

We have constructed solutions of $a_n^{(1)}$ affine Toda field theory
on the half-line by the method of images: We obtained solutions describing
the reflection of solitons from the boundary at $x=0$ by pairing the
solitons with antisolitons moving with the opposite velocities.
We found that this works for the Neumann boundary condition and for
boundary conditions of the form
\begin{equation}\label{dbc}
\partial_x\bphi|_{x=0}=\left.
\sum_{i=0}^n A_i\aroot_i e^{\aroot_i\cdot\bphi/2}\right|_{x=0}
\text{~~~with~~~}A_i=\pm 1,
\end{equation}
where in the case of $n$ odd the signs $A_i$ have to satisfy the
additional constraint
\begin{equation}\label{dc}
\prod_{i=0}^n A_i =1.
\end{equation}

This leaves several open problems:

1) Corrigan et.al. have shown \cite{Cor94a,Bow95} that the boundary
conditions \eqref{dbc} are integrable also if the constraint \eqref{dc}
is not satisfied. How can one construct classical soliton solutions 
on the half line for boundary conditions with
$\prod_{i=0}^n A_i =-1$? 

2) For algebras other than $a_n^{(1)}$ of $c_n^{(1)}$
the integrable boundary conditions
found by Corrigan et.al. are not of the form \eqref{dbc} but rather
of the form
\begin{equation}
\partial _x\bphi|_{x=0}=\left.\sum_{i=0}^n\ A_i\,
\sqrt{\frac{2\eta _i}{|\aroot_i|^2}}\,\aroot_i\ e^{\aroot_i\cdot 
\bphi/2}\right|_{x=0},  \label{dcbc}
\end{equation}
Solutions satisfying these boundary conditions can not be obtained
by the simple method of images because soliton-antisoliton pairs
do not even satisfy them asymptotically at $t=\pm\infty$ because
$\sum_{i=0}^n
\sqrt{2\eta _i/|\aroot_i|^2}\aroot_i\neq 0$.
It is conceivable that the method could still work after placing
stationary solitons at exactly the right places near the boundary
to soak up the difference between \eqref{dcbc} and \eqref{dbc}.
This needs to be investigated.

3) For non-simply laced algebras Corrigan et.al. also found some
integrable boundary conditions which contain continuous parameters.
Again these can not be treated by our naive method of images.

Coming back to the case of $a_n^{(1)}$ Toda theory with $n$ odd we note
the fact that there are two ways of obtaining solutions describing
solitons reflected off the boundary. One can either let all solitons meet
their mirror antisolitons before the boundary (the $\varepsilon=1$ case)
or behind the boundary (the $\varepsilon=-1$ case). Thus these theories
actually have two completely disjoint sectors. In one sector all
solitons experience time advances during the reflection, in the other
sector they all experience time delays. This is not really a problem,
just a surprise.

We would like to draw attention to the fact that solitons are
converted into their antisolitons during reflection. For the Neumann
boundary condition $\partial_x\bphi=0$ it is immediately clear that
the mirror particle has to be the antisoliton. For the other boundary
conditions which we study this is less obvious but comes out very quickly 
from the calculations. 

Having identified the solutions for $a_n^{(1)}$ Toda theory satisfying
the boundary conditions \eq{dbc} we went on to find  those which are
purely real and which are therefore candidates for the vacuum of real
coupling Toda theory. Under the assumption that there exists a soliton with
topological charge $\weight$ for every weight $\weight$ in the fundamental
representations we found exactly one family of real solutions 
to any boundary 
condition of the type \eqref{dbc} with the restriction
\begin{equation}\label{dcc}
\prod_{i=0}^n A_i=(-1)^{n+1}.
\end{equation}
Note that now we have a restriction also in the case of $n$ even and the
doubling of solutions for $n$ odd mentioned above goes away (only the
$\varepsilon=-1$ case gives a real solution). 

This leaves the problem of how to quantize the theories not satisfying
the condition \eqref{dcc} which don't have a vacuum solution. It also
raises the question what importance to attach to the fact that classical
soliton solutions have been found only for a small subset of the
weights in the fundamental representations. This problem,
as well as that of the interpretation of
the zero mode $\zeta$ in the vacuum solution, was described in
section \ref{ireal}.

Finally we used the time delay extracted from the classical solutions
to perform a semiclassical check on the conjectures of Gandenberger
for the soliton reflection matrices in $a_2^{(1)}$
Toda theory \cite{Gan98}. 
They pass. The existence
of these quantum reflection matrices satisfying all the axioms
of boundary quantum field theory and having the correct semiclassical
limit is a strong indication that imaginary coupling Toda theory in
the presence of the Neumann boundary or the attractive boundary ($
\varepsilon=-1$) makes sense in the quantum regime. Gandenberger
did not find a reflection matrix which could describe the soliton
reflection off the repulsive boundary. This agrees with the prediction
of Fujii and Sasaki \cite{Fuj} that the $\varepsilon=1$ boundary
induces instabilities and leads to a sick quantum theory.

Gandenberger also found reflection matrices which don't match with
any of the boundary conditions discussed in this paper. Furthermore
one can probably derive $a_n^{(1)}$ reflection matrices which don't
change particles into antiparticles \cite{Nep} by starting with
the diagonal solutions \cite{deV} of the boundary Yang-Baxter equation.
Can these too be related to 
$a_n^{(1)}$ Toda theory by studying more general boundary conditions,
perhaps those depending on time-derivatives of the field at the boundary
\cite{Bow95b}?

Even though we have performed most calculations in this paper 
explicitly only for the case of $\hat{g}=a_n^{(1)}$, our calculations
can be applied directly to the case of $\hat{g}=c_n^{(1)}$ because
the solutions of $c_n^{(1)}$ Toda theory are given by those solutions
of $a_{2n-1}^{(1)}$ Toda theory which satisfy $\aroot_i\cdot\bphi=
\aroot_{2n-1}\cdot\bphi$. These are those solutions in which every
$a_n^{(1)}$ soliton is paired with its antisoliton moving with the
same velocity. This means that a single $c_n^{(1)}$ soliton reflected
off the boundary is described by a 4-soliton solution of $a_{2n-1}^{(1)}$
Toda theory on the whole line consisting of the soliton, its antipartner
moving in the same direction, its mirror antisoliton moving in the opposite
direction and the antipartner of this mirror antisoliton. 
Stationary $c_n^{(1)}$ solitons on the
half line are described by the same two-soliton solution as the 
stationary solitons
of $a_{2n-1}^{(1)}$ Toda theory on the half line because the mirror
soliton already moves with the same zero velocity. 

In particular
the real solution to $c_n^{(1)}$ Toda theory with a boundary condition
specified by some given signs $A_0,\dots,A_n$ is given by the real
solution to the $a_{2n=1}^{(1)}$ Toda theory with boundary conditions
given by the same signs $A_0,\dots,A_n$ and additionally $A_{2n-i}=A_i$
for $i=1,\dots,n-1$. Thus the condition \eqref{dcc} for the existence
of real solutions to an $a_n^{(1)}$ Toda theory translates into the 
condition $A_0 A_n=1$ for the $c_n^{(1)}$ Toda theory.

\acknowledgments{

This work was initiated during discussions with Ed Corrigan at the
meeting of the TMR "Integrability, non-perturbative effects, and 
symmetry in quantum field theory"
in Santiago de Compostella in September 1997 which was funded by 
the EU contract FMRX-CT96-0012.

This paper was essentially finished during a visit to Durham University in
June 1998. I would like to thank the members of the Centre for Particle
Physics for their hospitality.

It is a pleasure to acknowledge the important input to the paper 
received from Georg Gandenberger on
the subject of quantum reflection matrices and from Peter Bowcock on
the subject of classical solutions for real coupling Toda theory.
I would like to thank Ed Corrigan and Patrick Dorey for discussions.

This research has been funded by an EPSRC advanced fellowship.
}

\appendix

\end{document}